\documentclass[prc,twocolumn,superscriptaddress,showpacs,preprintnumbers,amsmath,amssymb,floatfix]{revtex4}

\usepackage{graphicx}
\usepackage{dcolumn}
\usepackage{bm}
\usepackage{longtable}

\begin{document}

\title{New effective interaction for $pf$-shell nuclei
 and its implications for the stability of the $N$=$Z$=28 closed core}

\author{M. Honma}
\affiliation{
Center for Mathematical Sciences, University of Aizu,
 Tsuruga, Ikki-machi, Aizu-Wakamatsu, Fukushima 965-8580, Japan}
\author{T. Otsuka}
\affiliation{
Department of Physics and Center for Nuclear Study, University of Tokyo, Hongo, Tokyo 113-0033, Japan}
\affiliation{
RIKEN, Hirosawa, Wako-shi, Saitama 351-0198, Japan}
\author{B. A. Brown}
\affiliation{
National Superconducting Cyclotron Laboratory and Department 
of Physics and Astronomy,
Michigan State University, East Lansing, MI 48824-1321}
\author{T. Mizusaki}
\affiliation{
Institute of Natural Sciences, Senshu University, Higashimita, 
Tama, Kawasaki, Kanagawa 214-8580, Japan}

\date{\today}

\begin{abstract}
The effective interaction GXPF1 for shell-model calculations
 in the full $pf$ shell is tested in detail from various viewpoints
 such as binding energies, electro-magnetic moments and transitions,
 and excitation spectra.
The semi-magic structure is successfully described for
 $N$ or $Z=28$ nuclei, $^{53}$Mn, $^{54}$Fe, $^{55}$Co and $^{56,57,58,59}$Ni,
 suggesting the existence of significant core-excitations
 in low-lying non-yrast states as well as in high-spin yrast states.
The results of $N=Z$ odd-odd nuclei, $^{54}$Co and $^{58}$Cu, also
 confirm the reliability of GXPF1 interaction in the isospin dependent
 properties.
Studies of shape coexistence suggest an advantage of Monte Carlo
 Shell Model over conventional calculations
 in cases where full-space calculations
 still remain too large to be practical.
\end{abstract}

\maketitle

\section{Introduction}
\label{sec:intro}

The effective interaction is a key ingredient for the success of the
 nuclear shell model.
Once a reliable interaction is obtained,
 we can describe various nuclear properties
 accurately and systematically,
 which helps us to understand the underlying structure,
 and to make predictions for
 unobserved properties.
The $pf$ shell for orbitals 
$1p_{3/2}$, $1p_{1/2}$, $0f_{7/2}$ and $0f_{5/2}$ 
is a region where
 the shell model can play an indispensable role,
and is at the frontier of our computational abilities.
In the $pf$ shell one finds the interplay of collective
and single-particle properties,
 both of which the shell model can describe within a
 unified framework.
Since the protons and neutrons occupy the same major shell,
 the proton-neutron
 interaction is relatively strong and one can
study related collective effects such as $T=0$ pairing.
The $pf$-shell nuclei are also of special interest
 from the viewpoint of astrophysics, such
 as the electron capture rate in supernovae explosions.
For all these applications, a suitable effective interaction
 for $pf$-shell nuclei is required.

Because of the spin-orbit splitting,
 there is a sizable energy gap between the 
$f_{7/2}$ orbit
 and the other three orbits ($p_{3/2}$, $p_{1/2}$, $f_{5/2}$).
Thus there exists an $N$ or $Z=28$ ``magic'' number
inside the major shell with the oscillator quantum number $N_{\rm osc}=3$.
For shell-model calculations around this magic number,
 $^{56}$Ni has often been assumed as an ``inert'' core.
However, it has been shown that this core is rather soft\cite{mcsm}
and the closed-shell model for the magic number 28 provides a 
very limited description especially for nuclei 
near $N$ or $Z$=28 semi-magic.
This ``active-two-shell'' problem is a challenge
 to both nuclear models and effective interactions.
We shall discuss this point in this paper, and the word ``cross-shell'' 
 refers to the $N$ or $Z$=28 shell gap hereafter.  
Because of such cross-shell mixing, it is necessary 
to assume essentially the full set of $pf$ configurations
and the associated unified interaction in order to describe
the complete set of data and to have some predictive power.

The effective interaction can in principle be derived from the free 
 nucleon-nucleon interaction.
In fact such microscopic interactions have been proposed
 for the $pf$ shell \cite{kb,g-mat} with certain success particularly 
 in the beginning of the shell.
These interactions, however, fail
 in cases of many valence nucleons, e.g., 
 $^{48}$Ca \cite{mcgrory,g-mat} and $^{56}$Ni.
Especially in the latter, the ground state is predicted to
 be significantly deformed in the full $pf$-shell calculation,
 contrary to its known double-magic structure.

It has been shown that modifications in the monopole
 part of the microscopic interaction can greatly 
 improves the description
 of experimental data.
In fact, KB3\cite{kb3} and KB3G\cite{kb3g} interactions,
 which were obtained on the basis of the microscopic
 Kuo-Brown's G-matrix interaction\cite{kb} with
 various monopole corrections, are remarkably successful
 for describing lighter $pf$-shell nuclei ($A \leq$ 52).
However, as we have pointed out\cite{upf}, these
 interactions fail near $^{56}$Ni.
Therefore it is interesting to investigate to what extent
 the monopole modification is useful as a simple recipe for
 improvement of the microscopic interaction and to what
extent one has to go beyond this.

One feasible way of modifying the microscopic interaction 
 for practical use is to carry out
 an empirical fit to a sufficiently large body of
experimental energy data.
In fact such a method has been successfully applied
to lighter nuclei and has resulted in 
the ``standard'' effective interactions
of Cohen-Kurath \cite{ck} and USD \cite{usd}
 for the $p$ and $sd$-shells, respectively.

Along this line, we have recently 
developed a new effective interaction
 called GXPF1\cite{upf} for use in the $pf$ shell.
Starting from a microscopic interaction, a subset of the
 195 two-body matrix elements and four single-particle energies
 are determined by fitting to 699 energy data
 in the mass range $A=47\sim66$.
It has already been shown that GXPF1
 successfully describes energy levels of various $pf$-shell nuclei,
 such as the first $2^+$ states in even-even $Z\leq 28$ isotopes,
 low-lying states of $^{56,57,58}$Ni, and the systematics of the yrast 
 $1/2^-$, $3/2^-$ and $5/2^-$ states in Ni isotopes \cite{upf}.
Therefore it is now important to analyze the wave functions and
 examine some electromagnetic properties to confirm further the
 reliability of the interaction.

The aim of this paper is to present various results
 predicted by GXPF1 in order to clarify its
 applicability and limitation
 by comparing these results with experimental data.
We also discuss what modifications of the microscopic interaction
 are needed in addition to the monopole corrections
 in order to obtain a better description of nuclear properties
 over the wide range of nuclei in the $pf$ shell.

In the derivation and previous tests of GXPF1,
 the Monte-Carlo Shell Model (MCSM) \cite{qmcd,mcsm} played 
 a crucial role,
 since at that time it was the only feasible way to obtain
 shell-model eigenenergies
 for many states in the middle of the $pf$ shell.
On the other hand, in the present paper, most of the results have 
 been obtained
 by the conventional Lanczos diagonalization method,
 which is now feasible with reasonable accuracy for most
 $pf$-shell nuclei owing to recent developments of
 an efficient shell-model code and fast computers.
We will confirm our previous MCSM results by comparison with
those by such conventional calculations, but there
are still places where the MCSM is necessary.
The $pf$ shell is the current frontier of 
 conventional methods, while the MCSM is applicable and useful
 in much larger model spaces,
 as demonstrated in Refs. \cite{mcsm-n20,f-drip,ni56super,mcsm-ba}.

This paper is organized as follows.
In section \ref{sec:int}, the derivation of GXPF1 is reviewed
 and its general properties are analyzed in some detail.
In section \ref{sec:result} experimental data are compared with 
the results of large-scale shell-model calculations 
based on GXPF1 and some its possible modifications.
We analyze the structure of wave functions focusing on
 the core-excitations across the $N$ or $Z=28$ shell gap.
A summary is given in section \ref{sec:summary}.

\section{Derivation and properties of the effective interaction}
\label{sec:int}

In this section, we first sketch how we have derived
the GXPF1 interaction.
The properties of GXPF1 is then analyzed from various viewpoints.

\subsection{Derivation of GXPF1 interaction}
\label{sub:fit}

An effective interaction for the $pf$ shell can be specified uniquely 
 in terms of interaction parameters consisting of 
 four single-particle energies $\epsilon_a$
 and 195 two-body matrix elements $V(abcd;JT)$,
 where $a, b, \cdots$ denote single-particle orbits,
 and $JT$ stand for the spin-isospin quantum numbers.
The $\epsilon_a$'s include kinetic energies. We take the 
traditional approach of evaluating the interaction energy
in zeroth-order perturbation theory for $n$ nucleons outside
of a closed shell for $^{40}$Ca. 
We adjust the values of the interaction parameters
 so as to fit experimental binding energies and energy levels.
We outline the fitting procedure here, while
 details can be found in \cite{gf40a}.
For a set of $N$ experimental energy data $E_{\rm exp}^k$
 ($k=1$, $\cdots$, $N$),
 we calculate the 
corresponding shell-model eigenvalues $\lambda_k$'s.
 We minimize the quantity
 $\chi^2 = \sum_{k=1}^N ( E_{\rm exp}^k - \lambda_k )^2$
 by varying the values of the interaction parameters. 
Since this minimization is a non-linear process with respect to
 the interaction parameters, we solve it in an iterative way
 with successive variations of those parameters followed by
 diagonalizations of the Hamiltonian until convergence.

Experimental energies used for the fit are limited to those of ground 
 and low-lying states.  
Therefore, certain linear combinations (LC's) of 
 interaction parameters are sensitive to those data
 and can be well determined, 
 whereas the rest of the LC's are not.
We then adopt the so-called LC method \cite{lc}, where the well-determined
 LC's are separated from the rest:
 starting from an initial interaction,
 well-determined LC's are optimized by the fit,
 while the other LC's are kept unchanged
 (fixed to the values given by the initial interaction).

In order to obtain shell-model energies, both the conventional and MCSM 
 calculations are used.  
Since we are dealing with global features of the
 low-lying spectra for essentially all $pf$-shell nuclei, 
 we use a simplified version of MCSM: 
 we search for a few (typically three) most important basis states 
 (deformed Slater determinants) for each spin-parity, 
 and diagonalize the Hamiltonian matrix in the subspace
 spanned by these bases.
The energy eigenvalues are improved by assuming an empirical correction formula, 
 which is determined so that it reproduces the results of
 more accurate calculations for available cases.
This method, the
 few-dimensional approximation with empirical corrections (FDA*),
 actually yields a reasonable estimate of the energy eigenvalues 
 with much shorter computer time.

In the selection of experimental data for the fitting calculation, 
 in order to eliminate intruder states
 from outside the present model space,
 we consider nuclei with $A \geq 47$ and $Z \leq 32$.
As a result 699 data for binding and excitation energies
 (490 yrast, 198 yrare and 11 higher states) were taken
 from 87 nuclei : $^{47-51}$Ca, $^{47-52}$Sc, $^{47-52}$Ti, $^{47-53,55}$V,
 $^{48-56}$Cr, $^{50-58}$Mn, $^{52-60}$Fe, $^{54-61}$Co, $^{56-66}$Ni,
 $^{58-63}$Cu, $^{60-64}$Zn, $^{62,64,65}$Ga and $^{64,65}$Ge.
We assume an empirical mass dependence $A^{-0.3}$
 of the two-body matrix elements similar to that used for 
the USD interaction \cite{usd}, which is meant to take into account
the average mass dependence of a medium-range interaction \cite{spiten}.
We start from the microscopically derived effective interaction
 based on the Bonn-C potential \cite{g-mat},
 which is simply denoted G hereafter.
In the final fit, 70 well-determined LC's are varied,
 and a new interaction, GXPF1, was obtained 
 with an estimated rms error of 168 keV within FDA*.
The resultant single particle energies and two-body matrix elements
 are listed in Table \ref{tbl:gxpf1}.

\begin{table*}
\caption{Two-body matrix elements $V(abcd;JT)$ (in MeV) of GXPF1 interaction.
Single particle energies are taken to be 
 $ -8.6240$,   $-5.6793$,   $-1.3829$ and  $-4.1370$ MeV for the
 $f_{7/2}$, $p_{3/2}$, $f_{5/2}$ and $p_{1/2}$ orbit, respectively.
For calculations of mass $A$ nuclei,
 these two-body matrix elements should be multiplied by a
 factor $(A/42)^{-0.3}$.
\label{tbl:gxpf1}
}
\begin{ruledtabular}
\begin{tabular}{ccccccrcccccccrcccccccrcccccccr}
 $2j_a$ & $2j_b$ & $2j_c$ & $2j_d$ & $J$ & $T$ & $V$ \ \ \ & \  &   $2j_a$ & $2j_b$ & $2j_c$ & $2j_d$ & $J$ & $T$ & $V$ \ \ \ & \  &   $2j_a$ & $2j_b$ & $2j_c$ & $2j_d$ & $J$ & $T$ & $V$ \ \ \ & \   &   $2j_a$ & $2j_b$ & $2j_c$ & $2j_d$ & $J$ & $T$ & $V$ \ \ \ \\ \hline
 7& 7& 7& 7& 1 & 0 &$   -1.2838$ &  &  7& 5& 3& 5& 1 & 0 &$   -1.2721$ &  &  5& 1& 5& 1& 2 & 0 &$   -0.3174$ &  &  7& 5& 7& 1& 4 & 1 &$    0.1907$ \\ 
 7& 7& 7& 7& 3 & 0 &$   -0.8418$ &  &  7& 5& 3& 5& 2 & 0 &$   -0.5980$ &  &  5& 1& 5& 1& 3 & 0 &$   -1.4023$ &  &  7& 5& 3& 3& 2 & 1 &$    0.0717$ \\ 
 7& 7& 7& 7& 5 & 0 &$   -0.7839$ &  &  7& 5& 3& 5& 3 & 0 &$   -0.7716$ &  &  1& 1& 1& 1& 1 & 0 &$   -1.2431$ &  &  7& 5& 3& 5& 1 & 1 &$    0.0521$ \\ 
 7& 7& 7& 7& 7 & 0 &$   -2.6661$ &  &  7& 5& 3& 5& 4 & 0 &$   -0.6408$ &  &  7& 7& 7& 7& 0 & 1 &$   -2.4385$ &  &  7& 5& 3& 5& 2 & 1 &$   -0.4247$ \\ 
 7& 7& 7& 3& 3 & 0 &$   -0.8807$ &  &  7& 5& 3& 1& 1 & 0 &$   -1.4651$ &  &  7& 7& 7& 7& 2 & 1 &$   -0.9352$ &  &  7& 5& 3& 5& 3 & 1 &$   -0.0268$ \\ 
 7& 7& 7& 3& 5 & 0 &$   -0.4265$ &  &  7& 5& 3& 1& 2 & 0 &$   -0.7434$ &  &  7& 7& 7& 7& 4 & 1 &$   -0.1296$ &  &  7& 5& 3& 5& 4 & 1 &$   -0.2699$ \\ 
 7& 7& 7& 5& 1 & 0 &$    1.8998$ &  &  7& 5& 5& 5& 1 & 0 &$   -0.2735$ &  &  7& 7& 7& 7& 6 & 1 &$    0.2783$ &  &  7& 5& 3& 1& 1 & 1 &$    0.0552$ \\ 
 7& 7& 7& 5& 3 & 0 &$    1.0917$ &  &  7& 5& 5& 5& 3 & 0 &$    0.6378$ &  &  7& 7& 7& 3& 2 & 1 &$   -0.5160$ &  &  7& 5& 3& 1& 2 & 1 &$   -0.0153$ \\ 
 7& 7& 7& 5& 5 & 0 &$    1.2853$ &  &  7& 5& 5& 5& 5 & 0 &$    1.1302$ &  &  7& 7& 7& 3& 4 & 1 &$   -0.2969$ &  &  7& 5& 5& 5& 2 & 1 &$   -0.5022$ \\ 
 7& 7& 7& 1& 3 & 0 &$    0.8883$ &  &  7& 5& 5& 1& 2 & 0 &$    0.5447$ &  &  7& 7& 7& 5& 2 & 1 &$    0.2167$ &  &  7& 5& 5& 5& 4 & 1 &$   -0.2709$ \\ 
 7& 7& 3& 3& 1 & 0 &$   -0.4313$ &  &  7& 5& 5& 1& 3 & 0 &$    0.6262$ &  &  7& 7& 7& 5& 4 & 1 &$   -0.4999$ &  &  7& 5& 5& 1& 2 & 1 &$   -0.1537$ \\ 
 7& 7& 3& 3& 3 & 0 &$   -0.3415$ &  &  7& 5& 1& 1& 1 & 0 &$    0.1928$ &  &  7& 7& 7& 5& 6 & 1 &$   -0.5643$ &  &  7& 5& 5& 1& 3 & 1 &$    0.1105$ \\ 
 7& 7& 3& 5& 1 & 0 &$   -0.0907$ &  &  7& 1& 7& 1& 3 & 0 &$   -1.6968$ &  &  7& 7& 7& 1& 4 & 1 &$   -0.2096$ &  &  7& 1& 7& 1& 3 & 1 &$    0.4873$ \\ 
 7& 7& 3& 5& 3 & 0 &$    0.0752$ &  &  7& 1& 7& 1& 4 & 0 &$   -1.0602$ &  &  7& 7& 3& 3& 0 & 1 &$   -0.7174$ &  &  7& 1& 7& 1& 4 & 1 &$   -0.1347$ \\ 
 7& 7& 3& 1& 1 & 0 &$    0.3150$ &  &  7& 1& 3& 3& 3 & 0 &$    0.6411$ &  &  7& 7& 3& 3& 2 & 1 &$   -0.2021$ &  &  7& 1& 3& 5& 3 & 1 &$    0.3891$ \\ 
 7& 7& 5& 5& 1 & 0 &$    0.6511$ &  &  7& 1& 3& 5& 3 & 0 &$   -0.0354$ &  &  7& 7& 3& 5& 2 & 1 &$   -0.1725$ &  &  7& 1& 3& 5& 4 & 1 &$   -0.6111$ \\ 
 7& 7& 5& 5& 3 & 0 &$    0.4358$ &  &  7& 1& 3& 5& 4 & 0 &$   -1.3607$ &  &  7& 7& 3& 5& 4 & 1 &$   -0.2224$ &  &  7& 1& 5& 5& 4 & 1 &$   -0.2248$ \\ 
 7& 7& 5& 5& 5 & 0 &$    0.1239$ &  &  7& 1& 5& 5& 3 & 0 &$   -0.2621$ &  &  7& 7& 3& 1& 2 & 1 &$   -0.0367$ &  &  7& 1& 5& 1& 3 & 1 &$   -0.1586$ \\ 
 7& 7& 5& 1& 3 & 0 &$   -0.1082$ &  &  7& 1& 5& 1& 3 & 0 &$    0.4505$ &  &  7& 7& 5& 5& 0 & 1 &$   -1.3832$ &  &  3& 3& 3& 3& 0 & 1 &$   -1.1165$ \\ 
 7& 7& 1& 1& 1 & 0 &$    0.0271$ &  &  3& 3& 3& 3& 1 & 0 &$   -0.6308$ &  &  7& 7& 5& 5& 2 & 1 &$   -0.2038$ &  &  3& 3& 3& 3& 2 & 1 &$   -0.0887$ \\ 
 7& 3& 7& 3& 2 & 0 &$   -0.5391$ &  &  3& 3& 3& 3& 3 & 0 &$   -2.2890$ &  &  7& 7& 5& 5& 4 & 1 &$   -0.0331$ &  &  3& 3& 3& 5& 2 & 1 &$   -0.4631$ \\ 
 7& 3& 7& 3& 3 & 0 &$   -1.0055$ &  &  3& 3& 3& 5& 1 & 0 &$    0.2373$ &  &  7& 7& 5& 1& 2 & 1 &$   -0.1295$ &  &  3& 3& 3& 1& 2 & 1 &$   -0.6340$ \\ 
 7& 3& 7& 3& 4 & 0 &$   -0.3695$ &  &  3& 3& 3& 5& 3 & 0 &$    0.2276$ &  &  7& 7& 1& 1& 0 & 1 &$   -0.3800$ &  &  3& 3& 5& 5& 0 & 1 &$   -1.2457$ \\ 
 7& 3& 7& 3& 5 & 0 &$   -2.9670$ &  &  3& 3& 3& 1& 1 & 0 &$    1.8059$ &  &  7& 3& 7& 3& 2 & 1 &$   -0.6081$ &  &  3& 3& 5& 5& 2 & 1 &$    0.0719$ \\ 
 7& 3& 7& 5& 2 & 0 &$   -0.6381$ &  &  3& 3& 5& 5& 1 & 0 &$    0.0483$ &  &  7& 3& 7& 3& 3 & 1 &$    0.1561$ &  &  3& 3& 5& 1& 2 & 1 &$   -0.1923$ \\ 
 7& 3& 7& 5& 3 & 0 &$    0.2540$ &  &  3& 3& 5& 5& 3 & 0 &$   -0.0546$ &  &  7& 3& 7& 3& 4 & 1 &$   -0.1398$ &  &  3& 3& 1& 1& 0 & 1 &$   -1.4928$ \\ 
 7& 3& 7& 5& 4 & 0 &$    0.1951$ &  &  3& 3& 5& 1& 3 & 0 &$    0.1150$ &  &  7& 3& 7& 3& 5 & 1 &$    0.5918$ &  &  3& 5& 3& 5& 1 & 1 &$    0.3284$ \\ 
 7& 3& 7& 5& 5 & 0 &$    0.6743$ &  &  3& 3& 1& 1& 1 & 0 &$    0.7675$ &  &  7& 3& 7& 5& 2 & 1 &$    0.0959$ &  &  3& 5& 3& 5& 2 & 1 &$    0.3608$ \\ 
 7& 3& 7& 1& 3 & 0 &$    1.6850$ &  &  3& 5& 3& 5& 1 & 0 &$   -2.7262$ &  &  7& 3& 7& 5& 3 & 1 &$   -0.5230$ &  &  3& 5& 3& 5& 3 & 1 &$    0.3460$ \\ 
 7& 3& 7& 1& 4 & 0 &$    0.1706$ &  &  3& 5& 3& 5& 2 & 0 &$   -1.5110$ &  &  7& 3& 7& 5& 4 & 1 &$   -0.2486$ &  &  3& 5& 3& 5& 4 & 1 &$   -0.2584$ \\ 
 7& 3& 3& 3& 3 & 0 &$   -0.4309$ &  &  3& 5& 3& 5& 3 & 0 &$   -0.5859$ &  &  7& 3& 7& 5& 5 & 1 &$   -0.4810$ &  &  3& 5& 3& 1& 1 & 1 &$   -0.1076$ \\ 
 7& 3& 3& 5& 2 & 0 &$   -1.2708$ &  &  3& 5& 3& 5& 4 & 0 &$   -1.0882$ &  &  7& 3& 7& 1& 3 & 1 &$   -0.1048$ &  &  3& 5& 3& 1& 2 & 1 &$   -0.4545$ \\ 
 7& 3& 3& 5& 3 & 0 &$    0.5790$ &  &  3& 5& 3& 1& 1 & 0 &$   -0.9930$ &  &  7& 3& 7& 1& 4 & 1 &$   -0.3351$ &  &  3& 5& 5& 5& 2 & 1 &$   -0.0560$ \\ 
 7& 3& 3& 5& 4 & 0 &$   -0.7103$ &  &  3& 5& 3& 1& 2 & 0 &$   -0.4885$ &  &  7& 3& 3& 3& 2 & 1 &$   -0.3738$ &  &  3& 5& 5& 5& 4 & 1 &$   -0.3615$ \\ 
 7& 3& 3& 1& 2 & 0 &$   -0.6228$ &  &  3& 5& 5& 5& 1 & 0 &$    0.4770$ &  &  7& 3& 3& 5& 2 & 1 &$   -0.5436$ &  &  3& 5& 5& 1& 2 & 1 &$   -0.4043$ \\ 
 7& 3& 5& 5& 3 & 0 &$    0.1660$ &  &  3& 5& 5& 5& 3 & 0 &$    0.3200$ &  &  7& 3& 3& 5& 3 & 1 &$    0.1836$ &  &  3& 5& 5& 1& 3 & 1 &$    0.0600$ \\ 
 7& 3& 5& 5& 5 & 0 &$    0.0334$ &  &  3& 5& 5& 1& 2 & 0 &$    0.3540$ &  &  7& 3& 3& 5& 4 & 1 &$   -0.4546$ &  &  3& 1& 3& 1& 1 & 1 &$   -0.1594$ \\ 
 7& 3& 5& 1& 2 & 0 &$    1.0933$ &  &  3& 5& 5& 1& 3 & 0 &$    1.0151$ &  &  7& 3& 3& 1& 2 & 1 &$   -0.4262$ &  &  3& 1& 3& 1& 2 & 1 &$   -0.2938$ \\ 
 7& 3& 5& 1& 3 & 0 &$    0.7227$ &  &  3& 5& 1& 1& 1 & 0 &$    0.8137$ &  &  7& 3& 5& 5& 2 & 1 &$    0.0880$ &  &  3& 1& 5& 5& 2 & 1 &$    0.0600$ \\ 
 7& 5& 7& 5& 1 & 0 &$   -4.5802$ &  &  3& 1& 3& 1& 1 & 0 &$   -2.5068$ &  &  7& 3& 5& 5& 4 & 1 &$   -0.2146$ &  &  3& 1& 5& 1& 2 & 1 &$   -0.2490$ \\ 
 7& 5& 7& 5& 2 & 0 &$   -3.2520$ &  &  3& 1& 3& 1& 2 & 0 &$   -2.3122$ &  &  7& 3& 5& 1& 2 & 1 &$   -0.8030$ &  &  5& 5& 5& 5& 0 & 1 &$   -1.2081$ \\ 
 7& 5& 7& 5& 3 & 0 &$   -1.4019$ &  &  3& 1& 5& 5& 1 & 0 &$   -0.0337$ &  &  7& 3& 5& 1& 3 & 1 &$   -0.1814$ &  &  5& 5& 5& 5& 2 & 1 &$   -0.4621$ \\ 
 7& 5& 7& 5& 4 & 0 &$   -2.2583$ &  &  3& 1& 5& 1& 2 & 0 &$    0.6900$ &  &  7& 5& 7& 5& 1 & 1 &$   -0.0889$ &  &  5& 5& 5& 5& 4 & 1 &$   -0.1624$ \\ 
 7& 5& 7& 5& 5 & 0 &$   -0.6084$ &  &  3& 1& 1& 1& 1 & 0 &$    0.8490$ &  &  7& 5& 7& 5& 2 & 1 &$   -0.1750$ &  &  5& 5& 5& 1& 2 & 1 &$   -0.3208$ \\ 
 7& 5& 7& 5& 6 & 0 &$   -3.0351$ &  &  5& 5& 5& 5& 1 & 0 &$   -0.8551$ &  &  7& 5& 7& 5& 3 & 1 &$    0.6302$ &  &  5& 5& 1& 1& 0 & 1 &$   -0.8093$ \\ 
 7& 5& 7& 1& 3 & 0 &$   -0.4252$ &  &  5& 5& 5& 5& 3 & 0 &$   -0.5599$ &  &  7& 5& 7& 5& 4 & 1 &$    0.4763$ &  &  5& 1& 5& 1& 2 & 1 &$   -0.1519$ \\ 
 7& 5& 7& 1& 4 & 0 &$   -0.3789$ &  &  5& 5& 5& 5& 5 & 0 &$   -2.2816$ &  &  7& 5& 7& 5& 5 & 1 &$    0.7433$ &  &  5& 1& 5& 1& 3 & 1 &$    0.2383$ \\ 
 7& 5& 3& 3& 1 & 0 &$    0.8914$ &  &  5& 5& 5& 1& 3 & 0 &$   -0.6276$ &  &  7& 5& 7& 5& 6 & 1 &$   -0.9916$ &  &  1& 1& 1& 1& 0 & 1 &$   -0.4469$ \\ 
 7& 5& 3& 3& 3 & 0 &$    0.6264$ &  &  5& 5& 1& 1& 1 & 0 &$   -0.3161$ &  &  7& 5& 7& 1& 3 & 1 &$    0.3224$ &  &  & & & & &  &  \\ 
\end{tabular}
\end{ruledtabular}
\end{table*}

\subsection{Corrections to the microscopic interaction}
\label{sub:correc}

\begin{figure}
\includegraphics[width=85mm]{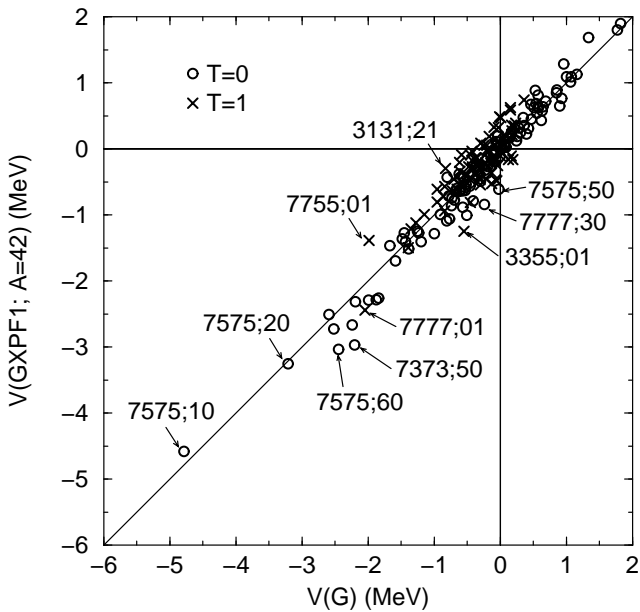}
\caption{Correlation of $V(abcd;JT)$ between G and GXPF1.
The matrix elements of $T$=$0$ and $T$=$1$ are shown by open circles and crosses,
 respectively.
For several matrix elements, corresponding quantum numbers
 are shown by using the notation $2j_a 2j_b 2j_c 2j_d;JT$.
\label{fig:gx1-vs-g}
}
\end{figure}

It is interesting to examine what changes have been made to the
 original G interaction by the empirical fit.
Figure \ref{fig:gx1-vs-g} shows
 a comparison between GXPF1 and G
 for the 195 two-body matrix elements.  
One finds a strong correlation. 
On average, the $T$=0 ($T$=1) matrix elements are modified to be more 
 attractive (repulsive). 
The most attractive matrix elements are $T$=0, with
the largest ones belong to $T$=0 7575 in both GXPF1 and G,
 as indicated in the figure,
 where the notation 7575 refers to the set of
 matrix elements $V(abab;JT)$ with $a=f_{7/2}$ and $b=f_{5/2}$.
It was stressed in \cite{magic,ykis} that this strong $T$=0
interaction between the orbits $j_> = l + 1/2$ and $j_< = l - 1/2$
is an important feature in all mass regions including the $sd$ and $p$
shells. In the $p$ shell variation of these type of matrix elements
relative to the Cohen-Kurath interaction \cite{ck} leads to improvement
\cite{magic,sfo}. One can see in  Fig. \ref{fig:gx1-vs-g} that the
the G and fitted (GXPF1) values for the $j_>$-$j_<$ (7575) 
interactions are very similar.

There are seven matrix elements in which
 the difference between GXPF1 and G is greater than 500 keV.
These matrix elements are listed in Table \ref{tbl:large-diff}.
It is remarkable that these largely-modified matrix elements are
 either of the diagonal $V(abab;JT)$ type which contributes
 to the monopole corrections or of
 the monopole pairing ($J=0$, $T=1$) type.
As for the former type, the 7373 and 7575, $T=0$ matrix elements
 with large $J$ are modified to be more attractive,
 contrasting to the relatively small corrections
 for small $J$ matrix elements.
The two monopole pairing matrix elements shown in the table
 are both related to the $f_{5/2}$ orbit.
As a result of the empirical fit,
 the pairing between $p_{3/2}$ and $f_{5/2}$ (3355) is modified to be
 strongly attractive,
 while that between $f_{7/2}$ and $f_{5/2}$ (7755) is made
 to be less attractive.

\begin{table}
\caption{Comparison of the two-body matrix elements
 $V(abcd;JT)$ (MeV) ($A=42$)
 for which the difference between G and GXPF1 is larger than 500 keV.
\label{tbl:large-diff}
}
\begin{ruledtabular}
\begin{tabular}{ccccccccc}
$2j_a$ & $2j_b$ & $2j_c$ & $2j_d$ & $J$ & $T$   & G & GXPF1 & difference \\ \hline
 7 & 3 & 7 & 3 & \ 5 & 0 &  $-2.2033$ & $-2.9670$ & $-0.7637$ \\
 3 & 3 & 5 & 5 & \ 0 & 1 &  $-0.5457$ & $-1.2457$ & $-0.7000$ \\
 7 & 7 & 7 & 7 & \ 3 & 0 &  $-0.2404$ & $-0.8418$ & $-0.6014$ \\
 7 & 5 & 7 & 5 & \ 6 & 0 &  $-2.4425$ & $-3.0351$ & $-0.5926$ \\
 7 & 5 & 7 & 5 & \ 5 & 0 &  $-0.0211$ & $-0.6084$ & $-0.5873$ \\
 3 & 1 & 3 & 1 & \ 2 & 1 &  $-0.8291$ & $-0.2938$ & $+0.5353$ \\
 7 & 7 & 5 & 5 & \ 0 & 1 &  $-1.9875$ & $-1.3832$ & $+0.6043$ \\
\end{tabular}
\end{ruledtabular}
\end{table}

\subsection{Monopole properties}
\label{sub:monopole}

In order to investigate basic properties of
 an effective Hamiltonian,
 it is convenient to decompose it
 into the monopole part and the multipole part \cite{dufour} as,
\begin{equation}
H = H_m + H_M.
\end{equation}
The monopole part $H_m$ plays a key role for describing
 bulk properties such as
 binding energies and shell-gaps \cite{kb3}, since it determines the average
 energy of eigenstates in a given configuration.
The monopole Hamiltonian is specified by the angular-momentum
 averaged two-body matrix elements:
\begin{equation}
V(ab;T)=\frac{\sum_J (2J+1)V(abab;JT)}{\sum_J (2J+1)},
\label{eqn:centroid}
\end{equation}
 where the summations run over all Pauli-allowed values of the
 angular momentum $J$.

\begin{figure}
\includegraphics[width=85mm]{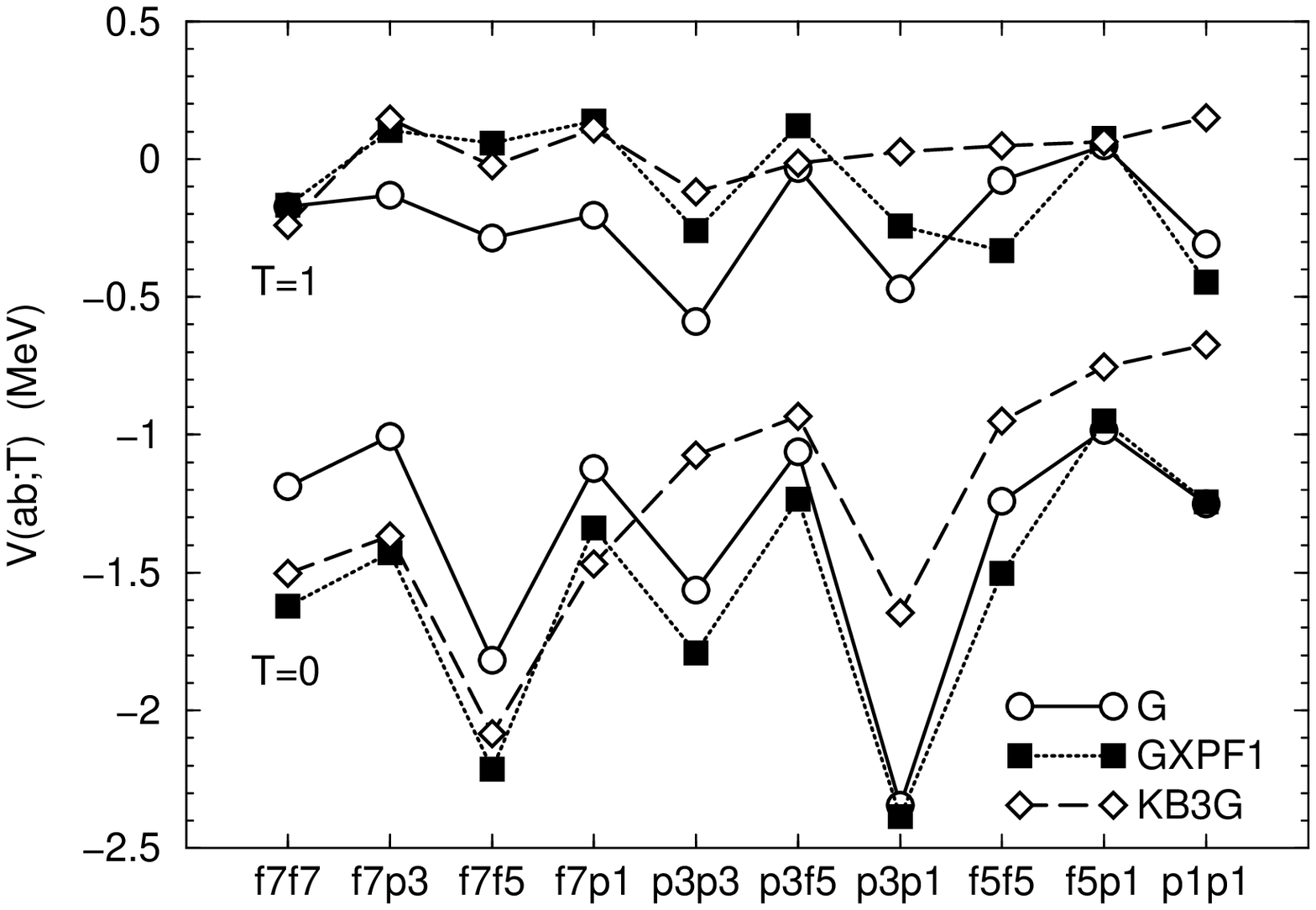}
\caption{Comparison of the monopole
 matrix elements $V(ab;T)$ ($A=42$)
 between G, GXPF1 and KB3G,
 which are shown by circles, squares and diamonds, respectively.
Lines are drawn to guide the eyes.
The orbit-pair label ``f5p3'' stands for $a=f_{7/2}$ and $b=p_{3/2}$,
 for example.
\label{fig:centroid}
}
\end{figure}

Figure \ref{fig:centroid} shows the matrix elements $V(ab;T)$.
As a reference, we consider also the KB3G \cite{kb3g} interaction.
Since this interaction gives an excellent description for $A\leq 52$ nuclei,
 it is expected that the monopole matrix elements of GXPF1 are similar
 to those of KB3G at least for those involving $f_{7/2}$.
In fact, both $T=0$ and $T=1$ average matrix elements of the
 f7f7, f7p3, f7f5 and f7p1 orbit pairs
 are rather close between GXPF1 and KB3G.
The $T=1$ matrix elements for p3p3, p3f5 and f5p1 are also
 not very different, which are
 important for describing $Z<28$, $N>28$ nuclei.
On the other hand, the similarity is lost in other matrix elements,
 especially for $T=0$ matrix elements between $p_{3/2, 1/2}$ orbits.
Therefore, GXPF1 and KB3G could give a very different description
 for nuclei with $Z, N>28$.

In order to confirm this observation, we have carried out
 the FDA* calculations by using KB3G
 for the similar set of nuclei included in the fitting calculations.
Table \ref{tbl:gx1-vs-kb3g} summarize the
 estimated rms difference between calculated energies and
 the experimental data for both GXPF1 and KB3G.
The data are classified into four groups 
 from the viewpoint of the location in the isotope table :
 (a) $Z, N<28$, (b) $Z$ or $N=28$, (c) $Z<28$, $N>28$ and
 (d) $Z, N>28$.
The rms deviations are estimated for each group, and
 the yrast and yrare states are considered separately.

It is clearly seen that the rms deviations
 by GXPF1 are almost the same in all groups.
On the other hand, those by KB3G 
 become larger for the group (b) and (d)
 in comparison to (a) and (c).
The deviation is especially large for the group (b),
where the major differences can be found in
 $^{56}$Ni, $^{55}$Co and $^{57}$Ni.
Since the property of the core-excitations
 appear directly in the low-lying spectra in these semi-magic nuclei,
 this result suggests that the core-excitations are not well
described by the KB3G interaction.

The rms deviations are in general larger for yrare states than the
 yrast states, especially in KB3G.
This also suggests the failure in the description of the
 core-excitation, which is expected to appear more directly
 in the yrare states than the yrast states.
Another possible reason is that the FDA*
 becomes less accurate for yrare states.

From this table we cannot infer that
 the description by GXPF1 is better than KB3G for group (a) and (c),
 because the rms deviations shown in the table are not
 exact values but the results of the FDA* with typical accuracy of
 about 200 keV.
Since GXPF1 is determined
from FDA*,
 the rms deviations of GXPF1 
 is naturally
smaller than those by other interactions within the FDA*.
In fact, in light $pf$-shell nuclei with more precise 
computations, we can find several examples
 where KB3G gives a better description than GXPF1.

\begin{table}
\caption{Comparison of rms deviations (MeV) between the 
experimental excitation energies and those calculated from
GXPF1 and KB3G.
Theoretical energies are estimated by the FDA*.
The numbers in the parentheses show the number of data
 included in the calculation.
\label{tbl:gx1-vs-kb3g}
}
\begin{ruledtabular}
\begin{tabular}{cccc}
group  & state & GXPF1 &  KB3G \\ \hline
(a) $Z, N<28$ & yrast & 0.154(136) & 0.235(129) \\
          & yrare & 0.201(45)  & 0.263(23)  \\
(b) $Z$ or $N=28$ & yrast & 0.184(92) & 0.647(87) \\
              & yrare & 0.195(57) & 0.802(44) \\
(c) $Z<28$, $N>28$ & yrast & 0.145(129) & 0.296(126) \\
               & yrare & 0.145(75)  & 0.302(55) \\
(d) $Z, N>28$ & yrast & 0.186(55) & 0.401(51) \\
          & yrare & 0.187(23) & 0.458(23) \\
\end{tabular}
\end{ruledtabular}
\end{table}

\subsection{Collective properties}
\label{sub:collective}

It has been pointed out \cite{dufour} that the multipole part of the
 Hamiltonian $H_M$ is dominated by several terms,
 such as the pairing and the quadrupole-quadrupole interactions,
 which determine collective properties of the effective interaction.
It is useful to investigate these collective aspects
 of GXPF1 and compare with that of other interactions.

According to the prescription in Ref. \cite{dufour},
 $H_M$ can be expressed in
 both the particle-particle (p-p) representation
 and the particle-hole (p-h) representation.
In the p-p representation, 
 $H_M$ is expressed in terms of particle-pair bases
 $[c^\dagger_a c^\dagger_b]^{JT}$,
 in which two nucleons are
 coupled to the good angular momentum $J$ and the isospin $T$.
The coefficient matrix is diagonalized for each $JT$ channel.
By using the resultant eigenvalues $E^{JT}_\alpha$, we obtain
\begin{equation}
H_M = \sum_{JT\alpha}E^{JT}_\alpha P^\dagger_{JT\alpha} P_{JT\alpha}
+ (\mbox{one-body terms}),
\end{equation}
 where $P^\dagger_{JT\alpha}$ denote particle-pair creation operators
 which are linear combinations of the particle-pair bases,
 and their structure
 is determined by the corresponding eigenvectors.
(Here, for simplicity, we omit the summations over $z$-components
 of both spin and isospin.)
The most important contributions come from the $JT=10$, $20$ and $01$ terms
 with large negative eigenvalues $E^{JT}$,
 which correspond to the $T=0$, small-$J$ pairing
 and the usual $T=1$ monopole pairing, respectively.

Similarly, in the p-h representation,
 by using the particle-hole (density) bases
 $[c^\dagger_a \tilde{c}_b]^{\lambda\tau}$
 with spin-isospin quantum numbers $\lambda\tau$, 
 the diagonalization of the coefficient matrix
 determines the structure of multipole operators $Q_{\lambda\tau k}$ and 
 the corresponding strengths $e^{\lambda\tau}_k$, leading to an expression
\begin{equation}
H_M = \sum_{\lambda\tau k}e^{\lambda\tau}_k 
Q_{\lambda\tau k} \, \cdot \, Q_{\lambda\tau k}.
\end{equation}
The symbol $\cdot$ stands for a scalar product with respect to
 both spin and isospin.
(We adopt here a slightly different definition
 of $e^{\lambda\tau}_k$ from that in Ref. \cite{dufour}
 by a phase factor $(-1)^{\lambda+\tau}$.)
Large eigenvalues appear for 
 $\lambda\tau=20$, $40$ and $11$,
 which are interpreted as the usual isoscalar quadrupole, hexadecapole and
 Gamow-Teller interactions, respectively.

\begin{table}
\caption{Comparison of the collective strengths (MeV) between
 G, GXPF1 and KB3G.
The mass number $A=42$ is assumed.
\label{tbl:collective}
}
\begin{ruledtabular}
\begin{tabular}{lccccccc}
Interaction & $E^{01}$ & $E^{10}$ & $E^{20}$ & & $e^{20}$ & $e^{40}$ & $e^{11}$ \\ \hline
G     &  $-4.20$ & $-5.61$ & $-2.96$ & & $-3.33$ & $-1.30$ & $+2.70$ \\
GXPF1 &  $-4.18$ & $-5.07$ & $-2.85$ & & $-2.92$ & $-1.39$ & $+2.67$ \\
KB3G  &  $-4.75$ & $-4.46$ & $-2.55$ & & $-2.79$ & $-1.39$ & $+2.47$ \\
\end{tabular}
\end{ruledtabular}
\end{table}

In Table \ref{tbl:collective}, the collective
 strengths, i.e. the largest (smallest) eigenvalues
 $E^{JT}$ or $e^{\lambda\tau}$ in each spin-isospin channel,
 are shown for G, GXPF1 and KB3G.
According to the comparison between G and GXPF1,
 in general, the empirical fit has reduced these strengths
 by at most 12\%
 except for the $\lambda\tau=40$ term.
Especially, the reductions in the $T=0$ terms
 $E^{10}$ and $e^{20}$ are relatively large,
 contrary to the general observation that
 $T=0$ matrix elements are on average modified to be
 more attractive by the empirical fit (see Fig. \ref{fig:gx1-vs-g}).
This means that the attractive modification has been
 applied mainly to the monopole terms 
(which do not enter into the
numbers of Table \ref{tbl:collective}).

For the comparison between GXPF1 and KB3G,
 it can be seen that
 the $T=0$ ($T=1$) pairing strength is stronger (weaker)
 in GXPF1.
Such a difference is expected to affect the description of the
 structure for $Z\sim N$ nuclei.
In GXPF1, $E^{10}$ is larger than $E^{01}$ by about 20\%,
 which is consistent with an estimate
 by the mean-field calculations \cite{satula} using
 the standard seniority pairing.
It has also been shown \cite{kb3g} that
 the $T=0$ pairing strength is larger than $T=1$
 in the density-dependent Gogny force.
Note that the situation is opposite in KB3G.
On the other hand, the multipole strengths $e^{\lambda\tau}$ are 
very similar for GXPF1 and KB3G, although
 the strengths of GXPF1 are slightly larger.

By using $e^{\lambda\tau}$,
 we can evaluate the collective quadrupole-quadrupole (QQ)
 strength between like-nucleons (proton-proton or neutron-neutron)
 and that between protons and neutrons.
For GXPF1, the strength of $Q_p \cdot Q_p$ or $Q_n \cdot Q_n$ is 
$-0.96$ MeV, while that of $Q_p \cdot Q_n$ is $-7.82$ MeV.
This result shows the dominance of the proton-neutron part
 in the collective QQ interaction, as in heavier nuclei \cite{mcsm-ba,otsuka1,otsuka2,pnqq}.
Such $Q_p \cdot Q_n$ dominance can be seen in other interactions, although
 the ratio of $Q_p \cdot Q_n$ to $Q_p \cdot Q_p$ (or $Q_n \cdot Q_n)$ is
 different (8.1 for GXPF1, 6.8 for G, 7.1 for KB3G).

\subsection{Spin-tensor decomposition}
\label{sub:spiten}

\begin{figure}[]
\includegraphics[width=75mm]{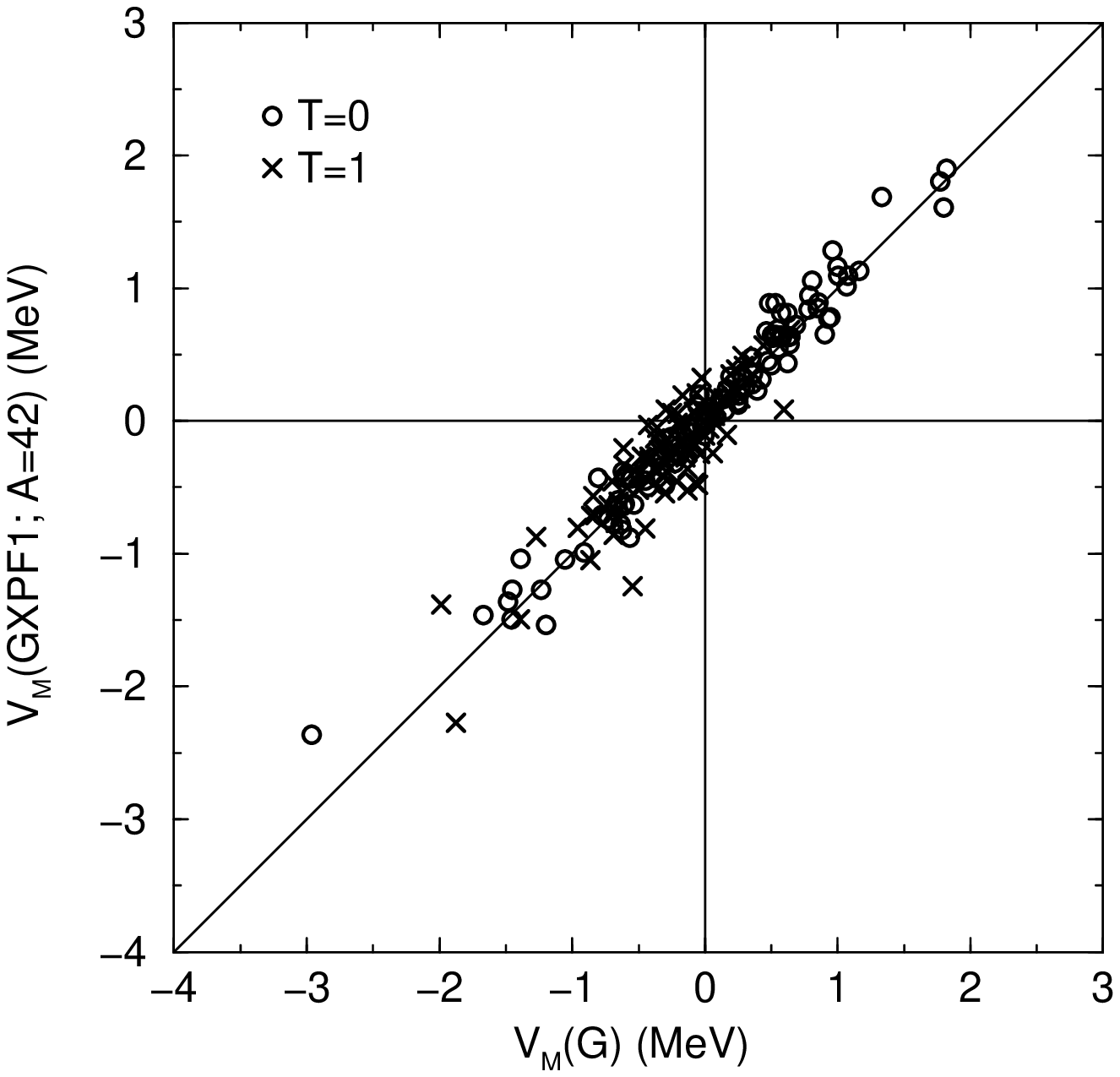}
\includegraphics[width=75mm]{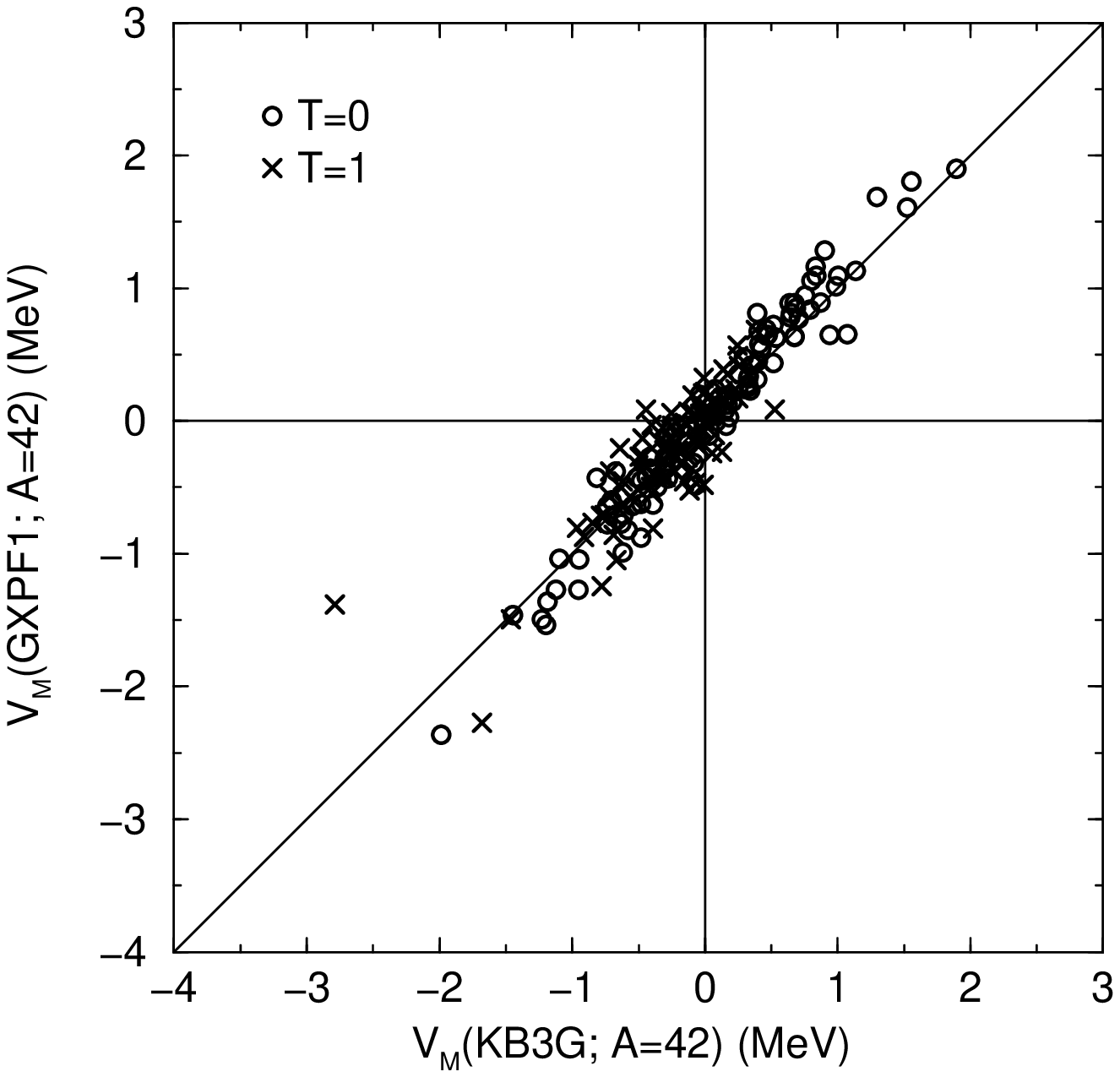}
\includegraphics[width=75mm]{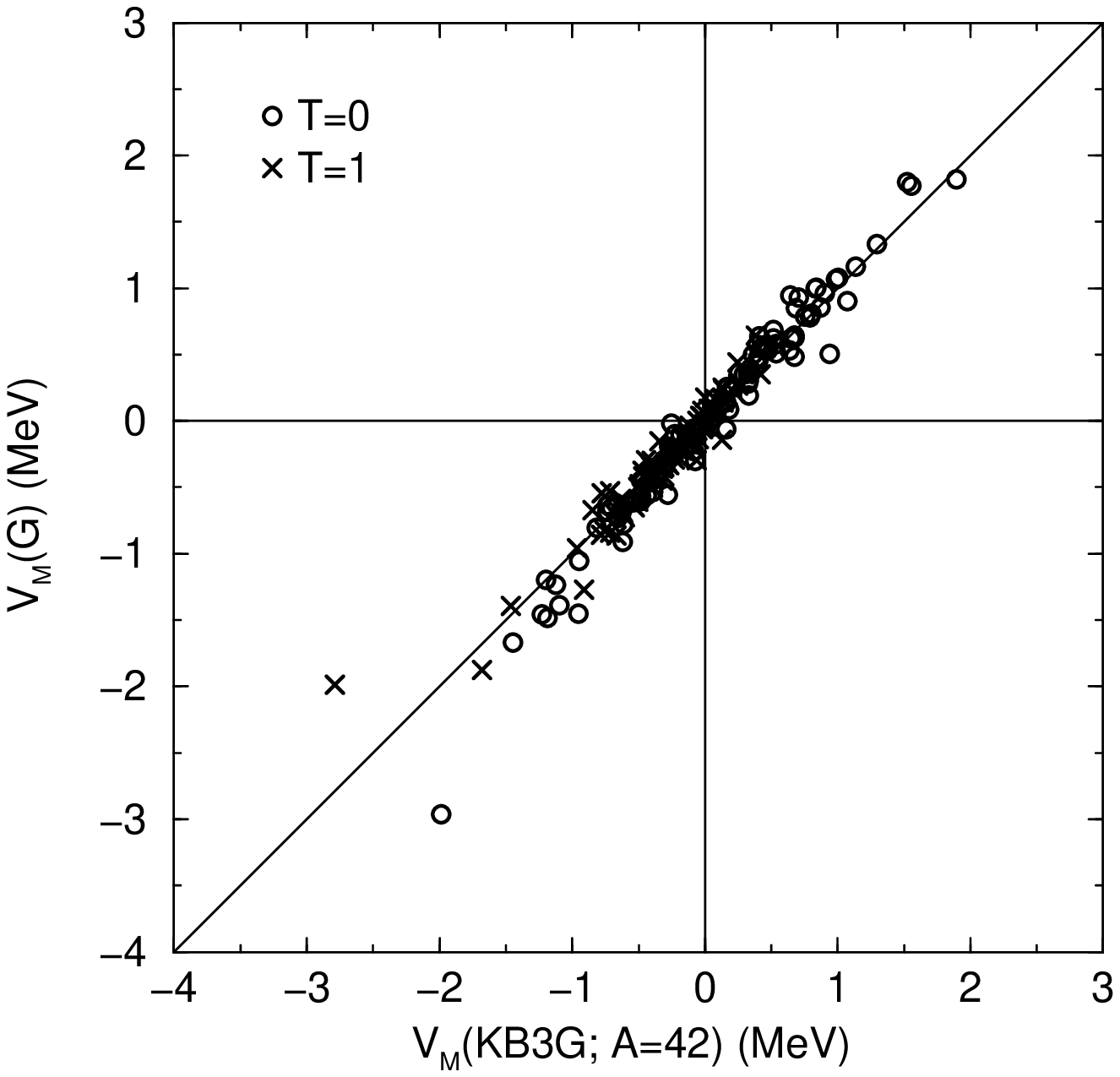}
\caption{Correlation of monopole-free two-body matrix elements $V_M$
 between (upper) GXPF1 and G, (middle) GXPF1 and KB3G
 and (lower) G and KB3G.
\label{fig:2bme-mf}
}
\end{figure}

In order to analyze the structure of an effective interaction,
 the spin-tensor decomposition \cite{spiten} is useful,
 since it gives physical insights from a different viewpoint.
In the following discussions, we consider the two-body
 interaction $V_M$
 in the multipole part $H_M$.
Therefore the results are free from the monopole effects.
We again take KB3G as a reference interaction.
It is essentially the same as the Kuo-Brown's renormalized
 G-matrix interaction \cite{kb} after the monopole subtraction.
(All of the results in this section are for the matrix elements
evaluated at $A=42$.)
An overview of the correlation in the $V_M$ between 
 these interactions are shown in Fig. \ref{fig:2bme-mf}.
It can be seen that the correction to the microscopic interaction
 imposed by an empirical fit contains sizable non-monopole components.
In addition, the microscopic interactions, G and KB3G ($\sim$ KB)
 are not identical even in the multipole parts,
 as seen in the lower part of Fig.\ref{fig:2bme-mf}.
The difference between G and KB3G looks not necessarily smaller
 than that between G and GXPF1, which 
 also suggests that the present correction to G 
 is in a reasonable magnitude.

We first transform the $jj$-coupled two-body matrix elements
 $\langle j_a j_b JT|V_M|j_c j_d JT \rangle$
 into the $LS$-coupled form
 $\langle l_a l_b LS JT|V_M|l_c l_d L'S' JT \rangle$,
 then carry out the spin-tensor decomposition of the two-body
 interaction as
\begin{equation}
V_M = \sum_k V_k = \sum_k U^k \cdot X^k,
\end{equation}
 where the operators $U^k$ and $X^k$ are irreducible tensors
 of rank $k$ in the space and spin coordinates, respectively.
The interaction components $V_k$ represent the central ($k=0$),
 spin-orbit ($k=1$) and tensor ($k=2$) parts.
The spin-orbit part includes both the normal part
 ($S=S'=1$) and the anti-symmetric spin-orbit ($S\neq S'$) part.

\begin{figure}
\includegraphics[width=85mm]{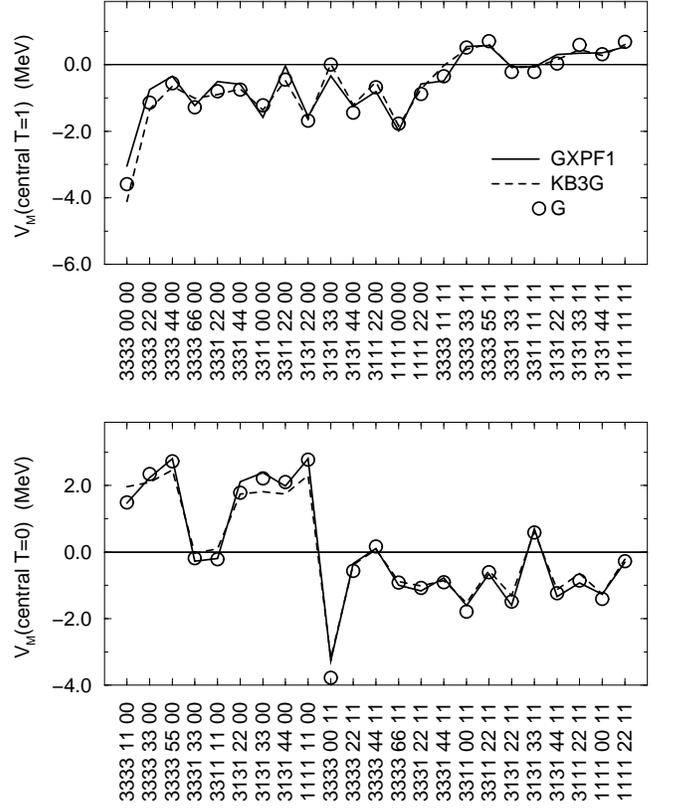}
\caption{Comparison of the central
 $T=0$ (lower panel) and $T=1$ (upper panel) components.
Results for GXPF1, KB3G and G are
 shown by solid lines, dotted lines and circles, respectively.
The quantum numbers $l_a l_b l_c l_d LL' SS'$ of
 the $LS$-coupled matrix elements
 $\langle l_a l_b LSJT|V|l_c l_d L'S'JT \rangle$
 are shown along the horizontal axis.
\label{fig:central}
}
\end{figure}

Figure \ref{fig:central} shows the central components.
For the $T=0$ matrix elements,
 G and GXPF1 are very similar to each other.
Sizable modifications of G by the empirical fit
 can only be seen in the $(l_al_bl_cl_d\ L)$=$(3333\ 0)$ 
 and $(3131\ 2)$ components,
 which are made more repulsive by
 about 0.4 and 0.3 MeV, respectively.
Other matrix elements of GXPF1 are very close to those of G.
On the other hand, GXPF1 deviates from G 
 in various $T=1$ matrix elements.
The modifications to G are in the repulsive direction
 for the matrix elements  ($3333\ 0$) and ($3333\ 2$)
 which are related to monopole and quadrupole pairing
 in the $f$ orbit, respectively,
 while the ($3311\ 0$) matrix element is
 made to be more attractive, which corresponds to
 the monopole pairing between $f$ and $p$ orbits.

For KB3G, it is remarkable that
 $T=0$, $S=1$ matrix elements are very close to those of GXPF1,
 including the ($3333\ 0$) matrix element
 which deviates from G significantly.
Considering that the origin of KB3G and GXPF1 is very different,
 this similarity is surprising.
However, there are large differences
 in the $T=0$, $S=0$ matrix elements,
 where the absolute values of the KB3G matrix elements
 are smaller in most cases than those of GXPF1 (and also G).
The only exception is the ($3333\ 1$) matrix element.
On the other hand, most of the $T=1$ matrix elements of KB3G are
 very close to those of G rather than GXPF1 for both $S=0$ and 1.
One can only find small deviations from G
 in the ($3333\ 0$) and ($3333\ 6$) elements, where
 the former is more attractive and the latter is more repulsive.
This similarity between G and KB3G indicates that 
the $T=1$ central part is converged in these two different G-matrix
calculations.

\begin{figure}
\includegraphics[width=85mm]{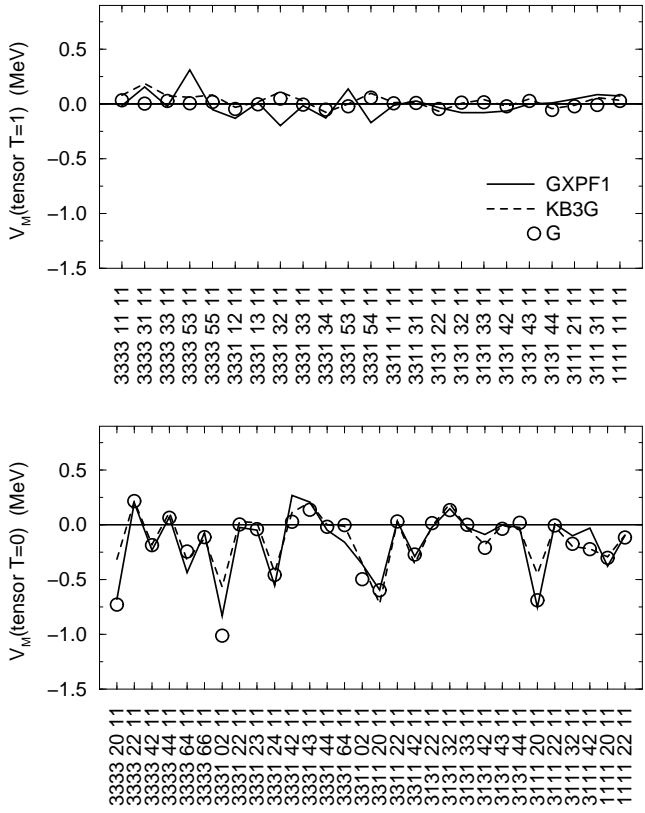}
\caption{Comparison of the tensor components.
Conventions are the same as in Fig. \ref{fig:central}.
\label{fig:tensor}
}
\end{figure}

In Fig. \ref{fig:tensor} the tensor components are compared.
The $T=0$ matrix elements are relatively large
for $\Delta L=2$.
The corrections to G are relatively large
 for the ($l_al_bl_cl_d\ LL'$)=($3331\ 02$), ($3331\ 42$), ($3311\ 02$),
 ($3131\ 42$) and ($3111\ 42$) matrix elements,
 which are all in the repulsive direction,
 while those in the attractive direction
 are found in ($3333\ 64$) and ($3331\ 64$) matrix elements.
These corrections are at most 0.2 MeV.
The $T=1$ components are in general very small compared to 
$T=0$. 
This is due to the fact that $T=0$ tensor interaction
is dominated
by a matrix element between two nucleons with $l=0$ to $l=2$
whereas $T=1$ must have odd $l$. 

The G and KB3G values show rather large differences
in cases with largest magnitudes, while the GXPF1 favors the G 
values with small corrections.
Namely, in the $T=0$ cases, ($3333\ 20$), ($3331\ 02$)
 and ($3111\ 20$) matrix elements,
 the attraction of KB3G is much weaker, 
 rather outstandingly,
 than that of G (and also GXPF1).

\begin{figure}
\includegraphics[width=85mm]{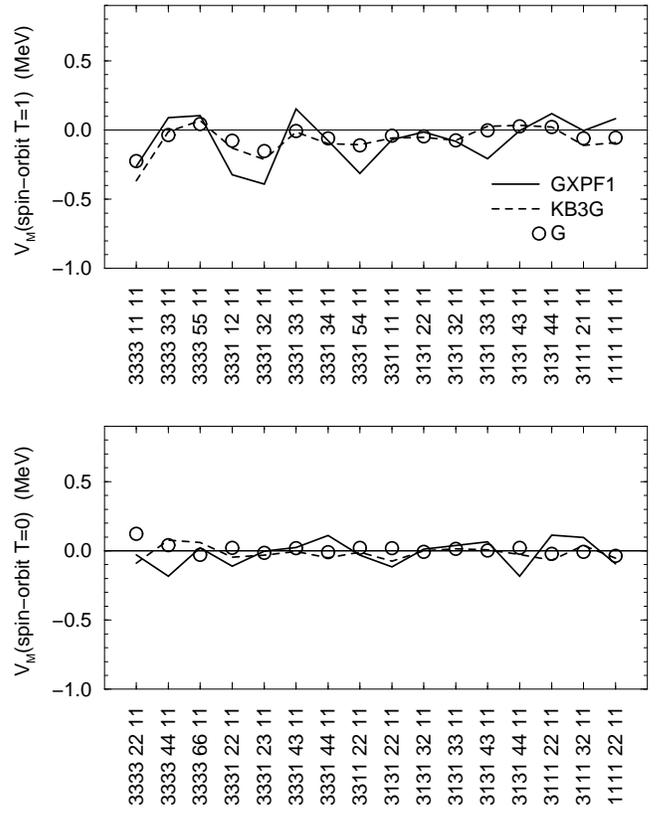}
\caption{Comparison of the spin-orbit components.
Conventions are the same as in Fig. \ref{fig:central}.
\label{fig:spin-orbit}
}
\end{figure}

The normal spin-orbit components are shown in Fig. \ref{fig:spin-orbit}.
For the $T=0$ elements,
 relatively large attractive corrections to G are found in
the ($l_al_bl_cl_d\ LL'$)=($3333\ 22$), ($3333\ 44$)
 and ($3131\ 44$) matrix elements.
Such corrections to the latter two matrix elements
 are not found in KB3G.
Similarly, large attractive corrections exist
 also in the $T=1$, ($3331\ 12$), ($3331\ 32$),
 ($3331\ 54$) and ($3131\ 33$) matrix elements,
 all of which are absent in KB3G.
It can be seen that the spin-orbit matrix elements
 of KB3G and G are very close in both cases of $T=0$ and 1.
Thus, we infer that there may be contributions to the
spin-orbit interaction that are not present in the 
two-nucleon G matrix, perhaps from an effective 
three-nucleon Hamiltonian.

\begin{figure}
\includegraphics[width=85mm]{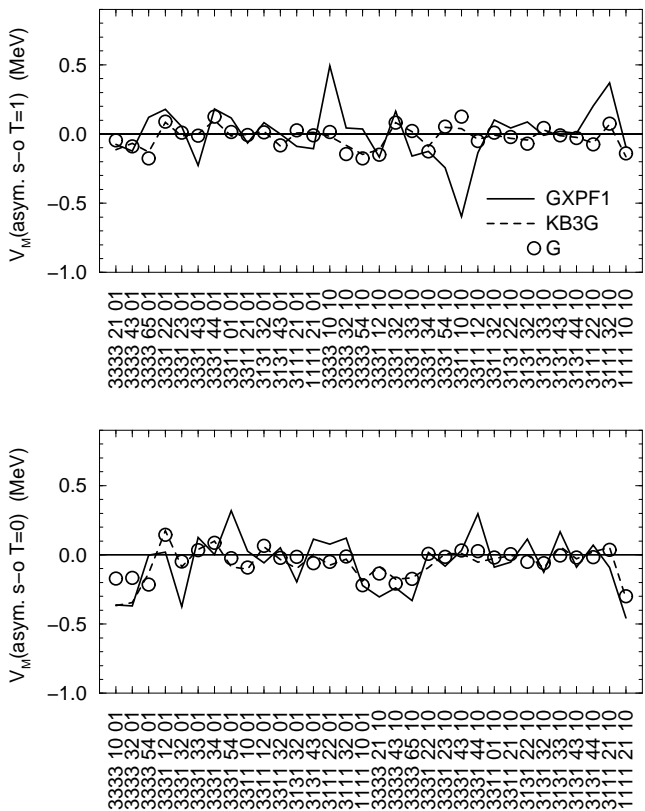}
\caption{Comparison of the anti-symmetric spin-orbit components.
Conventions are the same as in Fig. \ref{fig:central}.
\label{fig:asym-so}
}
\end{figure}

Figure \ref{fig:asym-so} shows the anti-symmetric spin-orbit components.
In general, matrix elements of G are small as in the case of
 the normal spin-orbit components.
Nevertheless, one can again see nearly perfect similarity 
between G and KB3G in almost all matrix elements.
The empirical fit has resulted in several large corrections to G
 which are inconsistent with KB3G, such as
 $T=0$,  ($l_al_bl_cl_d\ LL'\ SS'$)=
 ($3331\ 32\ 01$), ($3331\ 54\ 01$), ($3331\ 44\ 10$)
 and $T=1$ ($3333\ 10\ 10$), ($3311\ 10\ 10$), ($3111\ 32\ 10$).
Such relatively large matrix elements appear also
 in other empirical effective interactions such as
 FPMI3 \cite{fpd6} and TBLC8 \cite{tblc8}. 
Further investigation
is needed to find if these deviations are significant in
terms of the errors inherent in the fits to data.

Summarizing the results of the spin-tensor decomposition
of $V_M$,
 there are overall, reasonably good
 similarities between G and GXPF1 in relatively large
 matrix elements such as the central and the $T=0$ tensor components.
The corrections due to the empirical fit become sizable
 in several specific matrix elements,
 especially for $T=1$.
On the other hand, although many of matrix elements of KB3G are 
very close to those of G, 
one sees that several $T=S=0$ central and $T=0$ tensor matrix
elements are rather different.
This suggests significant changes in microscopic calculations
of the effective interaction or their input from the nucleon-nucleon
interaction that have evolved from
the original calculations of
Kuo-Brown to the more recent results of Ref. \cite{g-mat}. 

\subsection{Monopole fit}
\label{sub:monofit}

According to previous subsections, the empirical fit
 gives rise to sizable corrections to the microscopic interactions
 mostly in the monopole part
 and several specific matrix elements such as the monopole pairing.
Therefore we come to a natural question:
 to what extent can we improve the microscopic interaction
 for practical use
 with more restrictive corrections.
In order to assess this approach,
 we have carried out another fit
 by varying only the monopole parts, and the
 monopole-pairing ($T=1$, $J=0$)
 and quadrupole-pairing ($T=1$, $J=2$) matrix elements.
The number of parameters are 20, 10 and 36 for these components,
 respectively.
Thus in total 70 parameters are varied
 including 4 single particle energies.
Note that the $T=0$ components are varied
 only in the monopole part.
The mass dependence $A^{-0.3}$ is also assumed.
In the final fit, 45 best-determined LC's are taken,
 and the resultant interaction is referred to as GXPFM.
The estimated rms deviation is 226 keV within FDA* for 623 energy data.
Although this number sounds quite successful,
 GXPFM fails to describe $N$ or $Z$=28 semi-magic nuclei.
In fact, for these nuclei, the
 estimated rms deviation increases to 267 keV for yrast states (87 data)
 and to 324 keV for yrare states (44 data).

We first consider the multipole part of the Hamiltonian.
The collective strengths of the p-h interactions of GXPFM
 are found to have reasonable values.
For example, 
 $e^{20}=-2.85$, $e^{40}=-1.44$ and $e^{11}=+2.59$ (MeV),
 which are quite similar to those of GXPF1 (see Table \ref{tbl:collective}).
As for the p-p channel, since the multipole part of GXPFM is
 different from G only for the $T=1$ components,
 the strengths $E^{10}$ and $E^{20}$ are the same as those of G.
On the other hand, the monopole pairing strength
 is reduced to $E^{01}=-3.88$,
 which is much weaker than that of GXPF1
 and will compensate the large $T=0$ pairing strength.

\begin{figure}
\includegraphics[width=85mm]{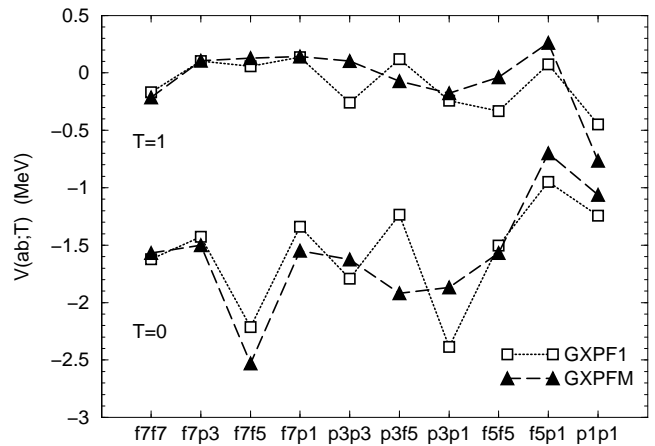}
\caption{Comparison of the monopole
 matrix elements $V(ab;T)$
 for the  GXPF1 and GXPFM interactions,
 which are shown by squares and triangles, respectively.
Conventions are the same as in Fig. \ref{fig:centroid}.
\label{fig:centroid-gxpfm}
}
\end{figure}

Figure \ref{fig:centroid-gxpfm} compares the
 monopole matrix elements $V(ab;T)$ for
 GXPF1 and GXPFM.
The $f_{7/2}$-related matrix elements of GXPFM take
 similar values to those of GXPF1 (and other interactions),
 although the $T=0$, f7f5 matrix element
 of GXPFM is more attractive than that of GXPF1 by 300 keV.
Note that the monopole Hamiltonian
 is specified by the linear combinations
 $\frac{3}{4}V(ab;1)+\frac{1}{4}V(ab;0)$ and $V(ab;1)-V(ab;0)$,
 where the former determine, roughly speaking,
 the mass dependence and the latter affects the isospin dependence.
The first combination is extremely well determined and
 almost identical between GXPF1 and GXPFM
 (differences are less than 50 keV).

Relatively large differences can be seen in
the  $T=1$, p3p3, f5f5 and p1p1 matrix elements,
 which are all related to the monopole pairing.
The former two of GXPFM are less attractive than
 those of GXPF1, which is required to
 keep the collective monopole pairing strength weaker
 than that of GXPF1, as mentioned above.
For the $T=0$ matrix elements,
 the p3f5 and p3p1 components are very different between GXPF1 and GXPFM.
Although the structure of Ga and Ge isotopes is rather sensitive
to these matrix elements, most of those isotopes cannot be included 
into the fit because $N$ is too close to 40.  This point may be 
clarified better in a future study including higher orbits.

\section{Results and Discussions}
\label{sec:result}

In this section, the results obtained from 
large-scale shell-model
 calculations with GXPF1 are given
 and compared with available experimental data.
In our previous studies of $pf$-shell
 nuclei \cite{qmcd,mcsm,ni56deform,semi},
 we took advantage of the MCSM,
 which was the only feasible way to evaluate
 the eigenvalues of the shell-model Hamiltonian
 in the middle of the $pf$ shell.
In the present study,
 most of shell-model calculations are carried out 
 in a conventional way by using
 the shell-model code MSHELL \cite{mshell}.
MSHELL enables calculations with 
$M$-scheme dimensions of up to $\sim 10^8$.
With this capability we can
handle essentially all low-lying states in the
$pf$-shell nuclei with a minimal truncation of the model
space.
We can also confirm the reliability of the previous
MCSM calculations by comparing the both results.
On the other hand, for non-yrast states in the
middle of the $pf$ shell, the MCSM it is still 
necessary.

In the following discussions,
 the truncation order $t$ denotes the
 maximum number of nucleons which are allowed to
 be excited from the $f_{7/2}$ orbit to higher three orbits 
 $p_{3/2}$, $f_{5/2}$ and $p_{1/2}$, relative to the
 lowest filling configuration.
The latter three orbits are expressed simply as $r$ hereafter.
Most results for Ca, Sc, Ti, V, Cr are exact (no truncation) while
 $t \geq 6$ for Mn, and $ t\geq5 $ or better for other isotopes.

In this paper we fully cover the data on magnetic and 
quadrupole  moments. Electro-magnetic
transitions are discussed for some representative nuclei. Gamow-Teller
beta decay will be covered in subsequent work.
Electro-magnetic transition matrix elements are
 calculated by using the effective g-factors
 and the effective charges adopted in sections \ref{sub:m-mom}
 and \ref{sub:q-mom}, respectively.

\subsection{Closed core properties \label{sub:core}}
\label{sub:coreprob}

First we consider the role of the $Z$ or $N=28$ 
closed shell.
This is important for the unified shell-model
 description of $pf$-shell nuclei,
 since it is convenient to base a truncation scheme on the
 stability of such a closed core.
It has been shown \cite{mcsm} that the $^{56}$Ni core
 is rather soft in comparison to the $^{48}$Ca core.
More precisely, the probability of the $(f_{7/2})^{16}$ configuration 
 in the ground-state wave function of $^{56}$Ni is
 much smaller than that of the $(f_{7/2})^8$ configuration
 in $^{48}$Ca.
It is attributed to the strong proton-neutron interaction.
The evolution of such a closed core
 in various isotope chains is of interest.

\begin{figure}
\includegraphics[width=85mm]{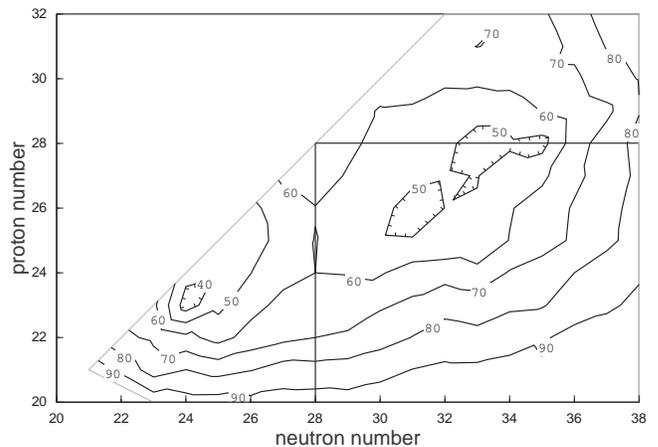}
\caption{The probability of $^{56}$Ni closed-shell configurations
 in the calculated ground-state wave functions.
\label{fig:prob}
}
\end{figure}

Hereafter, a group of configurations in which 
 both proton and neutron  $f_{7/2}$ orbits are maximally filled
 will be denoted by the $^{56}$Ni closed-shell configuration.
For each isotope with $Z_v$ valence protons and $N_v$ valence neutrons
 on top of the $^{40}$Ca core,
 such configurations are those given by 
  $\pi(f_{7/2})^m(r)^{Z_v-m}\nu(f_{7/2})^n(r)^{N_v-n}$
 with $m=\min\{Z_v, 8\}$ and $n=\min\{N_v, 8\}$.
Figure \ref{fig:prob} shows the probability of the 
 closed-shell configurations in the calculated ground-state
 wave function.
The lowest $T=0$ states are considered for odd-odd $N=Z$ isotopes,
 although they are not necessarily the ground states.
The truncation order
 is taken to be sufficiently large for describing these states.

The probability
 takes the minimum value ($\sim$40\%) around $^{48}$Cr,
 which can be understood as a result of the large deformation.
In the Ti isotopes, the probability becomes smallest
 around $N=24$ and then increases monotonously for larger $N$.
On the other hand, in the case of Fe and Ni,
 one can see a local maximum ($\sim$60\%) at $N=28$ and 
 the second minimum ($\sim$45\%) around $N=32\sim34$.
Thus it is not justified to assume an
 inert $^{56}$Ni core even for Ni isotopes.
As for Zn isotopes, the probability
 shows rather smooth behavior with moderate values $\sim$65\%,
 suggesting that the effects of core-excitations
 are similar over the wide range of $N$.


\subsection{Binding energy \label{sub:mass}}
\label{sub:binding}

Binding energies are obtained by adding suitable Coulomb energies 
 to the shell-model total energies.
In the present study the Coulomb energies are evaluated 
 by using an empirical formula:
\begin{equation}
E_{\rm C} =  V_{\pi \pi} \frac{\pi(\pi -1)}{2} - V_{\pi \nu} \pi \nu
 + e_{\pi} \pi,
\end{equation}
 where $\pi$ and $\nu$ denote the number of valence protons and neutrons, 
 respectively.
The adopted parameters are
 $V_{\pi \pi} = 0.264$,
 $V_{\pi \nu} = 0.038$,
 and $e_{\pi} = 7.458$ (MeV).
The same form was adopted in Ref. \cite{langanke}
 for describing light $pf$-shell nuclei
 with a different parameter set.
In the present approach, since we consider a wider mass region,
 the parameters are determined by fitting to the energy difference
 between 36 pairs of isobaric analog states with masses $A=47\sim74$.
In Ref. \cite{kb3-be}, the binding energies are
 also studied systematically by using the KB3 interaction for
 many $f_{7/2}$ shell nuclei up to $^{56}$Ni,
 and quite similar values of these parameters are proposed.
They have attained an rms deviation of 215 keV
 between theory and experiment for 70 nuclei of $A=42\sim56$.
However, the discrepancy in $^{56}$Ni (overbinding)
 is significantly large in comparison to that of neighboring nuclei.

\begin{figure}[h]
\includegraphics[width=85mm]{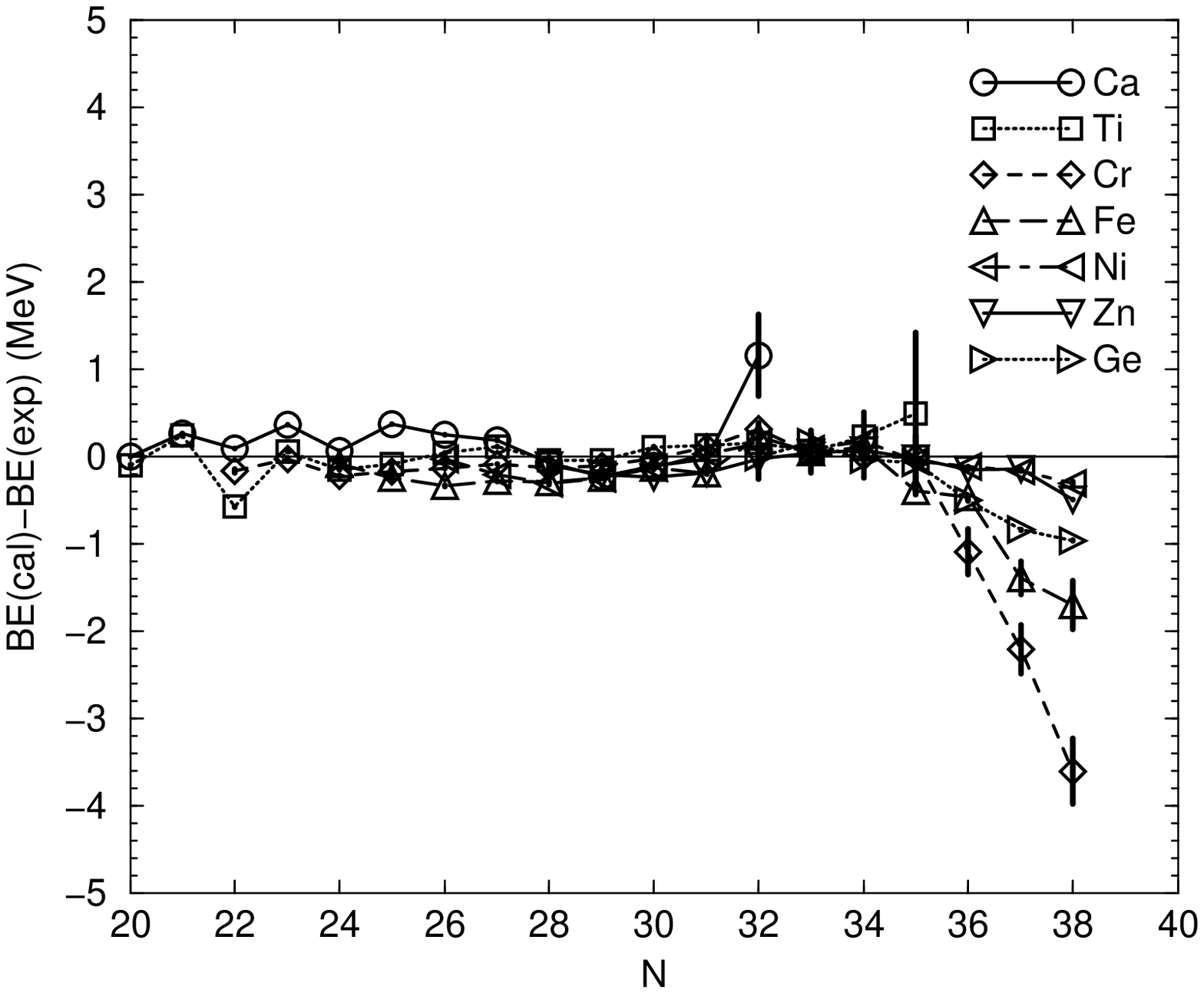}
\includegraphics[width=85mm]{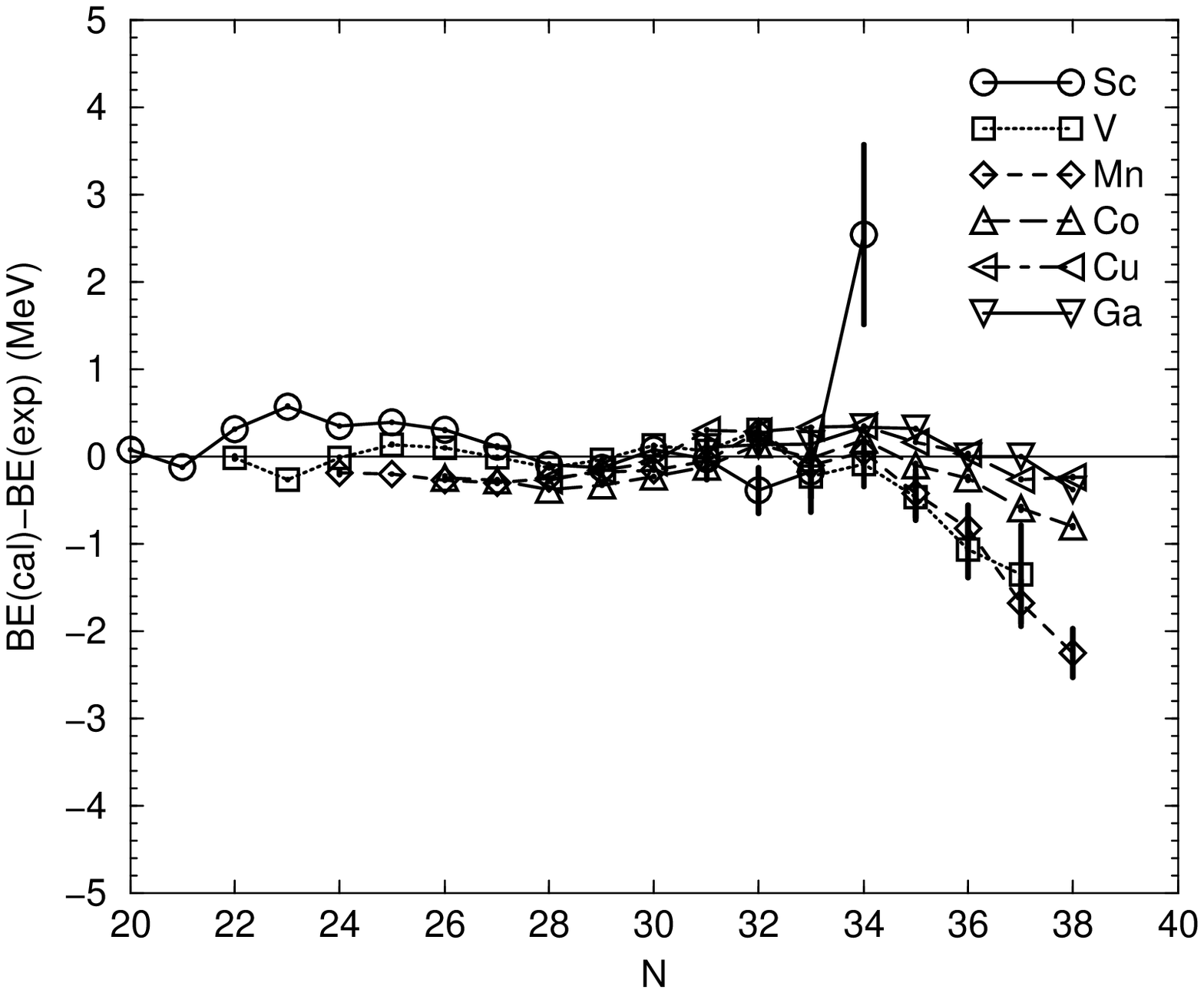}
\caption{Comparison of experimental binding energies with the 
 shell-model energies. 
The upper (lower) panel shows the results of
 even- (odd-) $Z$ isotope chains.
Data are taken from Ref. \cite{audi}.
\label{fig:be}
}
\end{figure}

In Fig. \ref{fig:be} the deviation of calculated binding energies from
 experimental values are shown for each isotope chain
 as a function of the neutron number $N$.
The theoretical values are obtained by GXPF1.
For the truncations,
all results are exact for Ca, Sc, Ti, V and Cr isotopes.
For other isotopes, the results are obtained in a subspace
 of $t=5$ or larger.
Judging from the convergence pattern as a function of $t$,
 the underbinding due to this truncation is typically smaller than 200 keV.
Note that the mass range included in the figure
 is much wider than that used in the fitting calculations
 for deriving GXPF1,
 and covers regions of nuclei which may not
be well described just by the $pf$ shell.
It can be seen that 
 the agreement between theory and experiment is quite good
 over the whole mass range included in the fit ($A=47-65$)
 and is also reasonable for many nuclei which were not included in the fit.

In the neutron-rich side,
 we can find relatively large deviations.
In Ca and Sc isotopes, the calculations give overbinding 
 at the end of the isotope chain, $N=32$ and 34, respectively, 
 with large error-bars in the experimental data.
Note that, for $^{52}$Ca, the experimental data was taken from
 Ref. \cite{audi}, where the value $Q_{\beta-}$=7.9(5) MeV is adopted.
On the other hand, in Ref. \cite{huck}, the measured
 value $Q_{\beta-}$=5.7(2) MeV is presented.
The prediction by GXPF1 is 5.9 MeV, in good agreement with the 
latter value.
It is important to have improved experimental data for 
this neutron-rich region of nuclei.
For the Ti isotopes, the agreement is satisfactory along the whole isotope chain.
In other isotopes,
 the deviation between theory and experiment becomes
 sizable around $N\sim 35$,
 and, for larger $N$, the calculation shows 
a systematic underbinding.
For a fixed $N$, the deviation is largest for Cr,
 and it is moderate for V, Mn, Fe, Ge,
 and it is small for Co, Ni, Cu, Zn, Ga.
We relate this to an increasing importance of the neutron $g_{9/2}$ orbit
which should have its maximal effect in the middle of proton $f_{7/2}$
shell (Cr) due to deformation.

\subsection{Magnetic dipole moments}
\label{sub:m-mom}

The magnetic dipole moments are calculated and
 compared with experimental data in Table \ref{tbl:mmom}.
The magnetic moment operator used in the present calculation is
\begin{equation}
\bm{\mu} = g_s \bm{s} + g_l \bm{l},
\end{equation}
 where $g_s$ and $g_l$ are the spin and the orbital g-factors,
 respectively.
By using the free g-factors
 $g_s=5.586$, $g_l=1$ for protons and $g_s=-3.826$, $g_l=0$ for neutrons,
 the agreement between
 the calculation ($\mu_{\rm th}^{\rm free}$)
 and the experiment ($\mu_{\rm exp}$) is in general quite good.

\begin{figure}[h]
\includegraphics[width=70mm]{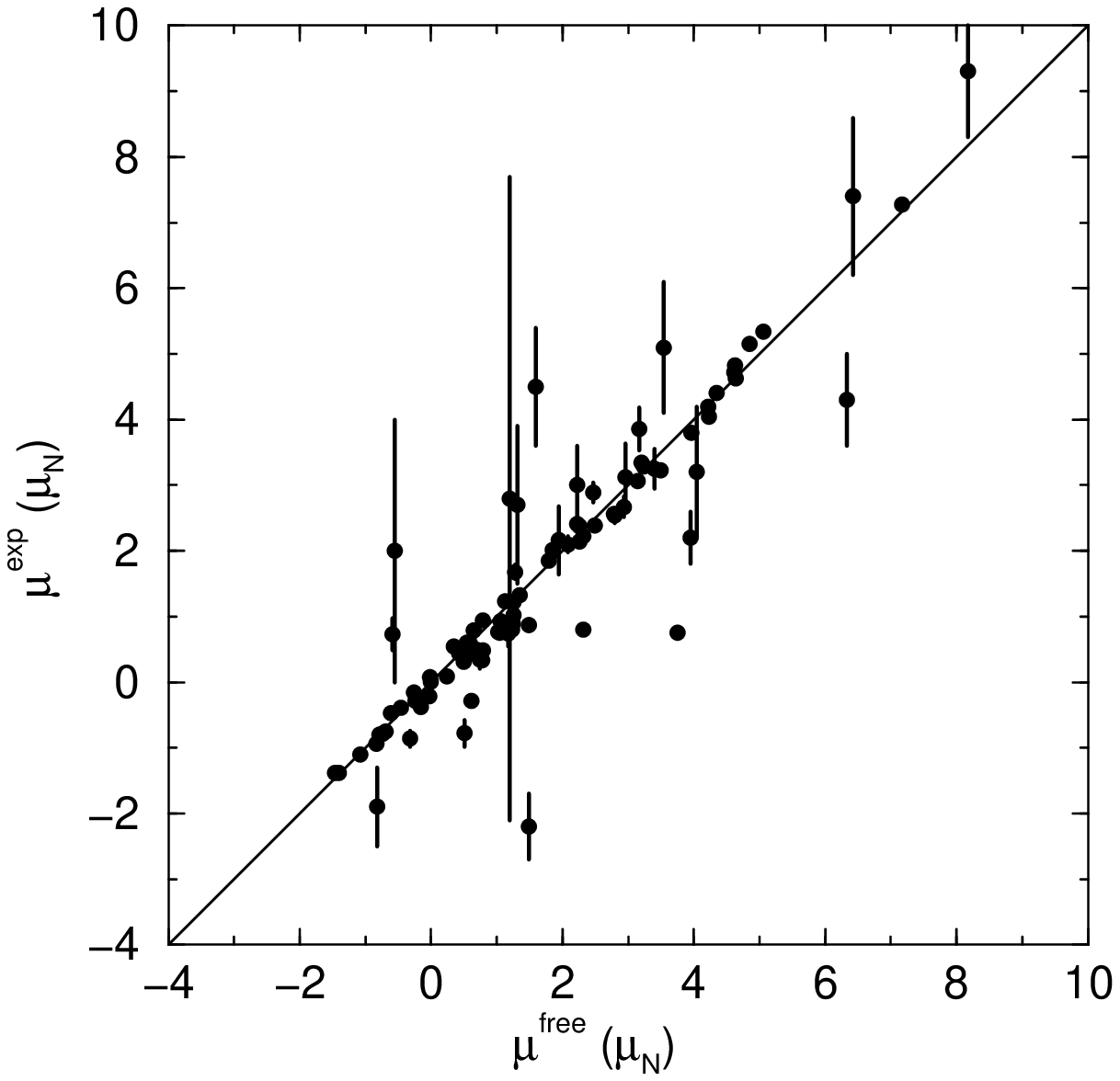}
\includegraphics[width=70mm]{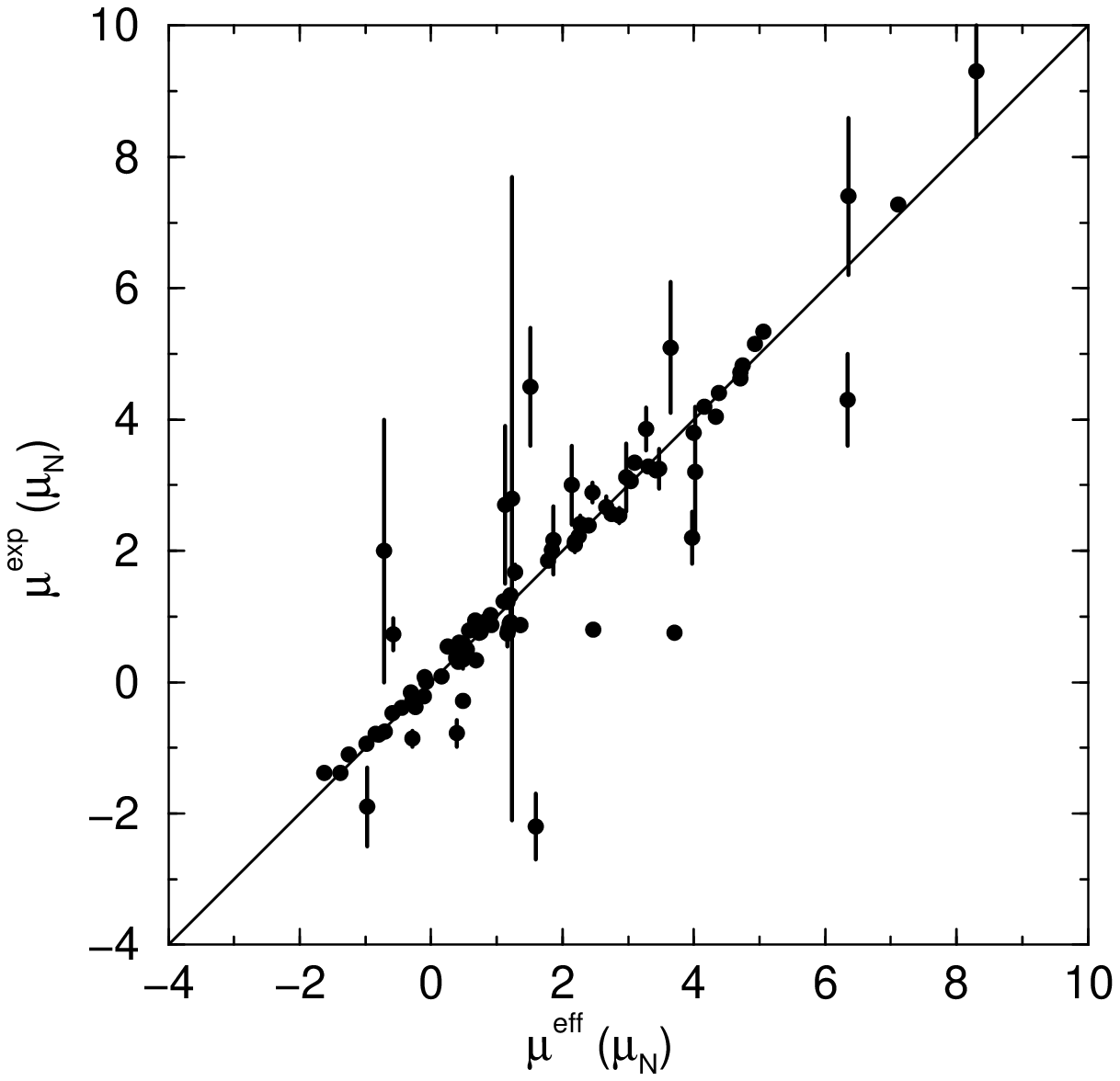}
\caption{
Comparison of experimental magnetic dipole moments
 with the shell-model results,
 which were obtained by using the free (upper panel)
 and effective (lower panel) nucleon g-factors.
All data in Table \ref{tbl:mmom}
 are included
 for which the sign is measured experimentally.
\label{fig:mmom}
}
\end{figure}

Figure \ref{fig:mmom} shows the comparison 
 of magnetic dipole moments between the experimental data and
 the shell-model predictions.
The agreement appears to be good except for several
 cases to be discussed.
As in the case of the $sd$-shell \cite{usd},
 the description is successful already by using
 the free-nucleon g-factors. It is well known that 
experimental magnetic moments deviate strongly from the
single-particle (Schmidt) values, as seen for example
on the left-hand side Fig. 37 of \cite{brown_review}.
For the $sd$ and $pf$ 
shells we find that the configuration mixing within the shells is
enough to fully account for observed magnetic moments.
This is in contrast to Gamow-Teller beta decay where the matrix
elements of the isovector spin factor, $g_s$,
are systematically reduced by factors of 0.77 in the
$sd$ shell \cite{usd} and 0.74 for the $pf$ shell \cite{fpga}.
Thus, the good agreement obtained for the magnetic moments with 
the free-nucleon operator is interpreted as cancellation
of the quenching observed in Gamow-Teller decay with enhancements
in the spin and orbital electromagnetic operators due to exchange 
currents \cite{usd}.

\vspace{2cm}

\begin{table}[h]
\caption{
Comparison of experimental magnetic dipole
 moments $\mu_{\rm exp}$ (in unit $\mu_N$) with
 theoretical values $\mu^{\rm free}_{\rm th}$
 and $\mu^{\rm eff}_{\rm th}$,
 which are calculated by using the free and
 the effective g-factors, respectively.
Most of the data are taken from Ref. \cite{nndc} except for
 $^{\rm a}$ from Ref. \cite{ernst},
 $^{\rm b}$ from Ref. \cite{speidel},
 $^{\rm c}$ from Ref.\cite{wagner},
 $^{\rm d}$ from Ref. \cite{fe59mmom},
 $^{\rm e}$ from Ref. \cite{kenn} and
 $^{\rm f}$ from Ref. \cite{kenn65}.
All results for Ca, Sc, Ti, V, Cr are exact, while
 $t \geq 6$ for Mn, and $t \geq 5$ for other isotopes.
\label{tbl:mmom}
}
\end{table}\addtocounter{table}{-1}
\vspace{-1cm}
\begin{longtable}{lcccrr}
 \\ \hline \hline
nuclei &  \ \ state  \ & $E_x$(MeV) & $\mu_{\rm exp}$ &  $\mu^{\rm free}_{\rm th}$ &   $\mu^{\rm eff}_{\rm th}$ \\ \hline
$^{47}$Ca &   7/2$^-$ &  0.000 & $   -1.380(24)              $ & $     -1.464 $ & $     -1.629 $  \\ 
$^{49}$Ca &   3/2$^-$ &  0.000 & $   -1.38(6)                $ & $     -1.403 $ & $     -1.376 $  \\ 
$^{47}$Sc &   7/2$^-$ &  0.000 & $   +5.34(2)                $ & $      5.054 $ & $      5.063 $  \\ 
$^{47}$Ti &   5/2$^-$ &  0.000 & $   -0.78848(1)             $ & $     -0.741 $ & $     -0.844 $  \\ 
          &   7/2$^-$ &  0.159 & $   -1.9(6)                 $ & $     -0.824 $ & $     -0.968 $  \\ 
$^{48}$Ti &     2$^+$ &  0.984 & $   +0.784(38)^{\rm a}      $ & $      0.650 $ & $      0.574 $  \\ 
          &     4$^+$ &  2.296 & $   +2.16(52)^{\rm a}       $ & $      1.949 $ & $      1.863 $  \\ 
$^{49}$Ti &   7/2$^-$ &  0.000 & $   -1.10417(1)             $ & $     -1.083 $ & $     -1.247 $  \\ 
$^{50}$Ti &     2$^+$ &  1.554 & $    2.89(15)^{\rm b}       $ & $      2.473 $ & $      2.455 $  \\ 
          &     6$^+$ &  3.199 & $   +9.3(10)                $ & $      8.170 $ & $      8.306 $  \\ 
$^{48}$ V &     4$^+$ &  0.000 & $    2.012(11)              $ & $      2.023 $ & $      1.934 $  \\ 
          &     2$^+$ &  0.308 & $    0.444(16)              $ & $      0.424 $ & $      0.410 $  \\ 
$^{49}$ V &   7/2$^-$ &  0.000 & $    4.47(5)                $ & $      4.335 $ & $      4.383 $  \\ 
          &   3/2$^-$ &  0.153 & $   +2.37(12)               $ & $      2.259 $ & $      2.286 $  \\ 
$^{50}$ V &     6$^+$ &  0.000 & $ +3.3456889(14)            $ & $      3.202 $ & $      3.097 $  \\ 
$^{51}$ V &   7/2$^-$ &  0.000 & $ +5.14870573(18)           $ & $      4.849 $ & $      4.931 $  \\ 
          &   5/2$^-$ &  0.320 & $   +3.86(33)               $ & $      3.165 $ & $      3.271 $  \\ 
$^{49}$Cr &   5/2$^-$ &  0.000 & $    0.476(3)               $ & $     -0.493 $ & $     -0.571 $  \\ 
          &(19/2$^-$) &  4.365 & $   +7.4(12)                $ & $      6.427 $ & $      6.354 $  \\
$^{50}$Cr &     2$^+$ &  0.783 & $   +1.238(52)^{\rm a}      $ & $      1.125 $ & $      1.103 $  \\ 
          &     4$^+$ &  1.881 & $   +3.1(5)^{\rm a}         $ & $      2.957 $ & $      2.976 $  \\ 
          &     6$^+$ &  3.164 & $   +3.2(10)                $ & $      4.044 $ & $      4.027 $  \\ 
          &     8$^+$ &  4.745 & $   +4.3(7)                 $ & $      6.333 $ & $      6.345 $  \\ 
$^{51}$Cr &   7/2$^-$ &  0.000 & $ (-)0.934(5)               $ & $     -0.829 $ & $     -0.989 $  \\ 
          &   3/2$^-$ &  0.749 & $   -0.86(12)               $ & $     -0.317 $ & $     -0.288 $  \\ 
$^{52}$Cr &     2$^+$ &  1.434 & $  +2.41(13)^{\rm a}        $ & $      2.220 $ & $      2.272 $  \\ 
$^{53}$Cr &   3/2$^-$ &  0.000 & $   -0.47454(3)             $ & $     -0.607 $ & $     -0.587 $  \\ 
          &   7/2$^-$ &  1.290 & $   +2.8(49)                $ & $      1.199 $ & $      1.233 $  \\ 
$^{54}$Cr &     2$^+$ &  0.835 & $    1.68(11)^{\rm c}       $ & $      1.281 $ & $      1.279 $  \\ 
$^{51}$Mn &   5/2$^-$ &  0.000 & $    3.5683(13)             $ & $      3.476 $ & $      3.503 $  \\ 
$^{52}$Mn &     6$^+$ &  0.000 & $   +3.063(1)               $ & $      3.149 $ & $      3.041 $  \\ 
          &     2$^+$ &  0.378 & $   +0.00768(8)             $ & $     -0.005 $ & $     -0.069 $  \\ 
$^{53}$Mn &   7/2$^-$ &  0.000 & $    5.024(7)               $ & $      4.746 $ & $      4.843 $  \\ 
          &   5/2$^-$ &  0.378 & $   +3.25(30)               $ & $      3.402 $ & $      3.467 $  \\ 
$^{54}$Mn &     3$^+$ &  0.000 & $   +3.2819(13)             $ & $      3.234 $ & $      3.311 $  \\ 
          &     2$^+$ &  0.055 & $  3.4^{+28}_{-16}          $ & $      3.762 $ & $      3.797 $  \\ 
          &     4$^+$ &  0.156 & $   +5.1(10)                $ & $      3.538 $ & $      3.650 $  \\ 
          &     5$^+$ &  0.368 & $  +38(21)                  $ & $      4.078 $ & $      4.188 $  \\ 
          &     6$^+$ &  1.073 & $    2.8(15)                $ & $      2.965 $ & $      2.818 $  \\ 
$^{55}$Mn &   5/2$^-$ &  0.000 & $    3.4532(13)             $ & $      3.387 $ & $      3.429 $  \\ 
          &   7/2$^-$ &  0.126 & $    4.4(7)                 $ & $      4.408 $ & $      4.480 $  \\ 
$^{56}$Mn &     3$^+$ &  0.000 & $   +3.2266(2)              $ & $      3.494 $ & $      3.420 $  \\ 
$^{53}$Fe &   3/2$^-$ &  0.741 & $   -0.386(15)              $ & $     -0.464 $ & $     -0.444 $  \\ 
$^{54}$Fe &     2$^+$ &  1.408 & $   +2.10(12)^{\rm b}       $ & $      2.087 $ & $      2.187 $  \\ 
          &     6$^+$ &  2.949 & $    8.22(18)               $ & $      7.848 $ & $      8.023 $  \\ 
          &    10$^+$ &  6.527 & $   +7.281(10)              $ & $      7.176 $ & $      7.110 $  \\ 
$^{55}$Fe &   5/2$^-$ &  0.931 & $   +2.7(12)                $ & $      1.314 $ & $      1.133 $  \\ 
          &   7/2$^-$ &  1.317 & $   +2(2)                   $ & $     -0.553 $ & $     -0.717 $  \\
          &   7/2$^-$ &  1.409 & $   -2.2(5)                 $ & $      1.493 $ & $      1.596 $  \\ 
$^{56}$Fe &     2$^+$ &  0.847 & $    +1.22(16)              $ & $      1.176 $ & $      1.183 $  \\ 
$^{57}$Fe &   1/2$^-$ &  0.000 & $   +0.09044(7)             $ & $      0.241 $ & $      0.162 $  \\ 
          &   3/2$^-$ &  0.014 & $   -0.1549(2)              $ & $     -0.266 $ & $     -0.312 $  \\ 
          &   5/2$^-$ &  0.137 & $   +0.935(10)              $ & $      1.052 $ & $      0.833 $  \\ 
$^{58}$Fe &     2$^+$ &  0.811 & $   +0.92(26)               $ & $      1.206 $ & $      1.209 $  \\ 
$^{59}$Fe &   3/2$^-$ &  0.000 & $   -0.3358(4)^{\rm d}      $ & $     -0.203 $ & $     -0.251 $  \\ 
$^{55}$Co &   7/2$^-$ &  0.000 & $   +4.822(3)               $ & $      4.630 $ & $      4.746 $  \\ 
$^{56}$Co &     4$^+$ &  0.000 & $    3.851(12)              $ & $      3.652 $ & $      3.774 $  \\ 
$^{57}$Co &   7/2$^-$ &  0.000 & $   +4.720(10)              $ & $      4.616 $ & $      4.704 $  \\ 
          &   3/2$^-$ &  1.378 & $   +3.0(6)                 $ & $      2.220 $ & $      2.140 $  \\ 
$^{58}$Co &     2$^+$ &  0.000 & $   +4.044(8)               $ & $      4.229 $ & $      4.332 $  \\ 
          &     4$^+$ &  0.053 & $   +4.194(8)               $ & $      4.218 $ & $      4.166 $  \\ 
          &     3$^+$ &  0.112 & $   +2.2(4)                 $ & $      3.954 $ & $      3.969 $  \\ 
$^{59}$Co &   7/2$^-$ &  0.000 & $   +4.627(9)               $ & $      4.637 $ & $      4.707 $  \\ 
          &   3/2$^-$ &  1.292 & $   +2.54(12)               $ & $      2.794 $ & $      2.868 $  \\ 
$^{60}$Co &     5$^+$ &  0.000 & $   +3.799(8)               $ & $      3.962 $ & $      3.996 $  \\ 
          &     2$^+$ &  0.059 & $   +4.40(9)                $ & $      4.349 $ & $      4.378 $  \\ 
$^{57}$Ni &   3/2$^-$ &  0.000 & $   -0.7975(14)^{\rm d}     $ & $     -0.789 $ & $     -0.802 $  \\ 
$^{58}$Ni &     2$^+$ &  1.454 & $   +0.076(17)^{\rm e}      $ & $     -0.017 $ & $     -0.096 $  \\ 
$^{59}$Ni &   5/2$^-$ &  0.339 & $   +0.35(15)               $ & $      0.744 $ & $      0.482 $  \\ 
$^{60}$Ni &     2$^+$ &  1.333 & $   +0.32(6)^{\rm e}        $ & $      0.496 $ & $      0.412 $  \\ 
$^{61}$Ni &   3/2$^-$ &  0.000 & $   -0.75002(4)             $ & $     -0.688 $ & $     -0.707 $  \\ 
          &   5/2$^-$ &  0.067 & $   +0.480(6)               $ & $      0.787 $ & $      0.516 $  \\ 
$^{62}$Ni &     2$^+$ &  1.173 & $   +0.33(6)^{\rm e}        $ & $      0.780 $ & $      0.686 $  \\ 
$^{63}$Ni &   5/2$^-$ &  0.087 & $   +0.752(3)               $ & $      1.042 $ & $      0.730 $  \\ 
$^{64}$Ni &     2$^+$ &  1.346 & $   +0.37(6)^{\rm e}        $ & $      0.510 $ & $      0.375 $  \\ 
$^{65}$Ni &   5/2$^-$ &  0.000 & $    0.69(6)                $ & $      1.101 $ & $      0.767 $  \\ 
$^{67}$Ni & (1/2$^-$) &  0.000 & $    0.601(5)               $ & $      0.547 $ & $      0.425 $ \\
$^{60}$Cu &     2$^+$ &  0.000 & $   +1.219(3)               $ & $      1.258 $ & $      1.159 $  \\ 
$^{61}$Cu &   3/2$^-$ &  0.000 & $   +2.14(4)                $ & $      2.258 $ & $      2.193 $  \\ 
$^{62}$Cu &     1$^+$ &  0.000 & $   -0.380(4)               $ & $     -0.157 $ & $     -0.236 $  \\ 
          &     2$^+$ &  0.041 & $   +1.32(3)                $ & $      1.350 $ & $      1.210 $  \\ 
          &     4$^+$ &  0.390 & $   +2.67(16)               $ & $      2.941 $ & $      2.663 $  \\ 
$^{63}$Cu &   3/2$^-$ &  0.000 & $   +2.22329(18)            $ & $      2.314 $ & $      2.251 $  \\ 
$^{64}$Cu &     1$^+$ &  0.000 & $   -0.271(2)               $ & $     -0.023 $ & $     -0.114 $  \\ 
$^{65}$Cu &   3/2$^-$ &  0.000 & $   +2.3817(3)              $ & $      2.496 $ & $      2.398 $  \\ 
          &   5/2$^-$ &  1.115 & $   +4.5(9)                 $ & $      1.592 $ & $      1.515 $  \\
$^{66}$Cu &     1$^+$ &  0.000 & $   -0.282(2)               $ & $      0.616 $ & $      0.490 $  \\ 
$^{62}$Zn &     2$^+$ &  0.954 & $   +0.74(20)^{\rm f}       $ & $      1.176 $ & $      1.161 $  \\
$^{63}$Zn &   3/2$^-$ &  0.000 & $   -0.28164(5)             $ & $     -0.243 $ & $     -0.282 $  \\ 
$^{64}$Zn &     2$^+$ &  0.992 & $   +0.89(9)^{\rm f}        $ & $      1.239 $ & $      1.200 $  \\
$^{65}$Zn &   5/2$^-$ &  0.000 & $   +0.7690(2)              $ & $      1.027 $ & $      0.753 $  \\ 
          &   3/2$^-$ &  0.115 & $   -0.78(20)               $ & $      0.511 $ & $      0.393 $  \\ 
          &   3/2$^-$ &  0.207 & $   +0.73(25)               $ & $     -0.583 $ & $     -0.579 $  \\ 
$^{66}$Zn &     2$^+$ &  1.039 & $   +0.80(8)^{\rm f}        $ & $      1.238 $ & $      1.171 $  \\
$^{67}$Zn &   5/2$^-$ &  0.000 & $   +0.87548(1)             $ & $      1.229 $ & $      0.914 $  \\ 
          &   1/2$^-$ &  0.093 & $   +0.587(11)              $ & $      0.595 $ & $      0.492 $  \\ 
          &   3/2$^-$ &  0.185 & $   +0.50(6)                $ & $      0.679 $ & $      0.543 $  \\ 
$^{68}$Zn &     2$^+$ &  1.077 & $   +0.87(9)^{\rm f}        $ & $      1.489 $ & $      1.363 $  \\
$^{70}$Zn &     2$^+$ &  0.885 & $   +0.76(8)^{\rm f}        $ & $      3.757 $ & $      3.704 $  \\
$^{66}$Ga &   (2)$^+$ &  0.066 & $    1.011(18)              $ & $      1.056 $ & $      0.874 $ \\ 
          &   (9$^+$) &  3.043 & $    4.2(9)                 $ & $      4.443 $ & $      4.422 $ \\ 
$^{67}$Ga &   3/2$^-$ &  0.000 & $   +1.8507(3)              $ & $      1.792 $ & $      1.786 $ \\ 
          &   5/2$^-$ &  0.359 & $    1.40(65)               $ & $      1.310 $ & $      1.584 $ \\ 
$^{68}$Ga &     1$^+$ &  0.000 & $    0.01175(6)             $ & $      0.188 $ & $      0.003 $ \\ 
$^{69}$Ga &   3/2$^-$ &  0.000 & $   +2.01659(4)             $ & $      1.850 $ & $      1.840 $ \\ 
$^{71}$Ga &   3/2$^-$ &  0.000 & $   +2.56227(2)             $ & $      2.782 $ & $      2.748 $ \\ 
$^{69}$Ge &   5/2$^-$ &  0.000 & $    0.735(7)               $ & $      1.063 $ & $      0.757 $ \\ 
$^{70}$Ge &     2$^+$ &  1.039 & $   +0.936(52)              $ & $      0.789 $ & $      0.671 $ \\ 
$^{71}$Ge &   1/2$^-$ &  0.000 & $   +0.547(5)               $ & $      0.346 $ & $      0.250 $ \\ 
          &   5/2$^-$ &  0.175 & $   +1.018(10)              $ & $      1.255 $ & $      0.909 $ \\ 
$^{72}$Ge &     2$^+$ &  0.834 & $   +0.798(66)              $ & $      2.314 $ & $      2.469 $ \\ 
  \hline \hline
\end{longtable}

However, for the Ni and Zn isotopes, calculated values 
 are systematically larger than the experimental ones
 by typically $0.2\sim 0.3 \mu_N$.
Such difficulties can be remedied to a certain extent
 by introducing the effective g-factors,
 as shown in the same table ($\mu_{\rm th}^{\rm eff}$).
In the present calculation, we took
 $g_s^{\rm eff}=0.9g_s^{\rm free}$, 
 $g_l=1.1$ and $-0.1$ for protons and neutrons, respectively,
 which were chosen from an estimate by the least-squares fit.
It can be seen that
 the description of odd-$A$ nuclei is systematically improved.

Nevertheless, deviations on order of $0.3\mu_N$ remain
for the $2^+$ state of Zn nuclei.
We have examined the effect of a 
more general effective M1 operator which contains the
 $[ Y_2 \bm{s} ]^{(1)}$ term, but it turns out that
 the deviations for the Zn nuclei could not be remedied.
It may be required to include the $g_{9/2}$ orbit
 to improve the description of these states,
 as also discussed in Ref. \cite{kenn65}. One may also
want to reexamine the systematic uncertainties which may
exist in the effective transient fields which are used to
deduce magnetic moments from the experimental data.

There are a few other cases where we find 
large differences between
theory and experiment.
Several of these can be interpreted as a consequence of 
 incorrect mixing of two closely-lying states.
For example, in $^{55}$Fe, the first and the second 7/2$^-$ states
 are separated only by 92 keV experimentally.
The GXPF1 predicts these states
 in the reversed order.
It is also the case for two 3/2$^-$ states of $^{65}$Zn,
 where the experimental energy separation is 92 keV.

We also find notable differences between theory and
 experiment for
 $^{50}$Cr $8^+$,
 $^{54}$Mn $5^+$,
 $^{58}$Co $3^+$, $^{47}$Ti $7/2^-$ and $^{51}$Cr $3/2^-$
 where experimental uncertainties are also large. More
precise measurements of these are required.
The deviations in $^{65}$Cu $5/2^-$ and $^{66}$Cu $1^+$
 may be attributed to the effect of $g_{9/2}$ orbit.
The deviations in $^{70}$Zn and $^{72}$Ge are
 naturally understood as a result of the insufficient model space.

\subsection{Electric quadrupole moments}
\label{sub:q-mom}

\begin{figure}[h]
\includegraphics[width=70mm]{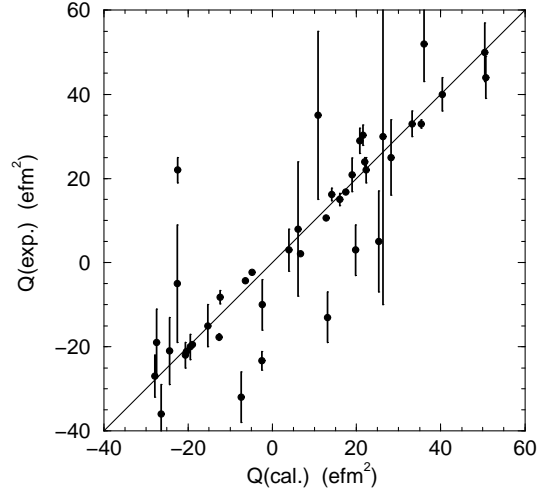}
\caption{
Comparison of experimental electric quadrupole moments
 with the shell-model results.
All data in Table \ref{tbl:qmom}
 are included
 for which the sign is measured experimentally.
\label{fig:qmom}
}
\end{figure}

The electric quadrupole moments are given in Table \ref{tbl:qmom}.
The effective charges $e_\pi = 1.5$, $e_\nu = 0.5$ are
 adopted in the present calculations.
The correlation between the calculated electric quadrupole moments 
 and the experimental data is shown in Fig. \ref{fig:qmom}.
One can find that, in general, the deviation of the theoretical
 prediction from the experimental data becomes large where the
 experimental error bar is also large,
 except for a few cases.
The sign of the calculated $Q(3/2^-)$ for $^{57}$Co is opposite
 to the experimental data.
This result suggests an incorrect mixing of the first and the second
 3/2$^-$ states in the calculation, which are close in energy
 (380 keV experimentally).
The calculated $3/2^-_2$ state lies 589 keV above $3/2^-_1$ state
 with $Q$=+0.138eb, which is consistent with this interpretation.
The description of $^{64}$Zn 2$^+$ is also very poor,
but the uncertainty in the data is large.
The description is unsuccessful for $^{70}$Zn, $^{70}$Ge and $^{72}$Ge,
 indicating the need for introducing the $g_{9/2}$ orbit.

\begin{table}[]
\caption{Comparison of experimental electric quadrupole
 moments $Q_{\rm exp}$ (in unit $e$fm$^2$) with
 theoretical values $Q_{\rm th}$, which are calculated
 by using the effective charges $e_\pi = 1.5$, 
 $e_\nu = 0.5$.
Data are taken from Ref. \cite{nndc}, except for
 $^b$ from Ref. \cite{dufek} and
 $^a$, which was evaluated based on $^b$ and a constraint 
 $Q$($^{54}$Fe; 10$^+$)/$Q$($^{57}$Fe; 3/2$^-$)=3.62$\pm$0.22 \cite{hass}
 from the analysis of M\"{o}ssbauer data.
\label{tbl:qmom}
}
\begin{ruledtabular}
\begin{tabular}{lcccr}
nuclei &  \ \ state  \ \ & $E_x$ (MeV) & \ \ \ \ \ \ $Q_{\rm exp}$  \ \ \ \ \ & \ \ \ \ \ \  $Q_{\rm th}$ \\ \hline
$^{47}$Ca &   7/2$^-$ &  0.000 & $  +2.1(4)   $ & $      6.7 $ \\ 
$^{47}$Sc &   7/2$^-$ &  0.000 & $  -22(3)    $ & $    -20.6 $ \\ 
$^{47}$Ti &   5/2$^-$ &  0.000 & $  +30.3(24) $ & $     21.6 $ \\ 
$^{48}$Ti &     2$^+$ &  0.984 & $  -17.7(8)  $ & $    -12.6 $ \\ 
$^{49}$Ti &   7/2$^-$ &  0.000 & $  +24(1)    $ & $     22.0 $ \\ 
$^{50}$Ti &     2$^+$ &  1.554 & $   +8(16)   $ & $      6.2 $ \\ 
$^{50}$ V &     6$^+$ &  0.000 & $  +20.9(40) $ & $     19.0 $ \\ 
$^{51}$ V &   7/2$^-$ &  0.000 & $   -4.3(5)  $ & $     -6.3 $ \\ 
$^{50}$Cr &     2$^+$ &  0.783 & $  -36(7)    $ & $    -26.4 $ \\ 
$^{52}$Cr &     2$^+$ &  1.434 & $   -8.2(16) $ & $    -12.3 $ \\ 
$^{53}$Cr &   3/2$^-$ &  0.000 & $  -15(5)    $ & $    -15.3 $ \\ 
$^{54}$Cr &     2$^+$ &  0.835 & $  -21(8)    $ & $    -24.4 $ \\ 
$^{51}$Mn &   5/2$^-$ &  0.000 & $   42(7)    $ & $     34.8 $ \\ 
$^{52}$Mn &     6$^+$ &  0.000 & $  +50(7)    $ & $     50.5 $ \\ 
$^{54}$Mn &     3$^+$ &  0.000 & $  +33(3)    $ & $     33.2 $ \\ 
$^{55}$Mn &   5/2$^-$ &  0.000 & $  +33(1)    $ & $     35.4 $ \\ 
$^{54}$Fe &     2$^+$ &  1.408 & $   -5(14)   $ & $    -22.6 $ \\ 
          &    10$^+$ &  6.527 & $   58(6)^{\rm a}    $ & $     53.5 $ \\ 
$^{56}$Fe &     2$^+$ &  0.847 & $  -19(8)    $ & $    -27.4 $ \\ 
$^{57}$Fe &   3/2$^-$ &  0.014 & $   16(1)^{\rm b}    $ & $     15.8 $ \\ 
$^{58}$Fe &     2$^+$ &  0.811 & $  -27(5)    $ & $    -27.9 $ \\ 
$^{56}$Co &     4$^+$ &  0.000 & $   +25(9)   $ & $     28.2 $ \\ 
$^{57}$Co &   7/2$^-$ &  0.000 & $   +52(9)   $ & $     36.0 $ \\ 
          &   3/2$^-$ &  1.378 & $   +22(3)   $ & $    -22.5 $ \\ 
          &   3/2$^-$ &  1.758 & $            $ & $     13.8 $ \\ 
$^{58}$Co &     2$^+$ &  0.000 & $   +22(3)   $ & $     22.3 $ \\ 
$^{59}$Co &   7/2$^-$ &  0.000 & $   +40(4)   $ & $     40.4 $ \\ 
$^{60}$Co &     5$^+$ &  0.000 & $   +44(5)   $ & $     50.7 $ \\ 
          &     2$^+$ &  0.059 & $   +30(40)  $ & $     26.3 $ \\ 
$^{58}$Ni &     2$^+$ &  1.454 & $  -10(6)    $ & $     -2.4 $ \\ 
$^{60}$Ni &     2$^+$ &  1.333 & $   +3(5)    $ & $      3.9 $ \\ 
$^{61}$Ni &   3/2$^-$ &  0.000 & $  +16.2(15) $ & $     14.2 $ \\ 
          &   5/2$^-$ &  0.067 & $  -20(3)    $ & $    -19.5 $ \\ 
$^{62}$Ni &     2$^+$ &  1.173 & $   +5(12)   $ & $     25.3 $ \\ 
$^{64}$Ni &     2$^+$ &  1.346 & $   +35(20)  $ & $     10.9 $ \\ 
$^{63}$Cu &   3/2$^-$ &  0.000 & $  -21.1(4)  $ & $    -20.4 $ \\ 
$^{65}$Cu &   3/2$^-$ &  0.000 & $  -19.5(4)  $ & $    -19.0 $ \\
$^{63}$Zn &   3/2$^-$ &  0.000 & $  +29(3)    $ & $     20.8 $ \\ 
$^{64}$Zn &     2$^+$ &  0.992 & $  -32(6)$ or $-26(6) $ & $     -7.4 $ \\ 
$^{65}$Zn &   5/2$^-$ &  0.000 & $   -2.3(2)  $ & $     -4.8 $ \\ 
$^{67}$Zn &   5/2$^-$ &  0.000 & $   15.0(15) $ & $     16.1 $ \\ 
$^{70}$Zn &     2$^+$ &  0.885 & $  -23.3(22) $ & $     -2.5 $ \\ 
$^{67}$Ga &   3/2$^-$ &  0.000 & $   19.5     $ & $     19.6 $ \\ 
$^{68}$Ga &     1$^+$ &  0.000 & $   2.77(14) $ & $     -1.3 $ \\ 
$^{69}$Ga &   3/2$^-$ &  0.000 & $  +16.8     $ & $     17.5 $ \\ 
$^{71}$Ga &   3/2$^-$ &  0.000 & $  +10.6     $ & $     12.8 $ \\ 
$^{69}$Ge &   5/2$^-$ &  0.000 & $    2.4(5)  $ & $      6.8 $ \\ 
$^{70}$Ge &     2$^+$ &  1.039 & $   +3(6)    $ & $     19.8 $ \\ 
$^{72}$Ge &     2$^+$ &  0.834 & $  -13(6)    $ & $     13.2 $ \\ 
\end{tabular}
\end{ruledtabular}
\end{table}

\subsection{Systematics of $2^+_1$ states}
\label{sub:2plus}

\begin{figure}
\includegraphics[scale=0.52]{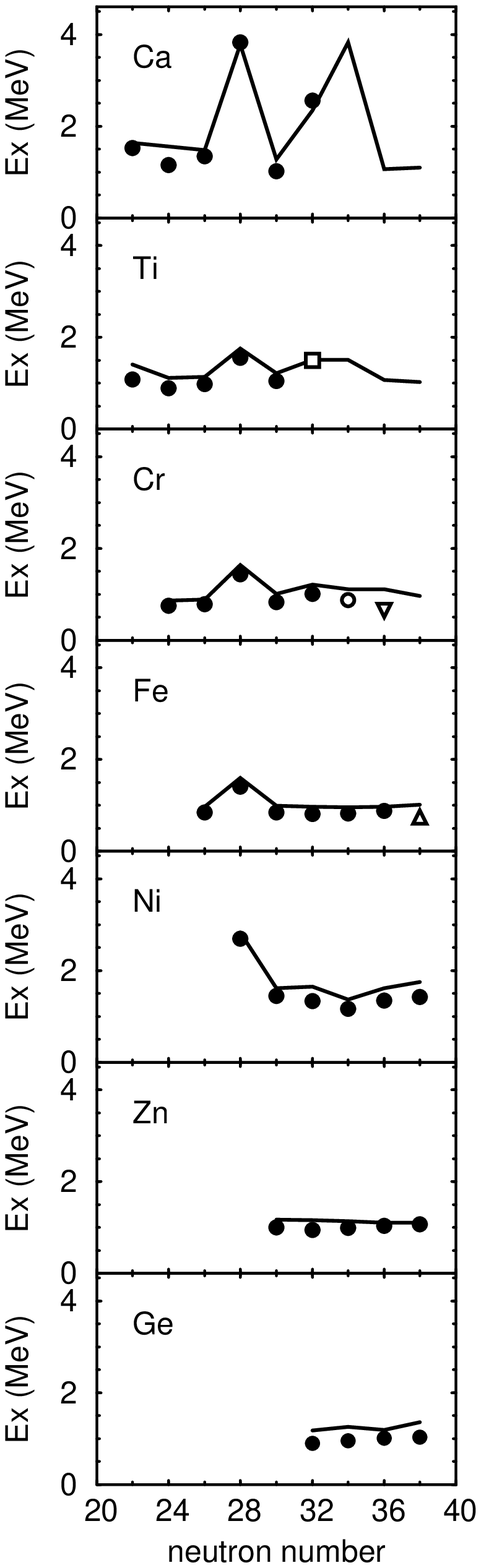}
\includegraphics[scale=0.52]{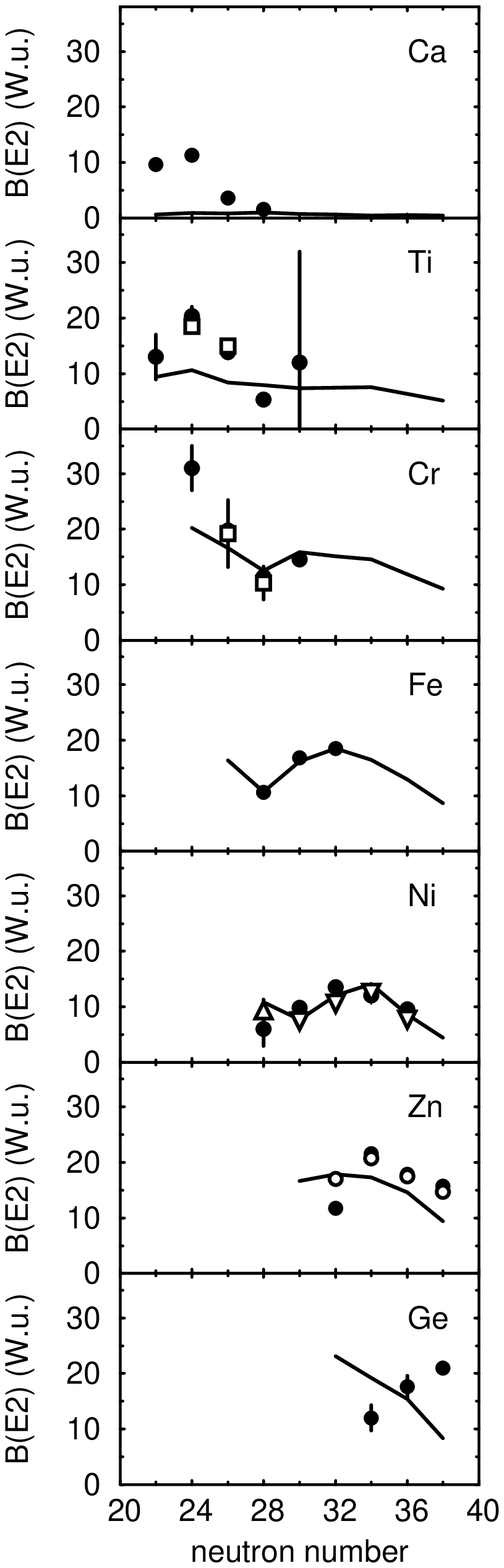}
\caption{(Left) First 2$^+$ energy levels as a function of
 the neutron number $N$. 
Experimental data are shown by
 filled circles \protect\cite{nndc},
 square \protect\cite{ti54},
 open circle \protect\cite{cr58},
 triangle(down) \protect\cite{cr60},
 and triangle(up) \protect\cite{hanna}.
Solid lines show results of shell-model calculations.
The truncation order $t$
 is 5 for $^{56,58}$Fe, $^{60}$Ni, $^{60,62,64}$Zn,
 6 for $^{56,58,62}$Ni and 7 for $^{52,54}$Fe.
The other results are exact.
(Right) The values of $B$(E2; $2^+_1 \rightarrow 0^+_1$).
Experimental data are shown by
 filled circles \protect\cite{nndc},
 squares \protect\cite{speidel},
 triangle (up) \protect\cite{kraus},
 triangles (down) \protect\cite{kenn},
 and open circles \protect\cite{kenn65}.
\label{fig:2plus}
}
\end{figure}

The first $2^+$ state of an even-even nucleus is
 a good systematic measure of the structure.
The left panel of Fig. \ref{fig:2plus}
 shows the excitation energies of the $2^+_1$ states for
 Ca, Ti, Cr, Fe, Ni, Zn and Ge isotopes.
The results except for Zn and Ge isotopes have already been
 discussed in our previous paper \cite{upf}.
The lightest nucleus in each isotope chain
 is taken to be  $N$=$Z$ (cases with $N < Z$ have mirror
nuclei similar properties).
The overall description of the $2^+$ energy levels 
 is reasonable for all these isotope chains,
 although the calculated energies are systematically
 higher than experimental ones by about 200 keV.
In all cases, the energy jump corresponding to $N=28$ shell closure
 is nicely reproduced for Ca to Ni isotopes.
In the case of Zn isotopes,
 $E_x(2^+_1)$ is almost constant, which is also well reproduced.
Recently measured value for $^{54}$Ti ($N$=32) \cite{ti54} 
 comes precisely on the prediction of GXPF1.
The new data of $^{58}$Cr ($N$=34) \cite{cr58}
 also follows the predicted systematics. 

Very recent
data for $^{56}$Ti gives a 2$^+$ energy of 1.13 MeV \cite{ti56}. This is
significantly lower than the GPFX1 prediction of 1.52 MeV. 
The KB3G interaction gives
0.89 MeV for the $^{56}$Ti 2$^+$, and experiment lies in between
GPFX1 and KB3G.
The dominant neutron component of this 2$^+$ state has
one neutron in the $f_{5/2}$ orbit. 
In $^{54}$Ti there are some high spin states whose
wave functions are dominated by the configuration with 
one neutron in the $f_{5/2}$ orbit, and their  
experimental energies are about 400 keV lower 
than GXPF1 and in between GPFX1 and KB3G \cite{ti54}.
The implication is that the effective single-particle
energy for the neutron $f_{5/2}$ orbit in $Z$=22 is about 800 keV too
high compared with GPFX1. The strong monopole 
interaction between the proton $f_{7/2}$ orbit and the 
neutron $f_{5/2}$ orbit \cite{magic}
is responsible for lowering the energy of the $f_{5/2}$
neutron energy (relative to $p_{3/2}$) as protons
are added to the $f_{7/2}$ orbit from $Z$=20 to 28 (see the right-hand side
of Fig. 1 in \cite{upf}). Thus to improve the agreement for $^{56}$Ti, one would
need to reduce the strength of these two-body matrix elements in a
way which is consistent with the entire fit. This will be one
of the considerations for a next generation interaction.
The $N$=34 gap between $p_{1/2}$ and $f_{5/2}$ is about 4 MeV.
If it is estimated too large by 800 keV by the GXPF1 interaction,
the real gap turns out be greater than 3 MeV.  Thus, this very new data seem
to support the appearance of $N$=34 gap, while details of the GXPF1
interaction may have to be improved.  One must be also aware of the
small separation energy of $f_{5/2}$ in some of the nuclei being
discussed here.  Such a small separation energy induces additional
relative lowering of $f_{5/2}$, whereas this effect becomes weaker
in nuclei with more protons.  This effect is not included in the
present fit, because it is not sizable in the nuclei used for the fit.

The 2$^+$ level for $^{60}$Cr ($N$=36) \cite{cr60}
 deviates toward lower energy as compared to the GXPF1 prediction.
This is correlated with the deviation in binding energy and both
are signatures of the more collectivity from mixing with $g_{9/2}$.
In the Fe isotopes, a similar deviation can be seen at $N$=38.
These results suggest the limitation of
the reliability of GXPF1 interaction for neutron-rich nuclei.
It is likely that one will need to explicitly introduce the
$g_{9/2}$ orbit into the model space in order to improve 
the calculations as one approaches $N$=40.

In the right panel of Fig. \ref{fig:2plus},
 the E2 transition matrix elements $B$(E2;$2^+_1 \rightarrow 0^+_1)$
 are shown. 
Experimental $B$(E2) are significantly larger than
theory for $ N\leq 24$, especially for the Ca isotopes,
 although the agreement in the excitation energy
 of $2^+_1$ state is reasonable.
These deviations show the large effects of the core excitation
that has long been known for nuclei like $^{42}$Ca \cite{flowers}.
The $^{40}$Ca core is significantly broken and
 we should take into account the excitation from the
 $sd$-shell explicitly in order to reproduce these $B$(E2) values.
On the other hand, the agreement between theory and
 experiment is quite good in the middle of the shell,
 especially for Fe and Ni isotopes.
The dependence on the neutron number is nicely reproduced
 including the $N=28$ magic number.
For Zn and Ge isotopes with $N\ge 34$, we again see
 the need of more collectivity in the model space, 
which also suggests the
 necessity of the $g_{9/2}$ orbit, consistently with
 the results of electro-magnetic moments
 discussed in the previous subsections.

\subsection{Semi-magic nuclei}
\label{sub:semi}

In this subsection, spectroscopic properties are studied in detail
 for several $N$ or $Z=28$ nuclei around $^{56}$Ni.
If we assume an inert closed core for $^{56}$Ni,
 these semi-magic nuclei are described
 by a few valence nucleons of one kind, i.e., only proton holes or
 only neutron particles.
Therefore, the low-lying level density is rather low and
 the effects of core-excitations are 
easy to interpret.
In fact, it has been pointed out in Ref. \cite{nakada} that,
 although 
 the shell-model calculation in a truncated $t\leq 2$ subspace is
 quite successful for describing most of the low-lying stats
 in $N$=28-30 nuclei, it is 
 impossible to reproduce those states which contain
 sizable broken-core components,
 such as the lowest excited states in $^{55}$Co,
 excited states above the lowest $3/2^-, 5/2^-, 1/2^-$ triplets in $^{57}$Ni,
 excited $0^+$ in $^{54}$Fe and $^{58}$Ni, etc.
Similar difficulties can be seen also in Ref. \cite{rud-epj},
 where the shell-model calculations in a truncated space
 have been carried out with a different effective interaction.
Note that the cause of this problem lies not only in the truncation
 of the model space
 but also in the effective interaction itself.
It is well known that the KB3 interaction (and also its descendants) gives
an excellent description for low-lying states of $A\leq 52$ nuclei,
 but KB3 
fails around $^{56}$Ni,
 even when the model space is sufficiently large for convergence.
Thus it is interesting to investigate whether GXPF1
 can properly describe these low-lying states which are sensitive to
 the core-excitation.
We consider semi-magic nuclei with $Z$=25, 26, 27 and $N$=29, 30, 31.

\subsubsection{$^{53}$Mn}
\label{sub2:mn53}

\begin{figure}
\includegraphics[width=85mm]{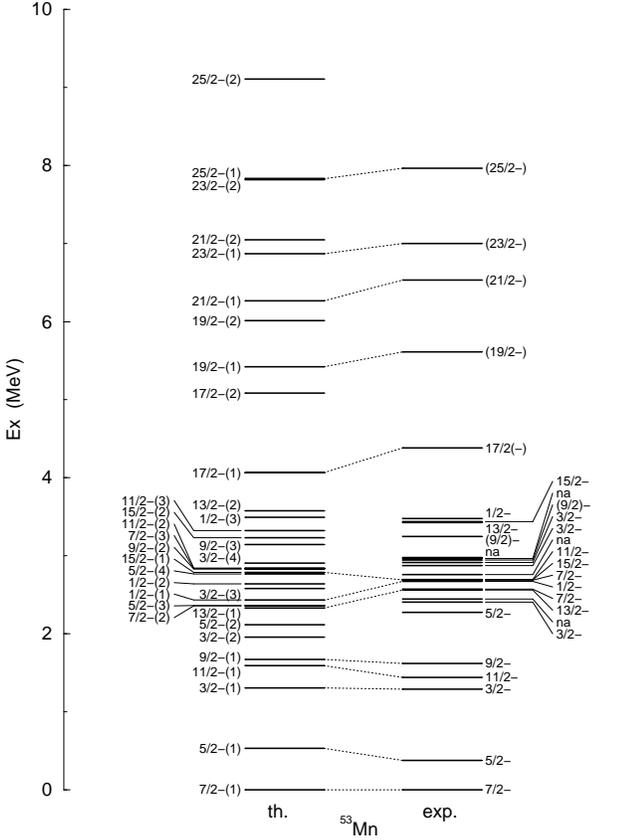}
\caption{Experimental (right) and calculated (left)
 energy levels of $^{53}$Mn.
Experimental data are taken from \protect\cite{nds-mn53}.
The label ``na'' indicates that the spin-parity is not assigned
 experimentally.
The calculated yrast states are
 connected with experimental counterparts
 (or their candidates) by dotted lines.
\label{fig:mn53}
}
\end{figure}

Figure \ref{fig:mn53} shows experimental and calculated energy
 levels of $^{53}$Mn.
Most of the theoretical energy levels are obtained
 in the $t=7$ subspace.
The exact ($t$=13) results are also obtained in some cases.
All experimental energy levels
 below 3 MeV excitation energy as well as
 the yrast states are shown with
 their theoretical counterparts and
 several additional states.

For the yrast states, the agreement between the 
 theory and the experiment is quite good.
Assuming an inert $^{56}$Ni core, this nucleus is
 described by three neutron holes in the $f_{7/2}$ orbit.
Under such an assumption,
 possible states are $J^\pi=3/2^-$, $5/2^-$, $7/2^-$,
 $9/2^-$, $11/2^-$ and $15/3^-$.
In the calculated wave functions,
 the lowest $(f_{7/2})^{13}$ configuration is in fact dominant
 (46$\sim$60\%)
 for the yrast states of these spin-parities.
Therefore, the truncation of the model space by a small $t\sim$2 is
 reasonable for these states.

Other yrast states, $1/2^-$, $13/2^-$, $17/2^-$, $19/2^-$,
 $21/2^-$, $23/2^-$ and $25/2^-$ consist mainly of
 $\pi(f_{7/2})^5\nu(f_{7/2})^7(p_{3/2})^1$ and
 $\pi(f_{7/2})^5\nu(f_{7/2})^7(f_{5/2})^1$ configurations.
These states are also well reproduced in the
 calculation, indicating that the effective
 single-particle gap above $N=28$ is reasonably
 reproduced in the present interaction.

In order to
 investigate the properties of the core in detail,
 it is important to consider non-yrast states
 which contain a large amount of multi-particle excitations
 from the $N=28$ core.
Above $E_x\sim$ 2.2 MeV, the experimental
 level density increases significantly,
 which is well reproduced in the calculation.
The largest difference between the experiment and the calculation
 is found for the $3/2^-_2$ state
 which lies at lower energy by 450 keV in the calculation,
 resulting in the inversion of $5/2^-_2$ and $3/2^-_2$.
Nevertheless, one can find reasonable
 one-to-one correspondence between the experimental and
 theoretical energy levels.

Focusing on the neutron configuration,
 the main configuration of these non-yrast states
 is $\nu(f_{7/2})^7(r)^1$
 [$r$ stands for ($p_{3/2}$,$f_{5/2}$,$p_{1/2}$)].
The typical probability for 
 the neutron 1p-1h configuration is $\sim$60\%.
These states contain also multi-neutron excitations as
 2p-2h ($\sim$25\%) and
 3p-3h ($\sim$7\%).
Therefore, taking into account the proton excitation, the
 $t\sim 4$ subspace is reasonable for describing these states.

\begin{table}
\caption{$B$(M1) and $B$(E2) for $^{53}$Mn.
The excitation energy ($E_x$) is shown in keV.
Experimental data are taken from \protect\cite{nds-mn53}.
\label{tbl:mn53}
}
\begin{ruledtabular}
\begin{tabular}{ccccr}
initial & final & multi- & exp. & th. \ \ \\ 
$J^{\pi}(E_x)$ & $J^{\pi}(E_x)$ & pole & (W.u) & (W.u) \\ \hline
 5/2$^-$(378)      &  7/2$^-$(0)        & M1 & $ 0.00254(23)          $ & $     0.0060 $ \\ 
                   &                    & E2 & $ 14(3)                $ & $       21.7 $ \\ 
 3/2$^-$(1290)     &  5/2$^-$(378)      & M1 & $ 0.023(3)             $ & $     0.0194 $ \\ 
                   &                    & E2 & $ 1.9(5)               $ & $        0.6 $ \\ 
                   &  7/2$^-$(0)        & E2 & $ 13.4(11)             $ & $        8.1 $ \\ 
11/2$^-$(1441)     &  7/2$^-$(0)        & E2 & $ 12.8(18)             $ & $        9.4 $ \\ 
 9/2$^-$(1620)     &  5/2$^-$(378)      & E2 & $ 3.7(6)               $ & $        3.6 $ \\ 
                   &  7/2$^-$(0)        & M1 & $ 0.0012(3)            $ & $     0.0002 $ \\ 
                   &                    & E2 & $ 7.0(9)               $ & $        5.4 $ \\ 
 5/2$^-$(2274)     &  3/2$^-$(1290)     & M1 & $ 0.0021(8)            $ & $     0.0000 $ \\ 
                   &                    & E2 & $ 3.0(15)              $ & $        0.0 $ \\ 
                   &  5/2$^-$(378)      & M1 & $ 0.0019(4)            $ & $     0.0010 $ \\ 
                   &                    & E2 & $ 0.61(14)             $ & $        1.6 $ \\ 
                   &  7/2$^-$(0)        & M1 & $ 0.0054(11)           $ & $     0.0023 $ \\ 
                   &                    & E2 & $ 0.07(4)              $ & $       0.02 $ \\ 
 3/2$^-$(2407)     &  3/2$^-$(1290)     & M1 & $ 0.064(24)            $ & $     0.0482 $ \\ 
                   &                    & E2 & $ 1.9(16)              $ & $        0.3 $ \\ 
                   &  5/2$^-$(378)      & M1 & $ >0.0038^{+22}_{-10}      $ & $     0.0058 $ \\ 
                   &                    & E2 & $ <0.43                $ & $        1.0 $ \\ 
                   &  7/2$^-$(0)        & E2 & $ 2.0(9)               $ & $        4.4 $ \\ 
13/2$^-$(2563)     & 11/2$^-$(1441)     & M1 & $ 0.00146(18)          $ & $     0.0022 $ \\ 
                   &                    & E2 & $                      $ & $        0.0 $ \\ 
 1/2$^-$(2671)\footnote[1]{1/2$^-_2$ in the calculation.}     &  5/2$^-$(378)      & E2 & $ 15(12)               $ & $        2.7 $ \\ 
 7/2$^-$(2686)     &  5/2$^-$(378)      & M1 & $ 0.004(4)             $ & $     0.0014 $ \\ 
                   &                    & E2 & $                      $ & $        0.1 $ \\ 
                   &  7/2$^-$(0)        & M1 & $ 0.010(9)             $ & $     0.0100 $ \\ 
                   &                    & E2 & $ 2.4(22)              $ & $        0.9 $ \\ 
15/2$^-$(2693)     & 13/2$^-$(2563)     & M1 & $ 0.21(4)              $ & $     0.0183 $ \\ 
                   &                    & E2 & $                      $ & $        4.7 $ \\ 
                   & 11/2$^-$(1441)     & E2 & $ 5.4(8)               $ & $        3.3 $ \\ 
 3/2$^-$(2876)     &  5/2$^-$(378)      & M1 & $ 2.5$E-$5(10)           $ & $     0.0040 $ \\ 
                   &                    & E2 & $                      $ & $        0.9 $ \\ 
                   &  7/2$^-$(0)        & E2 & $ 0.0008(3)            $ & $        0.2 $ \\ 
 9/2$^-$(2947)     &  7/2$^-$(0)        & M1 & $ 0.007(4)             $ & $     0.0315 $ \\ 
                   &                    & E2 & $ 1.8(8)               $ & $        3.9 $ \\ 
13/2$^-$(3426)     & 11/2$^-$(2698)     & M1 & $ 0.04(3)              $ & $     0.0144 $ \\ 
                   &                    & E2 & $ 1.6$E+$2(12)           $ & $       18.1 $ \\ 
15/2$^-$(3439)     & 15/2$^-$(2693)     & M1 & $ 0.23(4)              $ & $     1.0585 $ \\ 
                   &                    & E2 & $ 2.$E+1$(5)             $ & $        1.3 $ \\ 
19/2$^-$(5614)     & 17/2$^-$(4384)     & M1 & $ >0.22                $ & $     0.1183 $ \\ 
                   &                    & E2 & $                      $ & $       16.1 $ \\ 
21/2$^-$(6533)     & 17/2$^-$(4384)     & E2 & $ 2.8(16)              $ & $        7.0 $ \\ 
23/2$^-$(7004)     & 21/2$^-$(6533)     & M1 & $ 0.153(24)            $ & $     0.1099 $ \\ 
                   &                    & E2 & $                      $ & $        0.0 $ \\ 
\end{tabular}
\end{ruledtabular}
\end{table}

However, even in this energy region of $E_x\sim 2.5$ MeV,
 there are several states in which the $N=28$ core is
 more severely broken.
In $5/2^-_2$, $7/2^-_2$, $9/2^-_2$,
 the main neutron configuration is
 2p-2h type ($\sim$60\%), and
 also there are sizable broken-core components such as
 3p-3h ($\sim$17\%) and
 4p-4h ($\sim$6\%).
Evidently, the
 $t=4$ subspace is no longer sufficient
 and at least $t=6$ is needed to describe these states properly.
In fact, for example, the electric quadrupole moment for
 $5/2^-_2$ varies as $Q(5/2^-_2)$=$-0.25$, $+0.32$, $+0.36$ and $+0.37$eb 
 for $t$=4, 5, 6 and 13 (exact), respectively.
The change of the sign in going from $t=4$ to $t=5$
 is due to the crossing of $5/2^-_2$ and $5/2^-_3$.
A similar crossing is found in the case of $7/2^-_2$.
Although these neutron 2p-2h states
 are connected by relatively large (collective) M1 and E2 
 transition matrix elements as
 $B($E2$; 7/2^-_2\rightarrow 5/2^-_2)=34$ W.u. and
 $B($M1$; 7/2^-_2\rightarrow 5/2^-_2)=0.17$ W.u.,
 it may be difficult to observe such transitions 
 because of negligibly small gamma-ray energies
 in comparison to the transition to the yrast states.

In Table \ref{tbl:mn53}, calculated 
 electro-magnetic transition probabilities are
 compared with experimental data.
Reasonable agreement can be seen in most cases,
 including transitions between non-yrast states.
Note that
 the theoretical $1/2^-_2$ is assigned
 to the experimental $1/2^-$ at 2671 keV so that
 the transition data can be consistently described.

\subsubsection{$^{54}$Fe}
\label{sub2:fe54}

\begin{figure}
\includegraphics[width=85mm]{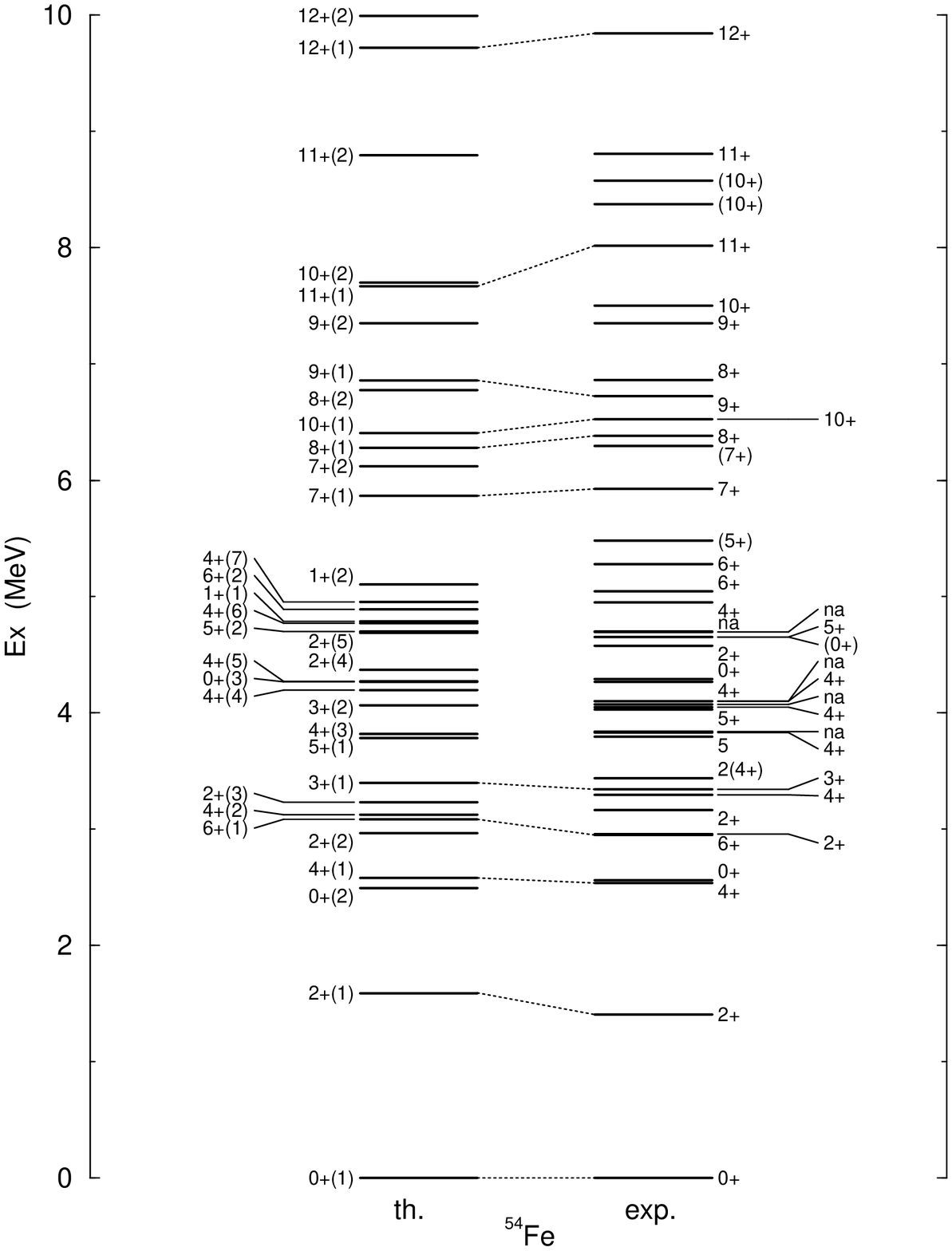}
\caption{Energy levels of $^{54}$Fe.
Experimental data are taken from
 \protect\cite{nds-fe54} and \protect\cite{rud-epj}.
Conventions are the same as in Fig. \ref{fig:mn53}.
\label{fig:fe54}
}
\end{figure}

In Fig. \ref{fig:fe54}, the energy levels of $^{54}$Fe are compared
 with experimental data.
All experimental energy levels below $E_x$=4 MeV and all 
 yrast states
 are shown with their theoretical counterparts.
Shell-model calculations have been carried out
 in the $t=7$ or larger subspace including the exact ($t$=14).
One can see remarkable one-to-one correspondence
 between theory and experiment for most of these levels.

The yrast band $0^+$-$2^+$-$4^+$-$6^+$ shows
 a typical proton two hole spectrum
 in the $f_{7/2}$ orbit under the influence of a short range interaction.
In fact the wave functions of these states are dominated
 by the $\pi(f_{7/2})^6 \nu(f_{7/2})^8$ configuration (50-60\%).
On the other hand, the yrast $8^+$ and $10^+$ states consist of
 core-excited configurations.
While the dominating configurations are still of
 neutron 1p-1h excitation type  
 from the $N=28$ core (36\% and 57\% for
 $8^+$ and $10^+$ states, respectively),
the wave functions can no longer be approximated by only one
 configuration.
The energy gap between $6^+$ and $8^+$ reflects the softness
 of the $^{56}$Ni core,
 which is determined by the effective interaction.
It is shown in Ref. \cite{brazil} that the
energy gap is too large with the KB3 interaction.
This is consistent with the fact that it also gives an energy
for the yrast $2^+$ energy in $^{56}$Ni which is too 
high compared with experiment.

It was pointed out in Ref. \cite{semi} that the FPD6 \cite{fpd6} interaction
 predicts the existence of deformed states
 at relatively low excitation energy ($\sim$3 MeV),
 which consist of neutron 2p-2h configurations.
In the present calculation, the candidates of such states appear
 as $0^+_2$ and $2^+_3$.
The neutron 2p-2h configuration is 35\% and 26\% in these
 states, respectively, which are the most dominant
 components in their wave functions.
However, the deformation of these states is much smaller
 than the FPD6 prediction.
The electric quadrupole moment for the ``2p-2h'' $2^+$ state is
 predicted to be $-0.30$eb \cite{semi}
 by the variation after the angular momentum projection (VAP) method
 with the FPD6 interaction.
On the other hand, in the present results, it is $Q(2^+_3)=-0.09$ eb, and
 the E2 transition matrix element is $B(E2; 2^+_3 \rightarrow 0^+_2)=11.3$ W.u.,
 which is slightly larger than that of the yrast in-band transition.
These values are
 inconsistent with the axially symmetric large prolate deformation.
In the present results, two neutrons are
 excited mainly to the $p_{3/2}$ orbit,
 while it is $f_{5/2}$ in the results of the FPD6 interaction,
 reflecting the fact that the
 effective single particle energy for the $f_{5/2}$ orbit
is too low for FPD6.

\begin{table}
\caption{$B$(M1) and $B$(E2) for $^{54}$Fe.
Experimental data are taken from \protect\cite{nds-fe54}.
\label{tbl:fe54}
}
\begin{ruledtabular}
\begin{tabular}{ccccr}
initial & final & multi- & exp. & th. \ \ \\ 
$J^{\pi}(E_x)$ & $J^{\pi}(E_x)$ & pole & (W.u) & (W.u) \\ \hline
   2$^+$(1408)     &    0$^+$(0)        & E2 & $ 10.6(4)              $ & $       10.7 $ \\ 
   4$^+$(2538)     &    2$^+$(1408)     & E2 & $ 6.3(13)              $ & $        5.5 $ \\ 
   0$^+$(2561)     &    2$^+$(1408)     & E2 & $ <16                  $ & $        3.8 $ \\ 
   6$^+$(2949)     &    4$^+$(2538)     & E2 & $ 3.25(5)              $ & $        3.3 $ \\ 
   2$^+$(2959)     &    2$^+$(1408)     & M1 & $ 0.051(9)             $ & $     0.0584 $ \\ 
                   &                    & E2 & $ 0.5(4)               $ & $        0.2 $ \\ 
                   &    0$^+$(0)        & E2 & $ 2.2(4)               $ & $        3.1 $ \\ 
   2$^+$(3166)     &    2$^+$(1408)     & M1 & $ 0.0035(21)           $ & $     0.0013 $ \\ 
                   &                    & E2 & $ 1.0(13)              $ & $        1.7 $ \\ 
                   &    0$^+$(0)        & E2 & $ 0.74(19)             $ & $        0.8 $ \\ 
   4$^+$(3295)     &    4$^+$(2538)     & M1 & $ <0.020               $ & $     0.2727 $ \\ 
                   &                    & E2 & $ <2.7                 $ & $        1.7 $ \\ 
                   &    2$^+$(1408)     & E2 & $ <0.15                $ & $        0.9 $ \\ 
   3$^+$(3345)     &    4$^+$(2538)     & M1 & $ <0.0086              $ & $     0.0000 $ \\ 
                   &                    & E2 & $ <0.022               $ & $        0.9 $ \\ 
                   &    2$^+$(1408)     & M1 & $ <0.00068             $ & $     0.0004 $ \\ 
                   &                    & E2 & $ <0.11                $ & $        0.0 $ \\ 
   4$^+$(3833)     &    4$^+$(3295)     & M1 & $ 0.011(12)            $ & $     0.0312 $ \\ 
                   &                    & E2 & $ 8.E+1(9)             $ & $        1.9 $ \\ 
                   &    4$^+$(2538)     & M1 & $                      $ & $     0.0032 $ \\ 
                   &                    & E2 & $ 18(7)                $ & $        0.0 $ \\ 
                   &    2$^+$(1408)     & E2 & $ 7.9(18)              $ & $        8.6 $ \\ 
   4$^+$(4031)\footnote[1]{4$^+_5$ in the calculation.}     &    4$^+$(3295)     & M1 & $ <0.033               $ & $     0.0112 $ \\ 
                   &                    & E2 & $ <1.3E2               $ & $        0.1 $ \\ 
                   &    4$^+$(2538)     & M1 & $ <0.0014              $ & $     0.0514 $ \\ 
                   &                    & E2 & $ <1.5                 $ & $        0.7 $ \\ 
   4$^+$(4048)\footnote[2]{4$^+_4$ in the calculation.}     &    3$^+$(3345)     & M1 & $ 0.15(7)              $ & $     0.0764 $ \\ 
                   &                    & E2 & $ 32^{+28}_{-24}           $ & $       16.3 $ \\ 
                   &    2$^+$(1408)     & E2 & $ 0.20(16)             $ & $        0.3 $ \\ 
   4$^+$(4268)     &    4$^+$(2538)     & M1 & $ 0.032(12)            $ & $     0.0052 $ \\ 
                   &                    & E2 & $ 6(5)                 $ & $        4.8 $ \\ 
                   &    2$^+$(1408)     & E2 & $ 0.60(17)             $ & $        0.5 $ \\ 
   0$^+$(4291)     &    2$^+$(1408)     & E2 & $ 4.3(14)              $ & $        4.3 $ \\ 
   2$^+$(4579)     &    2$^+$(1408)     & M1 & $ >0.067               $ & $     0.0170 $ \\ 
                   &                    & E2 & $                      $ & $        4.9 $ \\ 
                   &    0$^+$(0)        & E2 & $ >0.99                $ & $        0.7 $ \\ 
                   &                    & M1 & $ 0.0052(8)            $ & $     0.0000 $ \\ 
   8$^+$(6381)     &    6$^+$(2949)     & E2 & $ 0.86(22)             $ & $        3.3 $ \\ 
  10$^+$(6527)     &    8$^+$(6381)     & E2 & $ 1.69(4)              $ & $        2.0 $ \\ 
\end{tabular}
\end{ruledtabular}
\end{table}

In Table \ref{tbl:fe54}, the calculated electro-magnetic transition
 matrix elements are compared with experimental data.
The E2 transition matrix elements in the ground-state band
 are reproduced fairly well in the present calculations,
 although the $8^+\rightarrow 6^+$ transition is overestimated.
Such a deviation is commonly seen also
 in the results of KB3 and FPD6 interactions \cite{brazil}.
Other $B$(E2) values are reasonably reproduced
 except for several of the small ones.
As for the $B$(M1) values,
 the agreement between theory and experiment is basically good.
The exception is  $4^+(3294)\rightarrow 4^+(2538)$ transition,
 which is significantly overestimated by more than one order of magnitude
 in the present calculation.

\subsubsection{$^{55}$Co}
\label{sub2:co55}

\begin{figure}
\includegraphics[width=85mm]{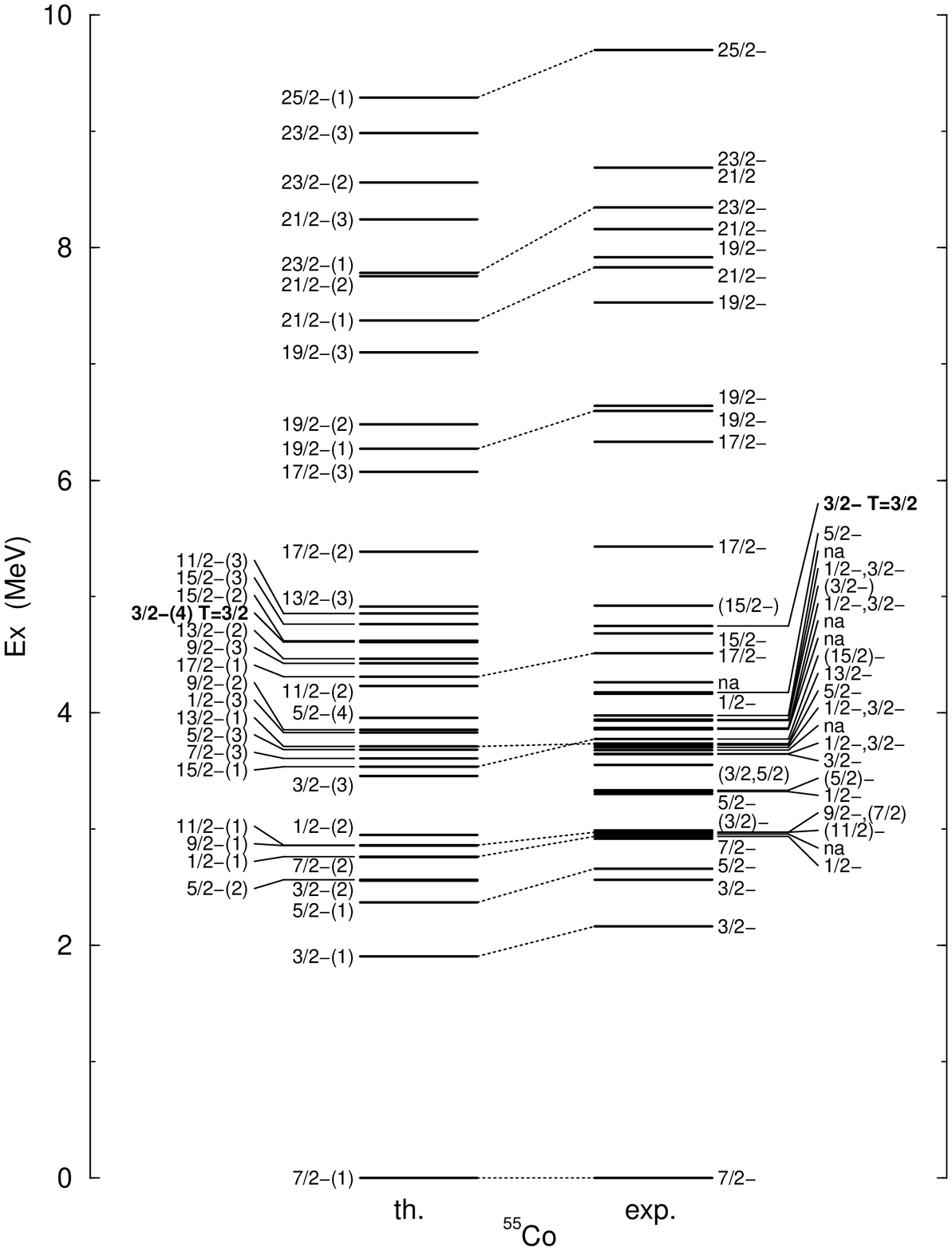}
\caption{Energy levels of $^{55}$Co.
Experimental data are taken from
 \protect\cite{nds-co55} and \protect\cite{rud-epj}.
Conventions are the same as in Fig. \ref{fig:mn53}.
\label{fig:co55}
}
\end{figure}

In Fig. \ref{fig:co55}, energy levels of $^{55}$Co are shown.
All experimental energy levels below $E_x$=4 MeV and all yrast states
 up to $J^{\pi}=23/2^-$ are compared with theoretical results.
There is a state with unknown spin-parity at $E_x$=2.960 MeV, 
 which can be interpreted tentatively as a $5/2^-$ state in the
 correspondence to the present calculation.
The overall agreement between theory and experiment
 is quite good.

If we assume an inert $^{56}$Ni core,
 this nucleus is described as one neutron hole in the $f_{7/2}$ orbit,
 corresponding to only one $7/2^-$ state.
In the calculated ground-state wave function,
 the probability of this 0p-1h configuration 
 relative to the $^{56}$Ni core
 is only 65\%,
 and there are sizable ($\sim$20\%) 2p-3h configurations.
All excited states consist of core-broken
 configurations even in the lowest order of the approximation.
Therefore
 the property of the core is expected to manifest clearly
 in the yrast spectrum.

The yrast states
 up to $J^{\pi}=17/2^-$ except for the ground state
 are described mainly by 1p-2h configurations
 (30$\sim$56\%).
Thus the truncated subspace with small $t \sim 2$
 is expected to be reasonable.
In fact the truncated $t=2$ subspace was used 
 in the shell-model calculations of Ref. \cite{rud-epj},
 where $13/2^-$ and higher spin states are described
 reasonably well.
However, the excitation energies of lower spin states
 $9/2^-$ and $11/2^-$ are predicted to be too high by $\sim$1 MeV,
 and the results of
 $5/2^-$, $3/2^-$ and $1/2^-$ are not shown in Ref. \cite{rud-epj}.

A part of such a difficulty may be attributed
 to the severe truncation.
In the present $t=7$ results, the position of both these
 higher ($J^{\pi}\geq 13/2^-$)
 and lower spin states are successfully described.
In the lower spin states, there are sizable
 3p-4h ($\sim$21\%) and 4p-5h ($\sim$9\%) components,
 including both proton and neutron excitations almost equally.
On the other hand, the higher spin states
 $13/2^-$, $15/2^-$ and $17/2^-$
 are described mainly by neutron excitations only,
 and the probabilities of 3p-4h and 4p-5h components
 are much smaller ($\sim$16\% and 6\%, respectively).
When the  GXPF1 interaction is used in the $t=2$ subspace,
 the excitation energies of lower spin states
 becomes too large by about 0.4 MeV,
 while those of higher spin states are almost unchanged.

For constructing higher spin states,
 it is necessary to excite one more nucleon
 from the $f_{7/2}$ orbit.
In fact the most dominant component in the
 wave function of these states is the 2p-3h type ($\sim$45\%).
A large gap ($\sim$2 MeV) between $17/2^-$ and $19/2^-$
 reflects this core-excitation.
Note that, although it is possible to construct a $19/2^-$ state
 within the 1p-2h configuration space, this component
 is only 9\% in the $19/2^-_1$ state,
 similarly to the result in Ref. \cite{rud-epj}.
The present calculations successfully reproduce both 
 1p-2h and 2p-3h excitations.

\begin{table}
\caption{$B$(M1) and $B$(E2) for $^{55}$Co.
Experimental data are taken from \protect\cite{nds-co55}.
\label{tbl:co55}
}
\begin{ruledtabular}
\begin{tabular}{ccccr}
initial & final & multi- & exp. & th. \ \ \\ 
$J^{\pi}(E_x)$ & $J^{\pi}(E_x)$ & pole & (W.u) & (W.u) \\ \hline
 3/2$^-$(2166)     &  7/2$^-$(0)        & E2 & $ 9.8(8)               $ & $        8.1 $ \\ 
 3/2$^-$(2566)     &  7/2$^-$(0)        & E2 & $ 1.05(25)             $ & $        1.3 $ \\ 
 1/2$^-$(2939)     &  3/2$^-$(2566)     & M1 & $ 0.9^{+7}_{-3}        $ & $     0.1130 $ \\ 
                   &                    & E2 & $                      $ & $       25.9 $ \\ 
                   &  3/2$^-$(2166)     & M1 & $ 0.27^{+20}_{-8}      $ & $     0.1371 $ \\ 
                   &                    & E2 & $ 34^{+48}_{-28}       $ & $        0.2 $ \\ 
 5/2$^-$(3303)\footnote[1]{5/2$^-_3$ in the calculation.}     &  7/2$^-$(0)        & M1 & $ 0.0089(20)           $ & $     0.0132 $ \\ 
                   &                    & E2 & $ 0.15(6)              $ & $       0.03 $ \\ 
 1/2$^-$(3323)     &  3/2$^-$(2566)     & M1 & $ 0.47(12)             $ & $     0.3100 $ \\ 
                   &                    & E2 & $                      $ & $        7.0 $ \\ 
                   &  3/2$^-$(2166)     & M1 & $ 0.17(4)              $ & $     0.2514 $ \\ 
                   &                    & E2 & $                      $ & $        9.7 $ \\ 
 5/2$^-$(3725)     &  3/2$^-$(2566)     & M1 & $ 0.092(21)            $ & $     0.0398 $ \\ 
                   &                    & E2 & $ <0.4                 $ & $        0.3 $ \\ 
                   &  7/2$^-$(0)        & M1 & $ 0.0055(11)           $ & $     0.0263 $ \\ 
                   &                    & E2 & $ 0.03(3)              $ & $       0.00 $ \\ 
 1/2$^-$(4164)     &  3/2$^-$(2566)     & M1 & $ 0.157(20)            $ & $     0.2194 $ \\ 
                   &                    & E2 & $ 9(3)                 $ & $        3.5 $ \\ 
\end{tabular}
\end{ruledtabular}
\end{table}

The structure of non-yrast states is more complicated.
Since the present interest is the low-lying 
 core-excited states, we can focus on 
 $1/2^-_2$, $3/2^-_3$, $7/2^-_3$ and $5/2^-_3$.
These states are all below $E_x$=4 MeV and contain more than 45\%  
 neutron 2p-2h components.
A similar character can also be seen in $3/2^-_2$, although
 the probability of the neutron 2p-2h component is slightly smaller (37\%).
These states are connected by relatively large E2
 transition matrix elements such as
 $B$(E2; $7/2^-_3\rightarrow 3/2^-_2$)=11 W.u. and
 $B$(E2; $5/2^-_3\rightarrow 1/2^-_2$)=18 W.u.
It will be a good test if it is possible to
 observe such transitions.

The calculated $3/2^-_4$ has isospin $T=3/2$
 and is the isobaric analog state of the $^{55}$Fe ground state.
The correct position of this state, as seen in comparison to experiment 
in Fig. \ref{fig:co55}, implies the
 proper isospin structure of the present interaction.

In Table \ref{tbl:co55}, electro-magnetic transition 
 matrix elements are shown for $^{55}$Co.
We can find reasonable agreement between
 theory and experiment.
In general, the deviations are large where
 the experimental errors are also large.

\subsubsection{$^{56}$Ni}
\label{sub2:ni56}

\begin{figure}
\includegraphics[width=85mm]{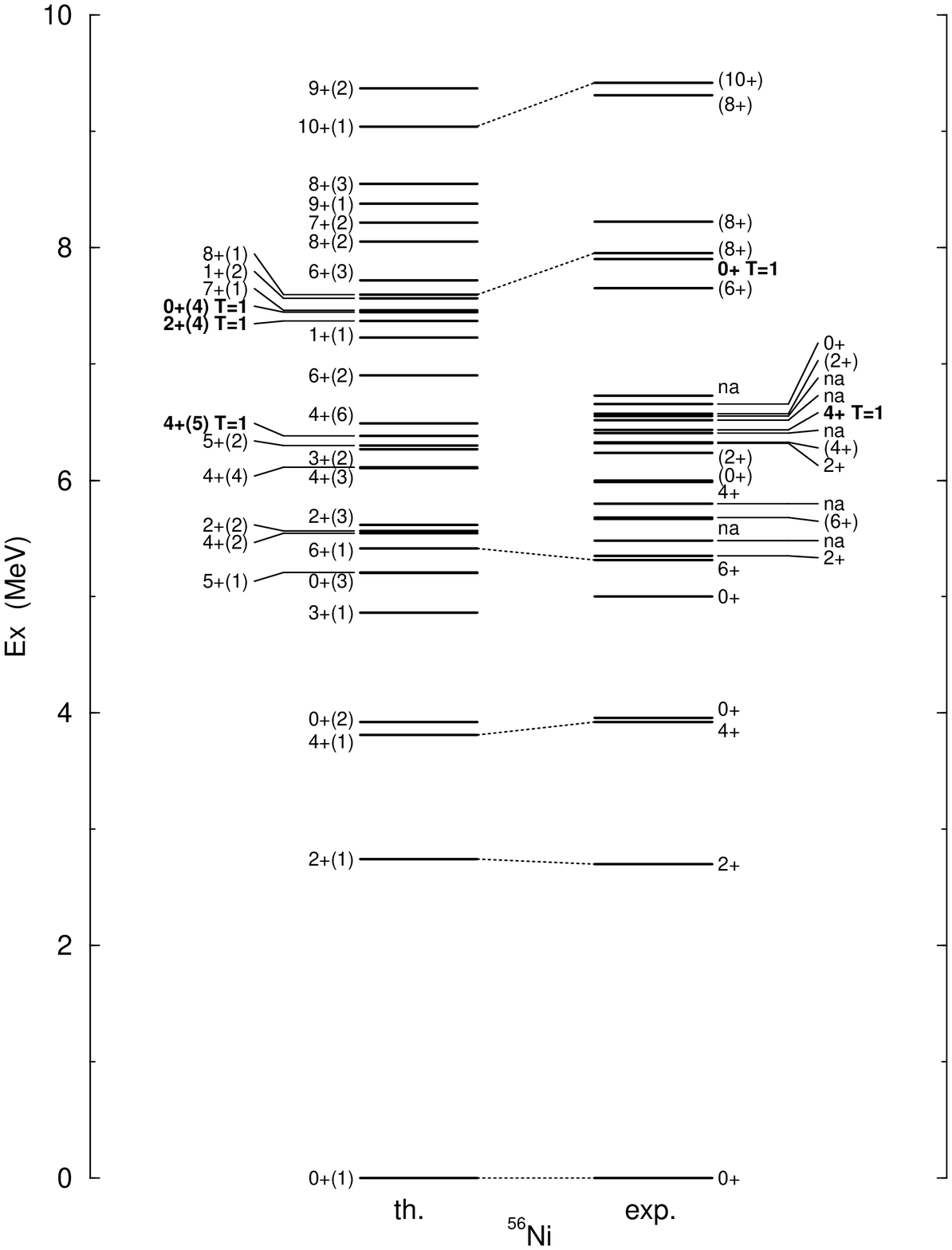}
\caption{Energy levels of $^{56}$Ni.
Experimental data are taken from
 \protect\cite{nds-ni56} and \protect{\cite{rud-epj}}.
Conventions are the same as in Fig. \ref{fig:mn53}.
\label{fig:ni56}
}
\end{figure}

In Fig. \ref{fig:ni56}, calculated energy levels of $^{56}$Ni
 are compared with experimental data.
All calculated states up to $E_x$=6 MeV as well as the yrast states
 are shown.
The shell-model results were obtained in the $t=7$ subspace.
The agreement between the experiment and the calculation is
 basically good especially for even spin yrast states.
However, the yrast $3^+$ and $5^+$ states are not found
 in the experimental data, while no candidate for the experimental
 $(2^+)$, $(4^+)$ and $(6^+)$ can be seen
 in the calculation at least around $E_x\sim$6 MeV.
More precise experimental information is needed
 to discuss such discrepancies.
Similar results are shown in Fig. 2 of Ref. \cite{upf}
 for even spin states,
 which were obtained by the MCSM calculations,
 taking typically 13 bases per one eigenstate.
We thus confirm the reliability of the
 MCSM results for non-yrast states. 

For the yrast states,
 the ground state consists mainly of the
 $(f_{7/2})^{16}$ closed-shell configuration (68\%).
The excited states
 $2^+$, $3^+$, $4^+$, $5^+$ and $6^+$ are of the 1p-1h character
 (42$\sim$50\%)
 relative to the closed-shell configuration,
 and the excitation energy of $2^+_1$
 reflects the shell gap at $N$ or $Z=28$.
The 2p-2h configurations are dominant in
 $1^+$, $7^+$, $8^+$, $9^+$ and $10^+$ states (35$\sim$51\%).
One can find again an energy gap between these 1p-1h and 2p-2h groups.
The 3p-3h dominant states appear above $E_x\sim$10 MeV.
In the same way, most of the non-yrast excited states shown in the figure
 can be classified as either 1p-1h ($3^+_2$, $5^+_2$, $\cdots$)
 or 2p-2h ($2^+_2$, $6^+_2$, $\cdots$) dominant states.
These $n$p-$n$h states contain
 typically 20\% of $(n+2)$p-$(n+2)$h configurations.
Therefore, the 4p-4h configuration space is
 essential (but not enough) to describe these states.
 
\begin{figure}
\includegraphics[width=85mm]{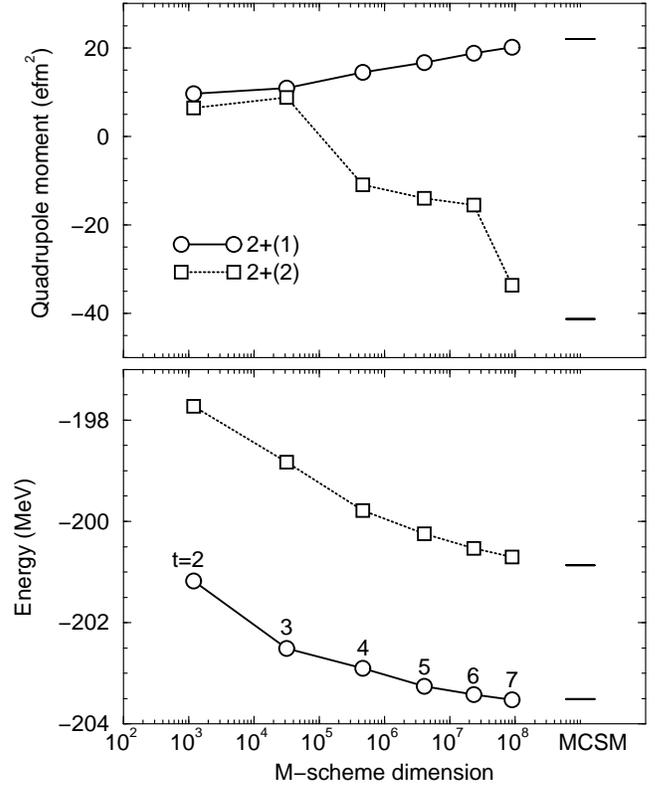}
\caption{Convergence of the
 calculated energies (lower panel) and quadrupole moments (upper panel)
 as functions of the truncation order $t$ for
 the lowest two 2$^+$ states of $^{56}$Ni.
Horizontal lines show the corresponding MCSM results.
\label{fig:ni56conv}
}
\end{figure}

As has already been discussed in Ref. \cite{upf}, GXPF1
 predicts the deformed 4p-4h band \cite{rudolph} as
 0$^+_3$, $2^+_2$, $\cdots$.
In the present results, $0^+_3$ consists mainly of
 4p-4h (48\%), 5p-5h (23\%) and 6p-6h (12\%) configurations.
The structure of $2^+_2$ is quite similar.
Special care should be taken concerning the convergence of
 the calculation for such a deformed 4p-4h band, since
 the present truncation scheme is obviously not suitable
 for describing strongly deformed states,
 as discussed in Ref. \cite{ni56super} for the case of
 the FPD6 interaction.
Figure \ref{fig:ni56conv} shows the calculated
 energies and quadrupole moments
 of the lowest two $2^+$ states
 as a function of the truncation order $t$.
It can be seen that the
 4p-4h state $2^+_2$ is not completely converged
 even in the $t=7$ subspace,
 which corresponds to the $M$-scheme dimension of 89285264.
In the same figure, the MCSM results are also shown,
 where 25 bases were searched for each eigenstate.
The MCSM energy of the yrast 2$^+$ state
 is almost the same as that of the $t=7$ calculation,
 which shows reasonable convergence.
On the other hand, for $2^+_2$ states,
 the MCSM energy is lower than the $t=7$ result by about 0.2 MeV,
 and should be more accurate.
The electric quadrupole moment is $Q(2^+_2)=-41.2$efm$^2$
 in the MCSM calculation, while it is $-33.6$
 in the present $t=7$ result.
Similar results are obtained for the 4p-4h 0$^+_3$ state.
The $M$-scheme dimension of $t=8$ subspace
 reaches to 255478309, which is beyond our computational
limitations.
For the description of such deformed states where a lot of mixing 
among various spherical configurations occurs,
 the MCSM calculation is more efficient and suitable, because
 certain basis states can be sampled from and around local minima.

In the present results, we can find several $T=1$ states
 such as $0^+_4$, $2^+_4$ and $4^+_5$.
These states are interpreted as the isobaric analog states of $^{56}$Co.
The excited $0^+$ with $T=1$ is in fact
 found experimentally at excitation energy of 7.904 MeV,
 which is consistent with the present result.

Experimental data for the electro-magnetic transitions are rather limited.
We can only compare with the few transition strengths
shown in
 Table \ref{tbl:ni56}.

\begin{table}
\caption{$B$(M1) and $B$(E2) for $^{56}$Ni.\label{tbl:ni56}
Experimental data are taken from
 \protect{\cite{nds-ni56}} and \protect{\cite{kraus}}.
}
\begin{ruledtabular}
\begin{tabular}{ccccr}
initial & final & multi- & exp. & th. \ \ \\ 
$J^{\pi}(E_x)$ & $J^{\pi}(E_x)$ & pole & (W.u) & (W.u) \\ \hline
   2$^+$(2701)     &    0$^+$(0)        & E2 & $ 9.4(1.9)                 $ & $       11.1 $ \\ 
   4$^+$(3924)     &    2$^+$(2701)     & E2 & $ <24                  $ & $        8.4 $ \\ 
   6$^+$(5317)     &    4$^+$(3924)     & E2 & $                      $ & $        5.4 $ \\ 
   2$^+$(5351)     &    2$^+$(2701)     & E2 & $                      $ & $        4.1 $ \\ 
                   &    0$^+$(0)        & E2 & $                      $ & $        0.1 $ \\ 
   8$^+$(7956)     &    6$^+$(5317)     & E2 & $                      $ & $        0.1 $ \\ 
  10$^+$(9419)     &    8$^+$(7956)     & E2 & $                      $ & $        0.1 $ \\ 
\end{tabular}
\end{ruledtabular}
\end{table}

\subsubsection{$^{57}$Ni}
\label{sub2:ni57}

\begin{figure}
\includegraphics[width=85mm]{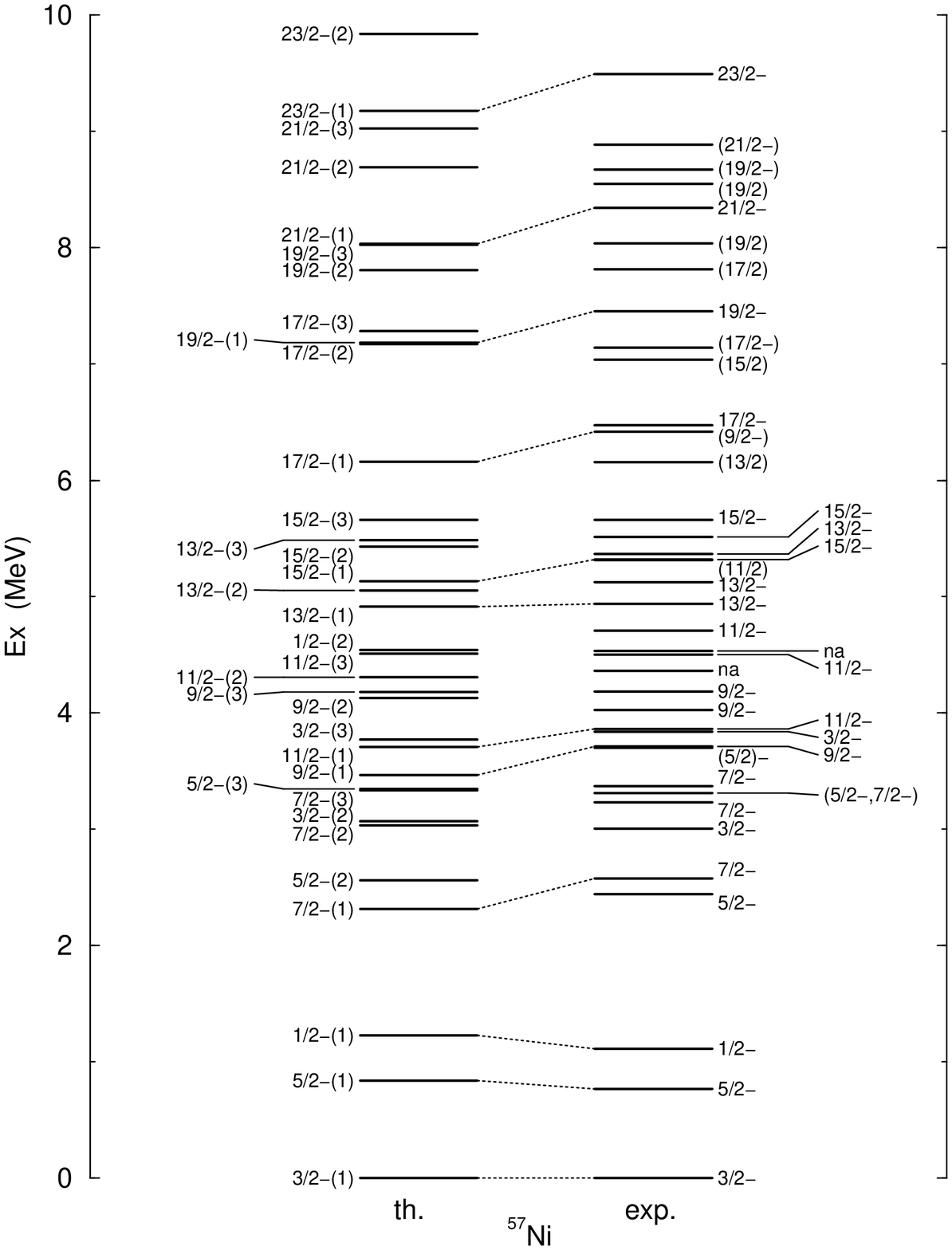}
\caption{Energy levels of $^{57}$Ni.
Experimental data are taken from
 \protect\cite{nds-ni57} and \protect\cite{rud-epj}.
Conventions are the same as in Fig. \ref{fig:mn53}.
\label{fig:ni57}
}
\end{figure}

Energy levels of $^{57}$Ni are shown in Fig. \ref{fig:ni57}.
All experimental yrast levels as well as 
 non-yrast states up to 4 MeV excitation energy are
 compared with the theoretical predictions.
The shell-model calculations were carried out
 in the $t=6$ subspace.
The agreement between experiment and theory 
 is basically good up to high spin states $23/2^-$,
 although there are several inversion of the order
 of two closely lying states, such as 
 $7/2^-_1$ and $5/2^-_2$.

Comparing the present results with the previous MCSM ones
 in Ref. \cite{upf},
 we can again confirm the reliability of the MCSM calculations
 for non-yrast states as well as the yrast states
 of odd-mass nuclei.
The root mean square difference of the excitation energies
 between the present
 and the previous MCSM results is 154 keV for 19 states
 shown in Fig. 2 of Ref. \cite{upf}.

If we assume an inert $^{56}$Ni core, $^{57}$Ni is 
 described by only one neutron.
Then, three states
 $3/2^-$, $5/2^-$ and $1/2^-$ are possible,
 corresponding to the occupation of $p_{3/2}$, $f_{5/2}$,
 and $p_{1/2}$ orbits by one neutron.
According to the present calculations,
 the lowest three states show such a ``single particle'' character,
 although there is a sizable mixing of the broken-core configurations.
In fact, the pure ``single particle'' configurations
 are 62, 58 and 46\%, respectively,
 in the calculated wave functions.
Since the closed core configuration in the
 ground state of $^{56}$Ni is 69\% \cite{upf},
 the core is further broken in the ground state 
 of $^{57}$Ni, as can also be seen in Fig. \ref{fig:prob}.

The bare single particle energy
 of the $f_{5/2}$ orbit relative to the $p_{3/2}$ orbit is
found to be 4.3 MeV in GXPF1.
 (see Table \ref{tbl:gxpf1}).
The effective single particle energy (ESPE) \cite{kb3,mcsm-n20}
 of the $f_{5/2}$ orbit
 relative to the $p_{3/2}$ orbit
 decreases rapidly as the proton $f_{7/2}$ orbit is occupied,
 and reaches 1.1 MeV at $^{57}$Ni.
The mechanism for this behavior of the $f_{5/2}$ orbit
 is explained by the strong attraction
 between the protons in the $f_{7/2}$ orbit and 
 neutrons in the $f_{5/2}$ orbit
 due to a strong $j_>$ - $j_<$ coupling
term in the nuclear force, as discussed in Ref. \cite{magic}.
Thus the proper contribution of the two-body interaction,
 especially the monopole part, is needed to reproduce
 the experimental order of the lowest three states in $^{57}$Ni.
The FPD6 interaction is too strong in this respect,
 giving an ESPE for the  
$f_{5/2}$ orbit which is too low for the Ni isotopes.
Note that, the ESPE's
 of the $f_{5/2}$ and the $p_{1/2}$ orbit are still
 too large by 0.3 and 0.9 MeV, respectively,
 in comparison to the experimental 
 excitation energies of $5/2^-$ and $1/2^-$ states,
since they are defined in terms of a closed-shell structure 
for $^{56}$Ni.
Further energy gain by the mixing of
 the broken-core components is needed for the better reproduction
 of the experimental data.

There is an energy gap ($\sim$ 1.3 MeV)
 above the lowest three states, which 
 corresponds to the 2p-1h core-excitation.
The doublets $5/2^-_2$ and $7/2^-_1$ consist of
 two dominant configurations,
 $\pi(f_{7/2})^7(p_{3/2})^1\nu(f_{f7/2})^8(p_{3/2})^1$ and
 $\pi(f_{7/2})^8\nu(f_{f7/2})^7(p_{3/2})^2$,
 sharing almost equal probabilities ($\sim$20\%).
Above these doublets, most of the states up to $E_x\sim$5 MeV
 are also of the 2p-1h nature,
 although one can no longer find single dominant configuration
 which exhausts more than 20\% probability.
In this 2p-1h regime, as exceptional cases, 
the $3/2^-_3$ and $1/2^-_2$ states 
are comprised mainly of higher configurations:
 the 3p-2h configuration
 takes the largest weight (25 and 29\%, respectively)
 and even the 5p-4h configuration can be seen
 with a non-negligible probability ($\sim$5\%).
In the yrast $15/2^-$ and most of the higher spin states,
 the 3p-2h configuration becomes dominant,
 carrying more than 40\% probability.
We have not identified 4p-3h or
 more significantly core-broken states
 such as those of  the deformed 4p-4h band in the $^{56}$Ni
 in the present results for $^{57}$Ni.

Electro-magnetic transition 
strengths are shown in Table \ref{tbl:ni57}.
Because of the large ambiguities in the experimental data,
 it is difficult to draw a definite conclusion.
Most of the calculated M1 transitions are consistent
 with the experimental data.
A notable deviation
 is found in the $B$(M1; $5/2^-_1\rightarrow 3/2^-_1$),
 where the calculated value is too small by a factor of 8.
As for the E2 transitions, it can be seen that
 the calculated $B$(E2) values
 are in general smaller than the experimental data.
The exception is the transition from $5/2^-$(2443).
However, no experimental error is indicated in the data and 
 only the upper bound is given for the associated M1 transition.
The calculated $B$(E2) value does not 
 contradict with the experimental lifetime data.

\begin{table}
\caption{$B$(M1) and $B$(E2) for $^{57}$Ni.
Experimental data are taken from
 \protect\cite{nds-ni57}.
\label{tbl:ni57}
}
\begin{ruledtabular}
\begin{tabular}{ccccr}
initial & final & multi- & exp. & th. \ \ \\ 
$J^{\pi}(E_x)$ & $J^{\pi}(E_x)$ & pole & (W.u) & (W.u) \\ \hline
 5/2$^-$(769)      &  3/2$^-$(0)        & M1 & $ 0.0144(18)           $ & $     0.0018 $ \\ 
                   &                    & E2 & $ 2.5(6)               $ & $        1.4 $ \\ 
 1/2$^-$(1113)     &  3/2$^-$(0)        & M1 & $ \le\,0.19            $ & $     0.1317 $ \\ 
                   &                    & E2 & $ \le\,3.0E+2          $ & $        7.3 $ \\ 
 5/2$^-$(2443)     &  3/2$^-$(0)        & M1 & $ <0.024               $ & $     0.0003 $ \\ 
                   &                    & E2 & $ 0.049                $ & $       10.7 $ \\ 
 7/2$^-$(2577)     &  3/2$^-$(0)        & E2 & $ 7.7(11)              $ & $        7.4 $ \\ 
 3/2$^-$(3007)     &  3/2$^-$(0)        & M1 & $ <0.13                $ & $     0.0229 $ \\ 
                   &                    & E2 & $ <27                  $ & $        7.2 $ \\ 
11/2$^-$(3866)     &  7/2$^-$(2577)     & E2 & $ 42^{+24}_{-10}           $ & $        6.8 $ \\ 
15/2$^-$(5321)\footnote[1]{15/2$^-_2$ in the calculation.}      & 11/2$^-$(3866)     & E2 & $ 10^{+4}_{-2}             $ & $        4.5 $ \\ 
\end{tabular}
\end{ruledtabular}
\end{table}

\subsubsection{$^{58}$Ni}
\label{sub2:ni58}

\begin{figure}
\includegraphics[width=85mm]{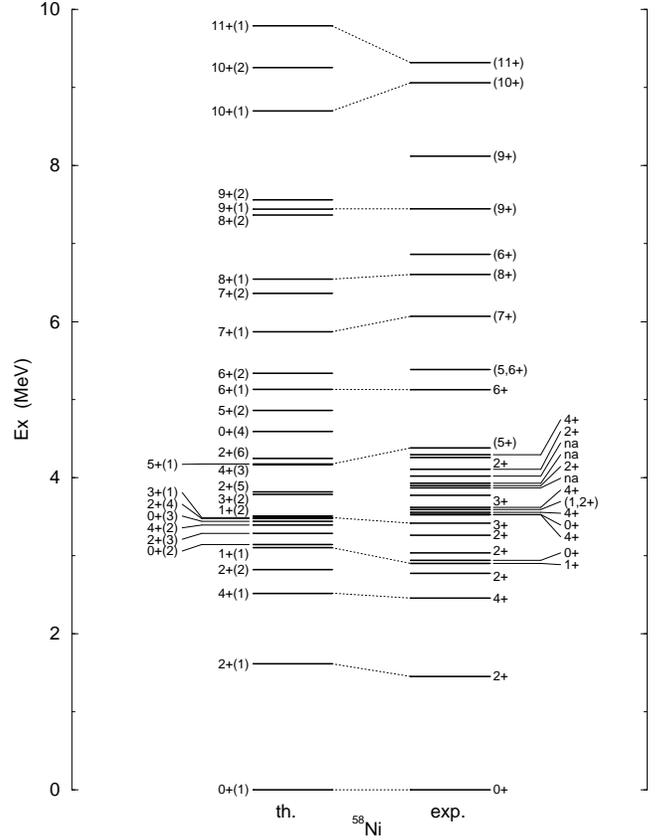}
\caption{Energy levels of $^{58}$Ni.
Experimental data are taken from
 \protect\cite{nds-ni58}.
Conventions are the same as in Fig. \ref{fig:mn53}.
\label{fig:ni58}
}
\end{figure}

Figure \ref{fig:ni58} shows calculated and
 experimental energy levels of $^{58}$Ni.
The yrast states up to 10 MeV excitation energy,
 non-yrast states below $E_x$=4 MeV,
 and several additional states
 are shown 
 with their theoretical counterparts.
The agreement between the experiment and the calculation is satisfactory.
The calculations were carried out in the $t$=6 
 truncated subspace.
The results are basically consistent with our previous
 MCSM calculations in Ref. \cite{upf},
 although the latter spectrum was slightly expanded
 for higher spin states
 due to the small number of basis states ($\sim$13 per one eigenstate).

In the calculated ground-state wave function, the total probability of
 the closed core configurations
 $\pi(f_{7/2})^{8}\nu(f_{7/2})^8(r)^2$
 is only 57\%,
 which is smaller than those in $^{56,57}$Ni.
As shown in Fig. \ref{fig:prob},
 this quantity decreases further for larger $N$
 and takes a minimum value at $N=34$ (52\%).
Thus the $^{56}$Ni core is soft in Ni isotopes.
Nevertheless, for understanding the basic structure,
 it is still useful to 
 assume an inert $^{56}$Ni core and to
 consider two neutrons
 in the $p_{3/2}$, $f_{5/2}$ and $p_{1/2}$ orbits.
Then, the allowed maximum spin is $J=4$,
 and core-excitation is needed to generate higher spin states.
In the calculated wave function, the leading configuration is
 $\nu(p_{3/2})^2$ for $0^+_1$ and $2^+_1$
 (37\% and 30\%, respectively),
 and $\nu(p_{3/2})^1(f_{5/2})^1$ for $1^+_1$ and $3^+_1$
 (42\% and 52\%, respectively).
There is a large energy gap ($\sim$ 2 MeV)
 in the yrast spectrum
 between $4^+_1$ and $5^+_1$, reflecting
 the core-excitation.
One-proton core-excitation 
 dominates the wave functions up to $J=9$,
 although there is no single configuration
 which exhausts more than 20\% weight in the calculated wave functions.
For $J=10$ and higher spin yrast states
one-proton one-neutron core-excitation
 becomes the most important mode.
Consistently,
 there is another energy gap ($\sim$ 1.6 MeV) above $9^+_1$
 in both calculated and experimental energy spectra.

The closed-core configurations
also play a major role in many non-yrast states.
Among the non-yrast states below $E_x\sim$ 4 MeV,
 the 3p-1h configurations associated with
 the one-proton core-excitation become dominant in
 $2^+_3$, $4^+_2$ and $3^+_2$.
Since the level density increases significantly
 above $E_x\sim$ 4 MeV,
 it is difficult to make a simple interpretation for
the properties of the core-excitations
 in higher states.

As discussed in Ref. \cite{upf}, 
 the core-excitation of two-protons is dominant
 in the wave function of $0^+_3$.
In the present calculation, we found that 
 $2^+_6$ shows a similar 4p-2h character.
The probability of such a 4p-2h configuration
 $\pi(f_{7/2})^6(r)^2\nu(f_{7/2})^8(r)^2$ is
 46 and 41\% for $0^+_3$ and $2^+_6$, respectively.
The calculated quadrupole moment is $Q(2^+_6)=-0.30$ eb,
 and $B$(E2; $2^+_6\rightarrow 0^+_3$)=18 W.u.
These values are consistent with those of the rotational model,
 corresponding to the $K=0$ band on top of
 the axially symmetric prolate rotor
 with an intrinsic quadrupole moment $Q_0$=110 eb.
The deformation of the present 4p-2h band
 appears to be slightly smaller than 
 the 4p-4h band in $^{56}$Ni \cite{upf}, which is consistent
 with $Q_0$=140 eb obtained by using the same effective charges.

\begin{table}
\caption{$B$(M1) and $B$(E2) for $^{58}$Ni.
Experimental data are taken from \protect\cite{nds-ni58}.
\label{tbl:ni58}
}
\begin{ruledtabular}
\begin{tabular}{ccccr}
initial & final & multi- & exp. & th. \ \ \\ 
$J^{\pi}(E_x)$ & $J^{\pi}(E_x)$ & pole & (W.u) & (W.u) \\ \hline
   2$^+$(1454)     &    0$^+$(0)        & E2 & $ 9.8(5)               $ & $        7.7 $ \\ 
   4$^+$(2459)     &    2$^+$(1454)     & E2 & $ <43                  $ & $        4.1 $ \\ 
   2$^+$(2775)     &    2$^+$(1454)     & M1 & $ 0.011(4)             $ & $     0.1103 $ \\ 
                   &                    & E2 & $ 15(5)                $ & $        1.2 $ \\ 
                   &    0$^+$(0)        & E2 & $ 0.029(10)            $ & $        2.8 $ \\ 
   0$^+$(2942)     &    1$^+$(2902)     & M1 & $ 0.084(6)             $ & $     0.0068 $ \\ 
                   &    2$^+$(2775)     & E2 & $ 14.9(18)             $ & $       12.4 $ \\ 
                   &    2$^+$(1454)     & E2 & $ 0.00029(4)           $ & $        0.63 $ \\ 
   2$^+$(3038)     &    2$^+$(2775)     & M1 & $ 0.23(6)              $ & $     0.0275 $ \\ 
                   &                    & E2 & $ 6^{+20}_{-6}             $ & $        1.0 $ \\ 
                   &    2$^+$(1454)     & M1 & $ 0.060(12)            $ & $     0.0190 $ \\ 
                   &                    & E2 & $ 2.0(7)               $ & $       15.0 $ \\ 
   2$^+$(3263)     &    2$^+$(1454)     & M1 & $ 0.028(11)            $ & $     0.0004 $ \\ 
                   &                    & E2 & $ 8(7)                 $ & $        0.01 $ \\ 
   3$^+$(3420)     &    4$^+$(2459)     & M1 & $ 0.09(8)              $ & $     0.1101 $ \\ 
                   &                    & E2 & $ 0.07^{+23}_{-7}          $ & $        0.11 $ \\ 
   0$^+$(3531)     &    2$^+$(1454)     & E2 & $ 5.5                  $ & $        3.5 $ \\ 
   4$^+$(3620)     &    4$^+$(2459)     & M1 & $ 0.07^{+7}_{-4}           $ & $     0.0002 $ \\ 
                   &                    & E2 & $ 44^{+60}_{-36}           $ & $        1.3 $ \\ 
                   &    2$^+$(1454)     & E2 & $ 1.4^{+12}_{-6}           $ & $        6.7 $ \\ 
   3$^+$(3774)     &    3$^+$(3420)     & M1 & $ 0.33(17)             $ & $     0.0169 $ \\ 
                   &                    & E2 & $ 1.$E$+1^{+11}_{-1}         $ & $        0.1 $ \\ 
                   &    4$^+$(2459)     & M1 & $ 0.019(10)            $ & $     0.0001 $ \\ 
                   &                    & E2 & $ 0.8^{+12}_{-8}           $ & $        0.4 $ \\ 
   2$^+$(3898)     &    2$^+$(1454)     & M1 & $ 0.042^{+25}_{-11}        $ & $     0.0410 $ \\ 
                   &    0$^+$(0)        & E2 & $ 0.42^{+25}_{-11}         $ & $        0.81 $ \\ 
   2$^+$(4108)     &    2$^+$(2775)     & M1 & $ 0.0051(20)           $ & $     0.0089 $ \\ 
                   &    2$^+$(1454)     & M1 & $ 0.0032(8)            $ & $     0.0003 $ \\ 
                   &                    & E2 & $ 0.30(10)             $ & $        0.29 $ \\ 
                   &    0$^+$(0)        & E2 & $ 0.15(4)              $ & $        0.08 $ \\ 
\end{tabular}
\end{ruledtabular}
\end{table}

In Table \ref{tbl:ni58}, calculated transition strengths are
 compared with experimental data.
In general, the calculated M1 matrix elements are too small.
The exception is $B$(M1; $2^+_2\rightarrow 2^+_1$), which is
 too large by one order of magnitude.
As for the E2 matrix elements,
 the calculated values are almost consistent with experimental data
 in the case of $\Delta J=2$ transitions, i.e.,
 no mixing of M1 transitions.
One notable exception is 
 $B$(E2; $0^+_2\rightarrow 2^+_1$), which is too large
 by more than 3 orders of magnitude, but this is 
an extremely weak transition.

\subsubsection{$^{59}$Ni}
\label{sub2:ni59}

\begin{figure}
\includegraphics[width=85mm]{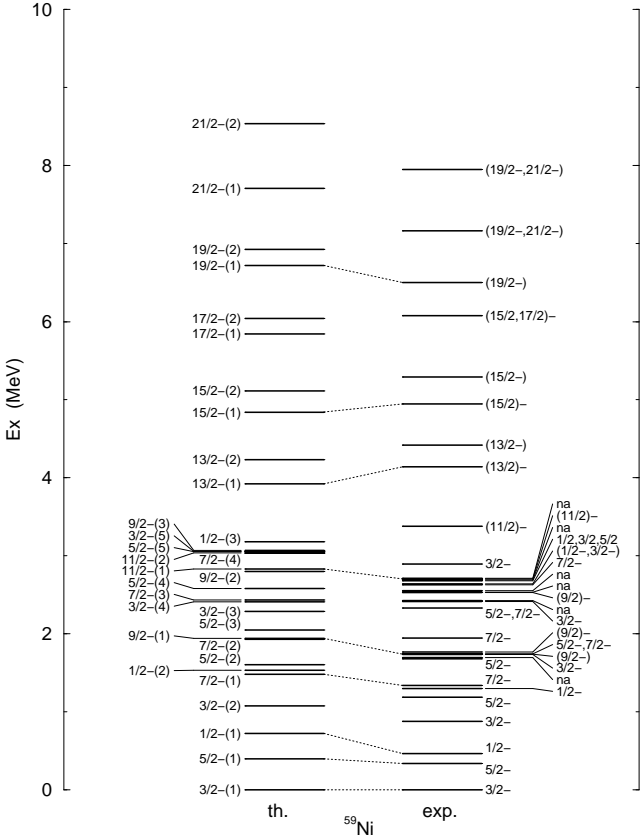}
\caption{Energy levels of $^{59}$Ni.
Experimental data are taken from
 \protect\cite{nds-ni59}.
Conventions are the same as in Fig. \ref{fig:mn53}.
\label{fig:ni59}
}
\end{figure}

Energy levels of $^{59}$Ni are shown in Fig. \ref{fig:ni59}.
All experimental data for yrast states as well as
 non-yrast states below 3 MeV excitation energy are 
 compared with the shell-model results obtained in the $t$=6 subspace.
It is found that the theoretical results 
 agree quite well with the experimental data.
The results of the lowest triplet $3/2^-$, $5/2^-$ and $1/2^-$
 have already been reported in Ref. \cite{upf}.
The present calculation has confirmed the
 validity of GXPF1 interaction also for higher excited states.

Assuming an inert $^{56}$Ni core, this nucleus is described
 by three valence neutrons in the upper three
 orbits $p_{3/2}$, $f_{5/2}$ and $p_{1/2}$.
Therefore various states are possible without any core-excitation.
In fact such closed-shell (3p-0h) configurations carry the maximum weight
 (31\%$\sim$56\%)
 in most of the calculated low-lying states below $E_x$= 3 MeV.
There are three exceptions, $7/2^-_2$, $7/2^-_3$ and $11/2^-_1$,
 where 4p-1h configurations take the largest weight.
Such 4p-1h configurations also dominate the wave functions of
 higher-lying excited states shown in the same figure.
The 5p-2h configurations appears to be the leading configurations
 in the yrast $19/2^-$ and $21/2^-$ states.

Like other nuclei, we would like to identify states with
large components of 5p-2h or higher configurations at
$E_x < \sim 3$ MeV.  This energy
boundary is set because the level density increases
very fast above it in $^{59}$Ni.  However, as seen from
the above discussions, there are no such candidate states
theoretically or experimentally.  
Therefore, the study of
5p-2h and higher states with lower angular momenta is not possible in
$^{59}$Ni. 
Thus, the discussion of the multi-particle multi-hole states which
is one of the main elements in our assessment of the effective
interaction is confined mainly
to a few nuclei around $^{56}$Ni.

Table \ref{tbl:ni59} shows the measured and calculated 
 $B$(E2) and $B$(M1).
Even below $E_x$=2 MeV, there are several uncertain levels
 in the experimental data at 1695 keV (no spin assignment),
 1739 keV (assigned to ($9/2^-$)),
 and 1746 keV (assigned to $5/2^-$,$7/2^-$).
In the present calculations, we have tentatively assigned
 no theoretical states to these three uncertain states,
 since no probable candidates appear around $E_x\sim$ 2 MeV.
More experimental information is needed for a detailed
 comparison between theory and experiment.

In general, the order of magnitude is well reproduced for
 most transitions.
More specifically, the calculated M1 transition matrix elements 
 are small in comparison to the experimental data in many cases,
 except for $1/2^-_1\rightarrow 3/2^-_1$ transition.
The calculated E2 matrix elements are also
 smaller than the experimental data.
However, one can clearly see the correlation 
 between the calculated values and the corresponding experimental data.
The agreement could be  improved if we introduced
 larger effective charges.

\begin{table}
\caption{$B$(M1) and $B$(E2) for $^{59}$Ni.
Experimental data are taken from \protect\cite{nds-ni59}.
\label{tbl:ni59}
}
\begin{ruledtabular}
\begin{tabular}{ccccr}
initial & final & multi- & exp. & th. \ \ \\ 
$J^{\pi}(E_x)$ & $J^{\pi}(E_x)$ & pole & (W.u) & (W.u) \\ \hline
 5/2$^-$(339)      &  3/2$^-$(0)        & M1 & $ 0.0083(10)           $ & $     0.0031 $ \\ 
                   &                    & E2 & $                      $ & $       0.26 $ \\ 
 1/2$^-$(465)      &  3/2$^-$(0)        & M1 & $ 0.011(2)             $ & $     0.0705 $ \\ 
                   &                    & E2 & $                      $ & $        1.7 $ \\ 
 3/2$^-$(878)      &  3/2$^-$(0)        & M1 & $ 0.068(13)            $ & $     0.0327 $ \\ 
                   &                    & E2 & $ 1.1(9)               $ & $        7.0 $ \\ 
 5/2$^-$(1189)     &  3/2$^-$(0)        & M1 & $ 0.043(9)             $ & $     0.0058 $ \\ 
                   &                    & E2 & $ 11(5)                $ & $       10.1 $ \\ 
 1/2$^-$(1301)     &  1/2$^-$(465)      & M1 & $ 0.036(8)             $ & $     0.0016 $ \\ 
                   &  5/2$^-$(339)      & E2 & $ 1.2(4)               $ & $        0.4 $ \\ 
 7/2$^-$(1338)     &  5/2$^-$(339)      & M1 & $ 0.0008(3)            $ & $     0.0002 $ \\ 
                   &                    & E2 & $ 30(7)                $ & $       10.7 $ \\ 
                   &  3/2$^-$(0)        & E2 & $ 2.9(7)               $ & $        0.8 $ \\ 
 5/2$^-$(1680)     &  1/2$^-$(465)      & E2 & $ 1.41(23)             $ & $       0.20 $ \\ 
                   &  3/2$^-$(0)        & M1 & $ 0.0009(7)            $ & $     0.0095 $ \\ 
                   &                    & E2 & $ 1.6(5)               $ & $        1.1 $ \\ 
 9/2$^-$(1768)     &  7/2$^-$(1338)     & M1 & $ 0.042(11)            $ & $     0.0210 $ \\ 
                   &                    & E2 & $ 4^{+5}_{-3}              $ & $        1.8 $ \\ 
                   &  5/2$^-$(339)      & E2 & $ 11(3)                $ & $        3.3 $ \\ 
 7/2$^-$(1948)     &  5/2$^-$(1189)     & M1 & $ 0.042(9)             $ & $     0.0368 $ \\ 
                   &                    & E2 & $ 10(10)               $ & $        0.9 $ \\ 
                   &  3/2$^-$(878)      & E2 & $ 3.6(10)              $ & $        2.7 $ \\ 
                   &  3/2$^-$(0)        & E2 & $ 5.3(10)              $ & $        7.2 $ \\ 
11/2$^-$(2705)     &  9/2$^-$(1768)     & M1 & $ 0.0040(13)           $ & $     0.0005 $ \\ 
                   &                    & E2 & $ 8(3)                 $ & $        1.7 $ \\ 
                   &  7/2$^-$(1338)     & E2 & $ 22(6)                $ & $        4.5 $ \\ 
\end{tabular}
\end{ruledtabular}
\end{table}

\subsection{$N$=$Z$ odd-odd nuclei}
\label{sub:neqz}

In the $pf$ shell, $N=Z$ odd-odd nuclei
 have been of special interest, where
 the lowest isospin $T$=0 and $T=1$ states are almost degenerate
 near the ground state.
In most cases $T=1$, $J=0$ is the ground state.
The only known exception is $^{58}$Cu, in which
 the ground state is $T=0$, $J=1$.
This reflects a detailed interplay of
 $T=0$ and $T=1$ interactions, and it is important to evaluate the
 effective interaction from this viewpoint.

\subsubsection{$^{54}$Co}
\label{sub2:co54}

\begin{figure}
\includegraphics[width=85mm]{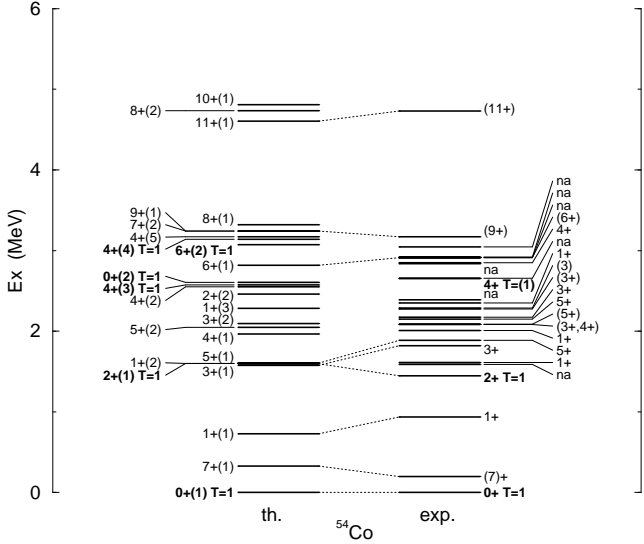}
\caption{Energy levels of $^{54}$Co.
Experimental data are taken from
 \protect\cite{nds-co54} and \protect\cite{rud-epj}.
Conventions are the same as in Fig. \ref{fig:mn53}.
\label{fig:co54}
}
\end{figure}

In Fig. \ref{fig:co54}, calculated energy levels of $^{54}$Co
 are compared with experimental data.
The spin-parity has not been assigned to most of the states
 above $E_x=$ 3 MeV, and there are several uncertain levels
 among low-lying states.
It can be seen that the present shell-model calculations
 in the $t=6$ subspace
 give a  reasonable description, although there are several small
 deviations such as the interchange
 of the order of $2^+_1$ and $3^+_1$.
The calculated $0^+_{1,2}$, $2^+_1$, $4^+_{3,4}$ and $6^+_2$ are $T=1$,
 corresponding to the isobaric analog states of $^{54}$Fe 
 (see Fig. \ref{fig:fe54}),
 while other calculated states shown in the figure are $T=0$.

In the calculated wave functions of the
 yrast odd-spin states $1^+_1$, $3^+_1$, $5^+_1$ and $7^+_1$,
 the lowest $\pi(f_{7/2})^7\nu(f_{7/2})^7$ configuration is dominant
 (59, 58, 39 and 62\%, respectively), but
 core-excited components are not small.
Such lowest 0p-2h configurations relative to $^{56}$Ni core
 are also dominant in even-spin $0^+_1$, $2^+_1$,
 $4^+_3$, and $6^+_2$, which are all $T=1$.

In the shell-model calculations \cite{co54}
 using the Surface Delta interaction in the
 ($f_{7/2}$, $p_{3/2}$) space with
 a restriction that the $p_{3/2}$ orbit can be occupied
 by one proton and one neutron,
 the $4^+$ and $6^+$ states of $T=1$
 appear as yrast states.
On the other hand, $4^+_{1,2}$ are isospin $T=0$
 in the present calculation.
If the calculated $4^+_1$ is assigned to the experimental state
 $(3^+,4^+)$ at 2083 keV, 
 the agreement between theory and experiment is good.
This state consists mainly of the
 1p-3h configurations (50\%).
A similar structure can be seen for $6^+_1$, $5^+_2$, $7^+_2$,
 and $8^+_{1,2}$, where one nucleon is
 excited across the shell gap mainly to the $p_{3/2}$ orbit.
The structure of $4^+_2$ is similarly of the 1p-3h type,
 although the excitation to the $f_{5/2}$ orbit is dominant.
Such an excitation is also large in $8^+_2$.

One can find a $1^+_2$ state in the calculated spectra
 at a reasonable excitation energy, which does not appear 
 in the shell-model results of Ref. \cite{co54}.
The most important configuration in this state is
 $\pi(f_{7/2})^6(p_{3/2})^1\nu(f_{7/2})^6(p_{3/2})^1$,
 but its probability is only 15\%.
Such 2p-4h configurations are 44\% in total,
 and there are other sizable core-excited components
 such as the 3p-5h (32\%) and 4p-6h (17\%) types.
Similar core-excitations also appear
in $2^+_2$, $3^+_2$ and $0^+_2$ ($T=1$).

The calculated M1 and E2 transition strengths are listed in
 Table \ref{tbl:co54}.
Since no experimental value is available,
 the present results are compared with those 
 of the shell-model calculation in Ref. \cite{co54}.
Both results agree within a factor of about 2
 for transitions between the states which consist mainly of
 the 0p-2h configurations.
The strong isovector M1 transitions
 predicted based on the ``quasideuteron'' picture \cite{qdc} are
 found also in the present large scale calculations.
Note that the effective spin g-factors $g_s^{\rm eff}=0.7g_s^{\rm free}$
 were used in Ref. \cite{co54}.
If we adopt the same g-factors in the present calculations,
 the $B$(M1) values are significantly reduced,
 for example, $B$(M1; $1^+_1\rightarrow 0^+_1$) becomes 1.25 W.u.
Thus the configuration mixing strongly affects the $B$(M1) value.

Several results related to the $1^+$ (1614), $3^+$ (2174) and $4^+$ (2852)
states are very different between two calculations,
 because the structure of the corresponding states are different
 as mentioned above.

\begin{table}
\caption{$B$(M1) and $B$(E2) for $^{54}$Co.
\label{tbl:co54}
}
\begin{ruledtabular}
\begin{tabular}{ccccr}
initial & final & multi- & Th-2. of Ref. \cite{co54} & th. \ \ \\ 
$J^{\pi}(E_x)$ & $J^{\pi}(E_x)$ & pole & (W.u) & (W.u) \\ \hline
   1$^+$(937)      &    0$^+$(0)        & M1 & $ 2.13                     $ & $     1.90 $ \\ 
   2$^+$(1446)     &    1$^+$(937)      & M1 & $ 2.33                     $ & $     2.36 $ \\ 
                   &                    & E2 & $ 0.033                     $ & $        0.078 $ \\ 
                   &    0$^+$(0)        & E2 & $ 10.6                     $ & $        7.7 $ \\ 
   3$^+$(1822)     &    2$^+$(1446)     & M1 & $ 2.36                     $ & $     1.92 $ \\ 
                   &                    & E2 & $ 0.091                     $ & $        0.155 $ \\ 
                   &    1$^+$(937)      & E2 & $ 10.6                     $ & $        4.2 $ \\ 
   3$^+$(2174)     &    2$^+$(1446)     & M1 & $ 0.016                     $ & $     0.0055 $ \\ 
                   &                    & E2 & $ 0.0012                     $ & $    0.0000 $ \\ 
                   &    1$^+$(1614)     & E2 & $ 1.1                     $ & $       16.0 $ \\ 
                   &    1$^+$(937)      & E2 & $ 0.29                     $ & $        0.06 $ \\ 
   4$^+$(2652)\footnote[1]{4$^+_3$ in the calculation.}     &    3$^+$(1822)     & M1 & $ 2.00                     $ & $     1.95 $ \\ 
                   &                    & E2 & $ 0.18                     $ & $        0.39 $ \\ 
                   &    2$^+$(1446)     & E2 & $ 7.8                     $ & $        3.6 $ \\ 
                   &    5$^+$(1887)     & M1 & $ 2.13                     $ & $     1.07 $ \\ 
                   &                    & E2 & $ 0.21                     $ & $        0.29 $ \\ 
   4$^+$(2852)\footnote[2]{4$^+_2$ in the calculation.}     &    3$^+$(1822)     & M1 & $ 0.002                     $ & $     0.0000 $ \\ 
                   &                    & E2 & $ 0.12                     $ & $       0.02 $ \\ 
                   &    5$^+$(1887)     & M1 & $ 0.004                     $ & $     0.0000 $ \\ 
                   &                    & E2 & $ 7.1                     $ & $        0.03 $ \\ 
\end{tabular}
\end{ruledtabular}
\end{table}

\subsubsection{$^{58}$Cu}
\label{sub2:cu58}

\begin{figure}
\includegraphics[width=85mm]{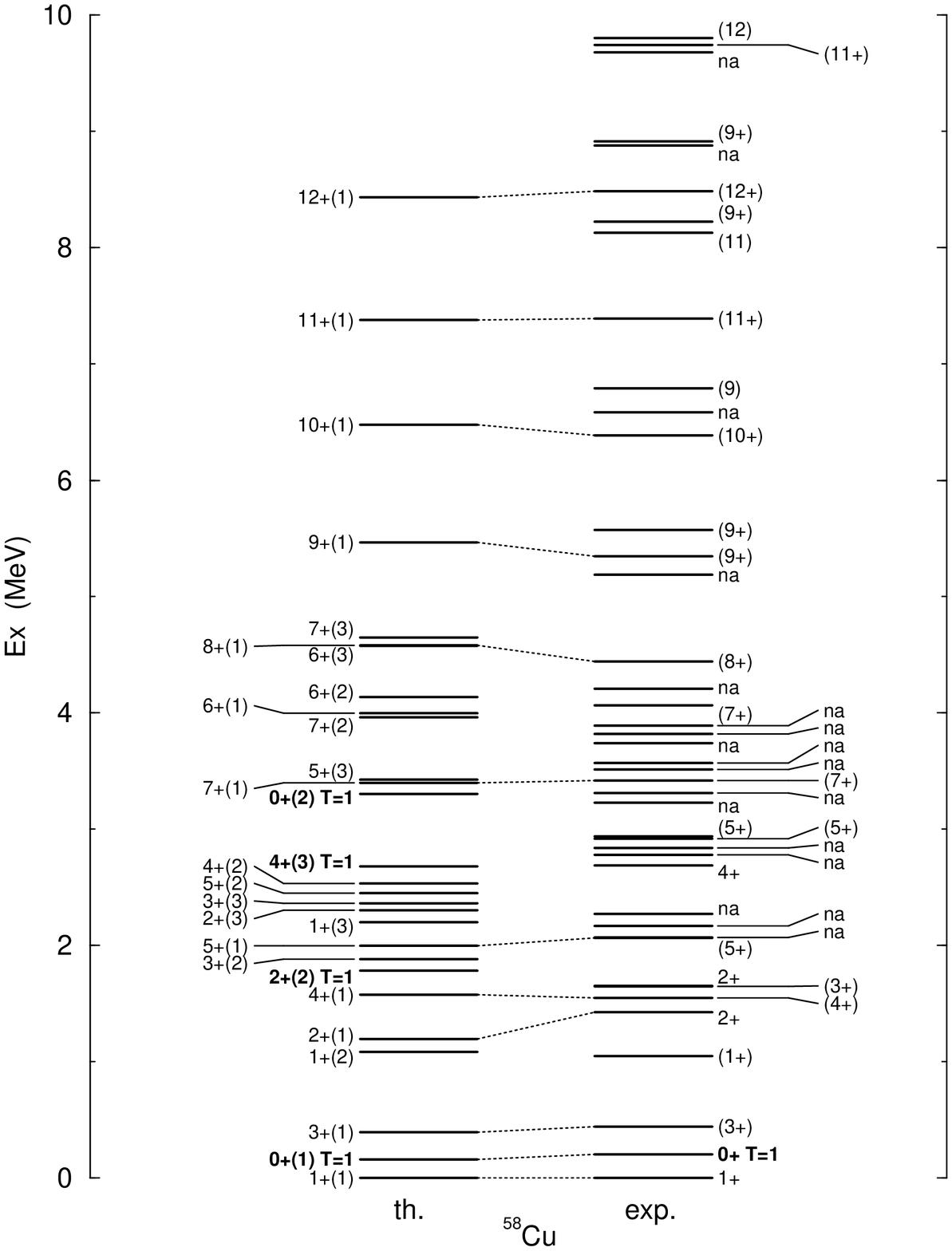}
\caption{Energy levels of $^{58}$Cu.
Experimental data are taken from
 \protect\cite{nds-cu58} and \protect\cite{rud-epj}.
Conventions are the same as in Fig. \ref{fig:mn53}.
\label{fig:cu58}
}
\end{figure}

In Fig. \ref{fig:cu58}, the calculated energy levels of $^{58}$Cu
 are compared with experimental data.
All experimental states up to 2.2 MeV as well as 
 the yrast states are shown.
The shell-model calculations have been carried out in the $t=6$
 truncated subspace.
The agreement between theory and experiment
 is satisfactory.
In the calculated spectrum, $0^+_1$, $2^+_2$, $4^+_3$ and
 $0^+_2$ are isospin $T=1$.
The $T_z=1$ members of the multiplets can in fact be seen
 in the calculated spectrum of $^{58}$Ni (see Fig. \ref{fig:ni58}).
The $T=0$ ground state $1^+_1$ is successfully reproduced.

Assuming an inert $^{56}$Ni core, one proton and
 one neutron are valence particles in the 
 $p_{3/2}$, $f_{5/2}$ and $p_{1/2}$ orbits.
Therefore $J\leq 5$ states are possible.
In the calculated wave functions, the yrast states
 up to $5^+$ are in fact dominated by such closed core configurations
 with probabilities of $44\sim 63\%$.
The $6^+_1$ state consists mainly of 3p-1h configurations (45\%).
The 4p-2h configurations are dominant in $7^+_1$, $8^+_1$, $9^+_1$ and $10^+_1$
 (41$\sim$59\%),
 and the 5p-3h configurations become the major components in 
 $11^+_1$ and $12^+_1$ ($\sim$40\%).
Thus it is natural that the $t=2$ shell-model results in Ref. \cite{rud-epj}
 show apparent discrepancies from experiment for $11^+$ and $12^+$.

In the present calculation, all states with $J\leq 10$ are
 dominated by either 3p-1h or 4p-2h configurations.
Therefore, shell-model calculations even in the severely truncated ($t\sim 2$)
 subspace give reasonable results at least near the yrast line.
In order to obtain further information on 
 the core-excitation, it is necessary to study higher spin states
 or highly excited states far from the yrast line.
From this viewpoint, recent precise measurements \cite{cu58gt} of the
 Gamow-Teller strength distribution in this nucleus
 up to high excitation energy provide quite important information.
The GXPF1 reproduces the measured strengths around $E_x\sim$4 MeV
 reasonably well, which are missing in the prediction of the KB3G interaction.
 
In the calculated wave function of the ground state $1^+_1$,
 the configuration $\pi(p_{3/2})^1\nu(p_{3/2})^1$ takes the largest weight (18\%) and
 the second largest components are $\pi(p_{3/2})^1\nu(p_{1/2})^1$
 and $\pi(p_{1/2})^1\nu(p_{3/2})^1$ (13\% for each).
This is not very far from the ``quasideuteron'' picture \cite{qdc},
 although the configuration $\pi(p_{3/2})^1\nu(p_{3/2})^1$ is distributed
 among several $1^+$ states such as $1^+_2$ (15\%) and $1^+_3$ (20\%).
However, it contradicts with the shell-model results in Ref. \cite{rud-epj},
 where $\pi(p_{3/2})^1\nu(f_{5/2})^1$ and $\pi(f_{5/2})^1\nu(p_{3/2})^1$
 are dominant (20\% for each).

The calculated $B$(M1) and $B$(E2) values are
 compared with experimental data in Table \ref{tbl:cu58}.
Although the experimental data are limited, 
 the agreement between theory and experiment is reasonable.
More detailed discussions are found in Ref. \cite{lis-cu58}.

\begin{table}
\caption{$B$(M1) and $B$(E2) for $^{58}$Cu.
Experimental data are taken from
 \protect\cite{cu58} and \protect\cite{nds-cu58}.
\label{tbl:cu58}
}
\begin{ruledtabular}
\begin{tabular}{ccccr}
initial & final & multi- & exp. & th. \ \ \\ 
$J^{\pi}(E_x)$ & $J^{\pi}(E_x)$ & pole & (W.u) & (W.u) \\ \hline
   0$^+$(203)      &    1$^+$(0)        & M1 & $                      $ & $     0.8473 $ \\ 
   1$^+$(1051)     &    0$^+$(203)      & M1 & $ 0.46(10)             $ & $     0.1358 $ \\ 
   2$^+$(1652)     &    0$^+$(203)      & E2 & $ 14.8(37)             $ & $       10.2 $ \\ 
                   &    3$^+$(444)      & M1 & $ 0.33(11)             $ & $     0.5220 $ \\ 
                   &                    & E2 & $                      $ & $        0.1 $ \\ 
                   &    1$^+$(1051)     & M1 & $ 0.17(6)              $ & $     0.1780 $ \\ 
                   &                    & E2 & $                      $ & $        0.0 $ \\ 
\end{tabular}
\end{ruledtabular}
\end{table}

\subsection{GXPF1 vs. GXPFM}
\label{sub:m-vs-1}

In the previous subsections, it has been shown that
 GXPF1 properly describes the low-lying core-excitations.
As discussed above, the effects of the core-excitations
 can be observed directly in the low-lying energy spectra
 of semi-magic nuclei,
 and therefore the description of such core-excited states
 plays a crucial role for the
 evaluation of the effective interaction.
From this viewpoint, in this subsection,
 we discuss a specific problem with the GXPFM interaction:
 an interaction derived by adding empirical corrections to the
 microscopic G interaction only in the
 monopole part and pairing matrix elements
 (see subsection \ref{sub:monofit}).
With the FDA* estimate, 
 the description of energy data by GXPFM
 appears to be poor especially for the yrare states of semi-magic nuclei.

\begin{figure}
\includegraphics[width=85mm]{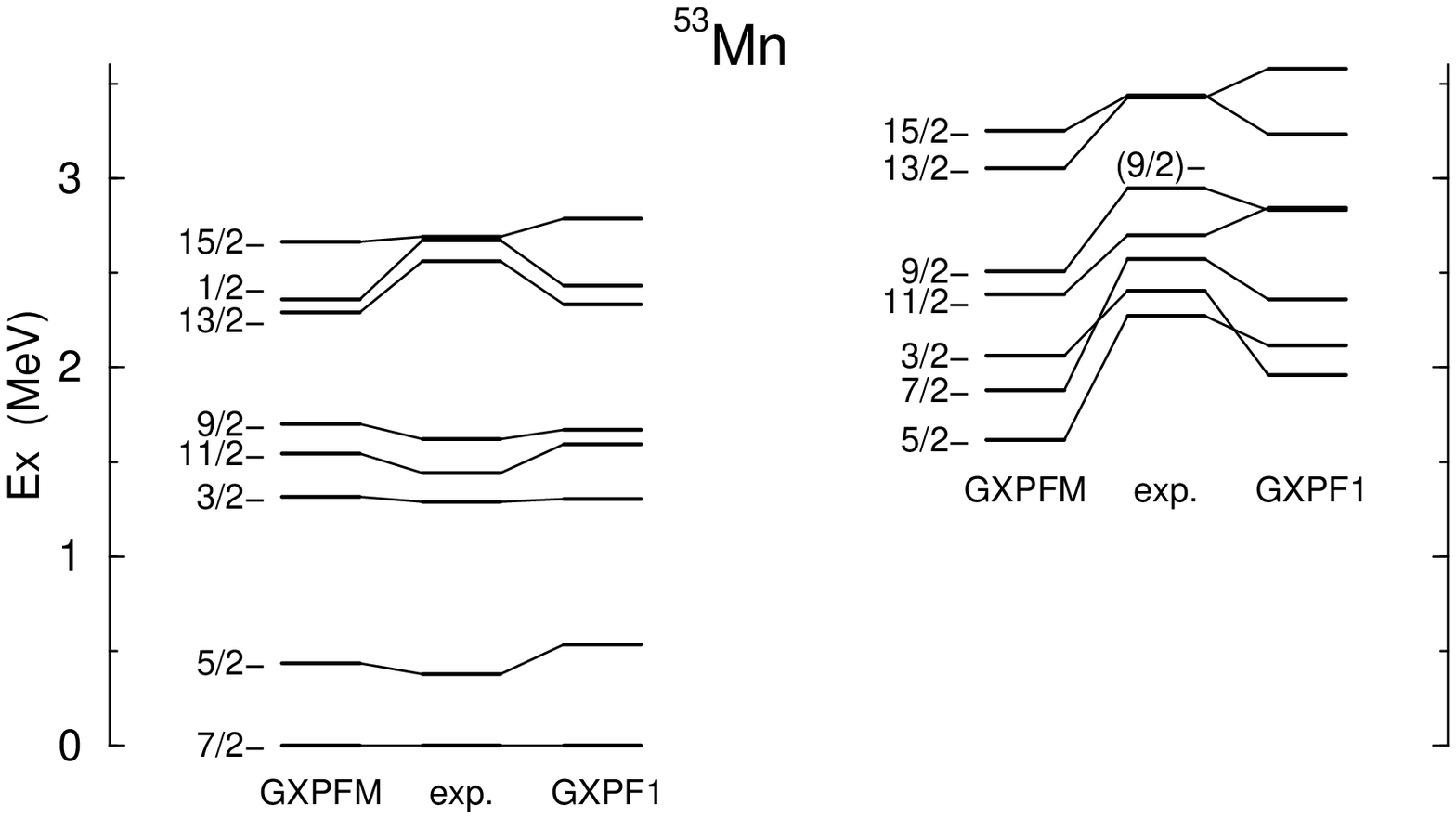}\vspace*{0.5cm}
\includegraphics[width=85mm]{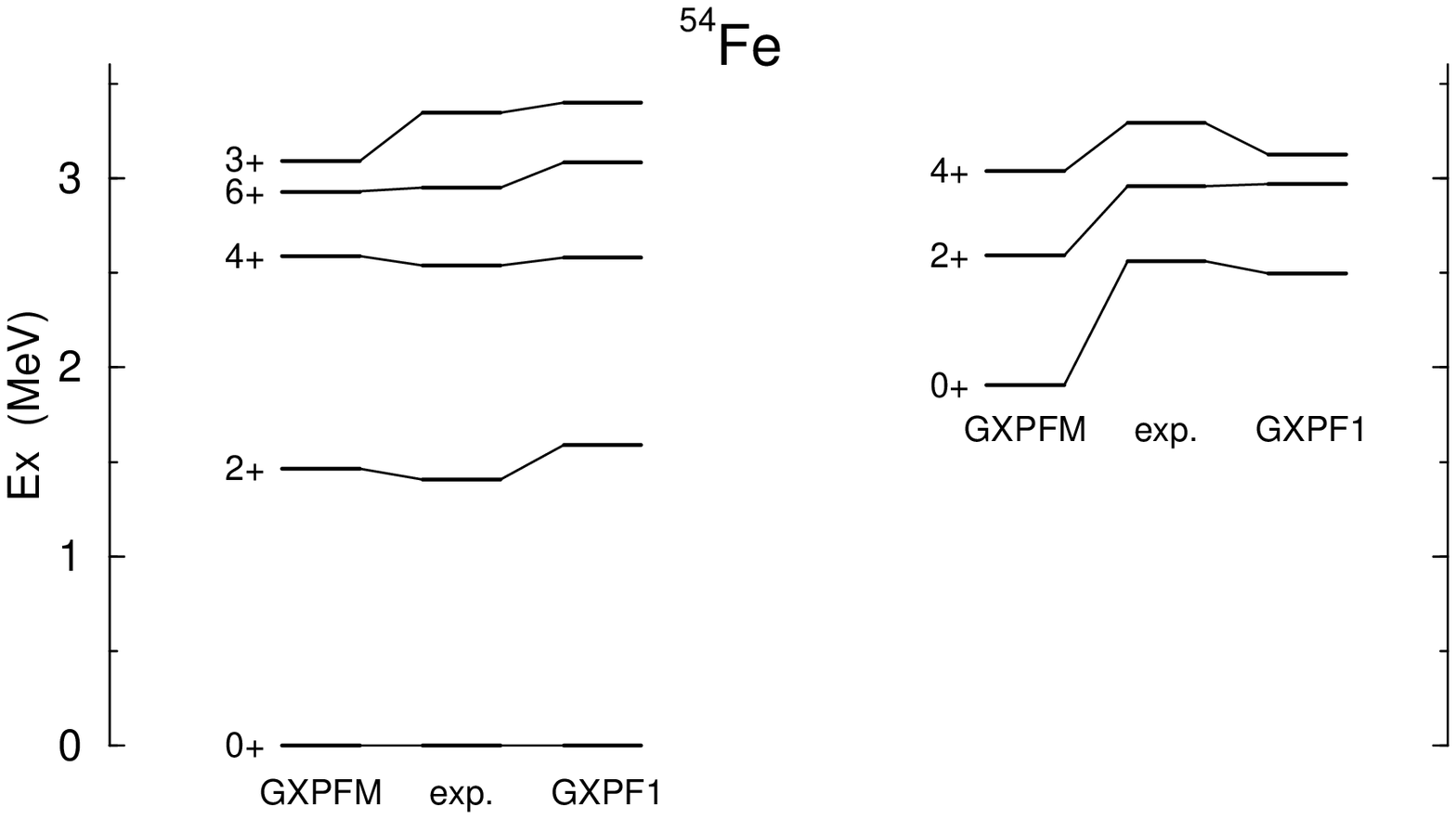}
\caption{Comparison of low-lying energy levels of
 (upper) $^{53}$Mn and (lower) $^{54}$Fe
 between the experimental data and the shell-model predictions by
 GXPFM and GXPF1.
\label{fig:gxm-lev}
}
\end{figure}

As examples, we consider the low-lying
 energy levels of $^{53}$Mn and $^{54}$Fe, which are
 shown in Fig. \ref{fig:gxm-lev} for
 yrast and yrare states.
Details of the shell-model calculations are
 the same as those described in subsection \ref{sub:semi}.
For a comparison, the results of GXPF1 are also shown.
It can be clearly seen that, in both nuclei,
 the yrast states are described nearly equally well by both
 GXPF1 and GXPFM
 at least in the region of $E_x \leq 3.5$ MeV.
On the other hand, several yrare states such as 
 $5/2^-_2$ and $7/2^-_2$ in $^{53}$Mn and $0^+_2$ in $^{54}$Fe
 are predicted to be too low by GXPFM.
According to the analysis of the shell-model wave functions,
 these states are dominated by neutron core-excited configurations.
Typically 40\% of these wave functions consist of
 neutron 2p-2h configurations relative to the $N=28$ core.
In the case of GXPF1, such states are predicted with a similar structure,
 while their excitation energies are closer to the experiment.

This fact suggests that the ``monopole + pairing'' correction
 works well for the description of the states in
 the lowest configuration and those with 1p-1h excitations relative to it,
 but it is not sufficient for treating the states which 
are dominated by the  2p-2h or more excited configurations.
This is natural because, in the former states,
 the structure is expected to be
 ``single-particle'' like, especially for the semi-magic nuclei,
 and therefore such states can be described well by adjusting
the monopole part only.
On the other hand, in the latter states,
 the contribution of the multipole part should be more important,
 and multipole parts must be modified for their proper descriptions.
Thus, the core-excitations provides us with a clue to
 investigate corrections to the interaction
that go beyond monopole plus pairing.

\subsection{Quadrupole corrections}
\label{sub:quad-cor}
In subsection \ref{sub:correc}, it has been shown that
 major modifications of the microscopic interaction can be
 found in the diagonal parts which are not necessarily
 of monopole character.
We have also shown in the previous subsection that
 the ``monopole + pairing'' correction is
 insufficient for describing the property of core-excitations.
Such insufficiency can be seen explicitly in the energy spectra
 of semi-magic nuclei.

\begin{figure}
\includegraphics[width=85mm]{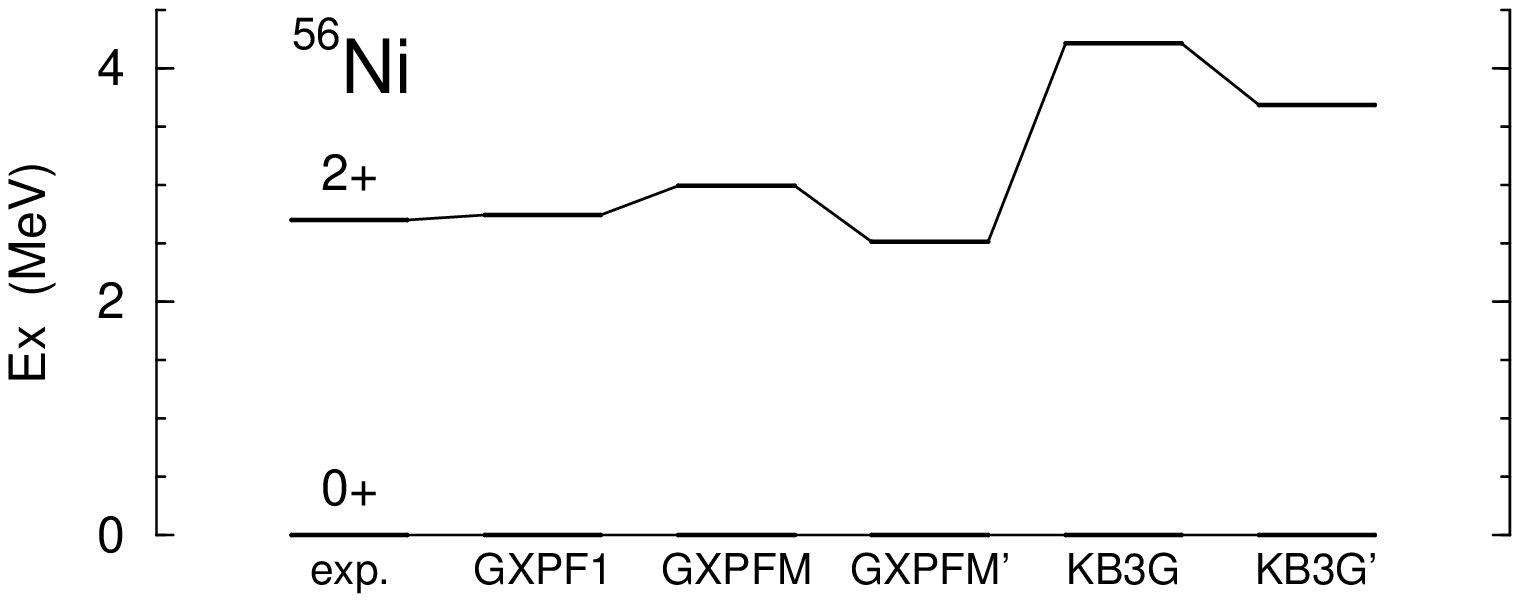}\vspace*{0.5cm}
\includegraphics[width=85mm]{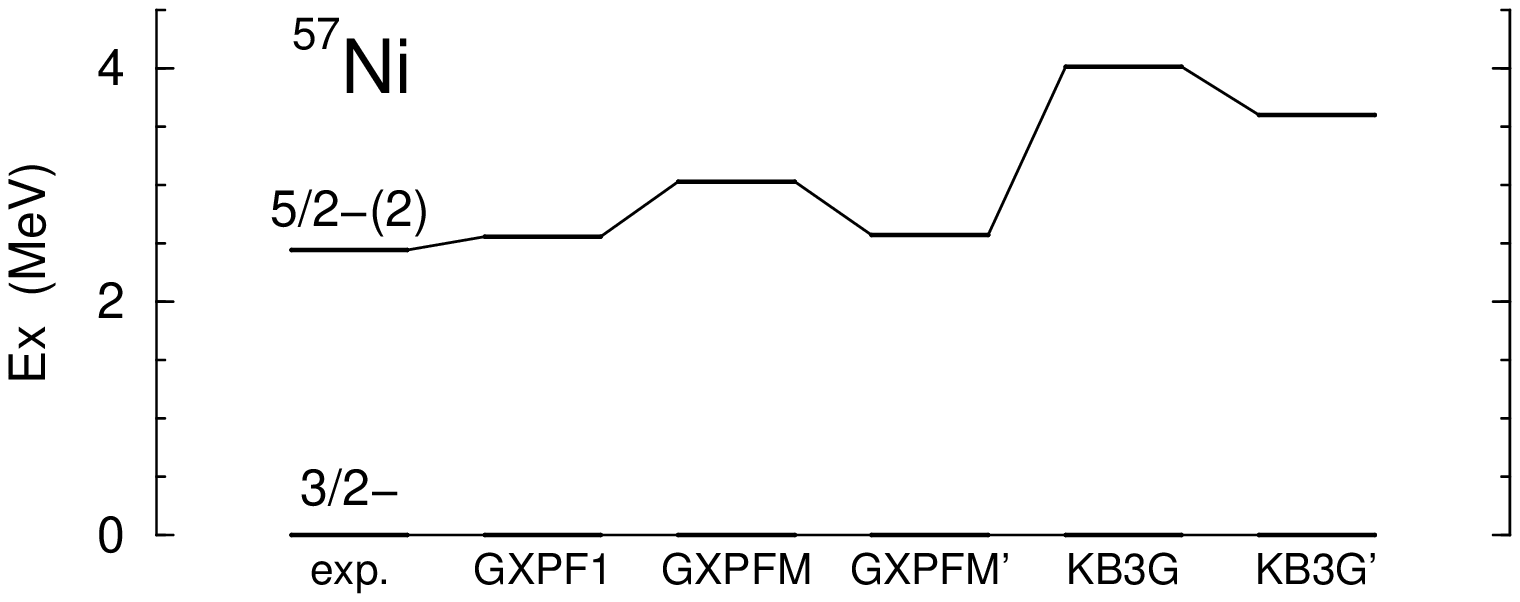}
\caption{
Comparison of the excitation energy of
 (upper panel) $2^+_1$ in $^{56}$Ni and
 (lower panel) $5/2^-_2$ in $^{57}$Ni
 between the experimental data and the results of
 various effective interactions.
The shell-model calculations were carried out
 in the $t=7$ and 6 truncated subspaces for 
 $^{56}$Ni and $^{57}$Ni, respectively.
\label{fig:ni5657}
}
\end{figure}

Figure \ref{fig:ni5657} shows a part of the low-lying energy levels
 of $^{56}$Ni and $^{57}$Ni, where the results of
 GXPF1, GXPFM and KB3G are compared.
The latter two interactions
 predict too high excitation energies for the $2^+_1$ state
in $^{56}$Ni
 and the $5/2^-_2$ state in $^{57}$Ni. Both of 
these excited states are dominated by the configuration with
one nucleon excited
 from the $f_{7/2}$ orbit to the $p_{3/2}$ orbit.

\begin{figure}
\includegraphics[width=85mm]{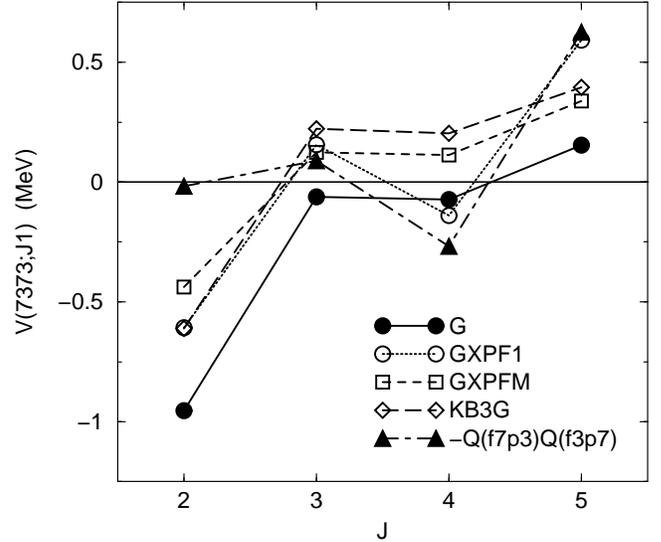}
\caption{
Comparison of two-body matrix elements $V(7373;J1)$
 between various effective interactions.
\label{fig:v7373j1}
}
\end{figure}

We have found that in this case the most relevant matrix elements are
 $V(7373;J1)$, which are compared
 in Fig. \ref{fig:v7373j1}
 for G, GXPF1, GXPFM and KB3G.
It can be seen that the difference between G
 and GXPFM is primarily of monopole character (constant shift).
Similar monopole corrections are present in KB3G,
 although its original microscopic interaction is not G
 but KB, reflecting that G and KB are 
 very close for these matrix elements.
On the other hand, GXPF1 shows a rather different $J$-dependence
 from these ``monopole-corrected'' interactions GXPFM and KB3G,
 especially for the $J=4$ and $J=5$ matrix elements.
The former is more attractive
 and the latter more repulsive by about 0.3 MeV. 
Note that these differences keep the monopole centroid
 quite similar value for these interactions
 as shown in Figs. \ref{fig:centroid} and \ref{fig:centroid-gxpfm}.

In order to examine to what extent
 such differences affect the energy spectra,
 similar calculations have been carried for
 both GXPFM and KB3G interactions,
 by modifying the $J=4$ matrix element to be more attractive
 by 0.300 MeV and $J=5$ to be more repulsive by 0.245 MeV
 so that the monopole centroid is kept unchanged.
The modified interactions are denoted as GXPFM' and KB3G', respectively.
The results are shown in Fig. \ref{fig:ni5657}.
It is clearly seen that
 the description is improved in both GXPFM' and KB3G',
 although it is not enough for the latter.
This fact highlights the importance of
modifications of the microscopic interaction that
go beyond the monopole type.

The above $J$-dependent modifications in GXPF1 can 
be understood to a good extent
in terms of a 
 quadrupole-quadrupole interaction 
 $[(f_{7/2})^\dagger(\tilde{p}_{3/2})]^{(2)}
 \cdot[(p_{3/2})^\dagger(\tilde{f}_{7/2})]^{(2)}$.
The matrix elements of this interaction are shown
 in Fig. \ref{fig:v7373j1},
 which give similar $J$-dependence to that of the difference 
between GXPF1 and GXPFM.
Indeed, the $V(7373;J1)$'s can be obtained 
by recoupling this interaction with the strength  
 $-0.81$ MeV for GXPF1 together with other multipoles with minor
contributions, while this strength turns out to be $-0.32$, $-0.34$, 
and $-0.38$ MeV for GXPFM, KB3G and G, respectively.
This quadrupole correction can arise from a core-polarization
diagram where the external lines are $f_{7/2}$ and $p_{3/2}$
with exchange between them and the bubble of $J^{\pi}=2^+$ is involved.
If this is the case, the coupling should be particularly strong in this
channel and/or the energy denominator should smaller, as compared to
what is assumed in the microscopic calculation of effective interaction.  
This is an intriguing problem of the
effective interaction, and further studies are needed.

\section{Summary}
\label{sec:summary}

Summarizing this paper,
 the effective interaction for $pf$-shell nuclei, GXPF1, 
 has been examined and tested from various viewpoints.
It was obtained by adding empirical corrections to
 the microscopic interaction derived from the nucleon-nucleon potential.
The corrections were determined
 through the systematic fitting to experimental energy data.
The most significant modifications are those 
for the monopole and pairing parts of the Hamiltonian
The monopole plus pairing  corrections give a substantial 
improvement, but are not sufficient for the description 
core-excitations
 over the $N$ or $Z=28$ shell gap. Other modifications 
are necessary including some which can be related to the
quadrupole-quadrupole component.

The analysis of the ground-state wave functions obtained by GXPF1
 shows that
 the assumption of an inert $^{56}$Ni core is
 not realistic even for the Ni isotopes.
Thus,
 the amount of core-excitations strongly depends on
 both valence proton and neutron numbers
 relative to the core.

The calculated binding energies agree with the
 experimental data quite well over a wide mass range
 even for many nuclei which were not included in the fit.
However, in neutron-rich nuclei ($N>34$) around $Z\sim$24,
 the difference between theory and experiment
 becomes larger, suggesting
 a need for including the $g_{9/2}$ orbit for describing
 large deformation.
Such a discrepancy can also be seen in
 the calculated $2^+_1$ excitation energy for $^{60}$Cr,
 which is predicted too high
 in comparison to the recent experimental data.

The description of the magnetic dipole and electric quadrupole
 moments is basically successful with a few exceptions.
The most remarkable difference between theory and experiment
 can be found in the $2^+_1$ states of Zn isotopes,
 which also suggests influence of the $g_{9/2}$ orbit.
As for the electric quadrupole moments,
 more precise experimental data are desired to discuss
 the quality of description especially for Ni and Zn isotopes.

The energy spectra and electro-magnetic transition matrix elements
 for $N$ or $Z=28$ semi-magic nuclei
 have been calculated
 and compared with experimental data.
The effects of the core-excitation across $N$, $Z=28$ shell gap
 can be seen already in
 the non-yrast states around $E_x\sim$ 3 MeV
 as well as high spin yrast states.
The present results show the reliability of GXPF1
 even in the cases that
 the core-excitation plays a crucial role.
Considering the current experimental and theoretical situations,
 it would be difficult to explore such significantly core-excited
 states in other nuclei with more ``valence'' particles or holes
 relative to $^{56}$Ni core,
 due to the explosive increase of
 the low-lying level density.
Therefore the description of $^{56}$Ni and neighboring nuclei
 is an important test of the effective interaction.

It has been shown that the reasonable description of $N=Z$ odd-odd nuclei
 can be obtained by GXPF1,
 indicating its proper isospin dependent structure,
 although the experimental data are limited especially
 for the transitions.
 
Most of shell-model calculations were done in the conventional
method so as to reach higher accuracy.  
The truncations were carefully examined to be accurate enough and
for the data we compare with.
In some shape coexistence cases, strongly deformed states appear
at lower energies, where the Monte-Carlo Shell Model still produces more
accurate results as compared to large-scale conventional calculations
with high (i.e., unrestrictive) truncations.  
Thus, one has to be cautious in using
conventional calculations for such cases. 
This may be an important lesson for further studies on
heavier nuclei with more developed collectivity.

Further applications to unexplored regimes of large proton/neutron numbers or
 high excitation energy is of great interest.
Future experiments
 will test the predictions
 and provide guidance for further improvements in the interaction.

\begin{acknowledgments}
This work has been supported in part 
by Grant-in-Aid for Specially Promoted Research  (13002001)
from the Japanese Ministry of Education, Science, Culture, Sport and
Technology.
Shell-model calculations were performed largely on
the parallel computer system at CNS and on the Alphleet computer of RIKEN.
The CNS system has been installed
by the above-mentioned grant, and 
these computers are operated under the joint large-scale
nuclear-structure calculation project by RIKEN and CNS. BAB acknowledges support
from NSF grants PHY-0244453 and INT-0089581.
\end{acknowledgments}

\end{document}